\RequirePackage{lineno}
\documentclass[twocolumn,showpacs,superscriptaddress,amsmath,amssymb,nofootinbib]{revtex4-2}
\usepackage{graphicx}
\usepackage{subcaption}
\usepackage{epsfig}
\usepackage{overpic}
\usepackage{dcolumn}
\usepackage{ulem}
\usepackage{bm}
\usepackage{color}
\usepackage{lineno}
\usepackage{xspace}
\usepackage{multirow}
\usepackage{epstopdf}
\usepackage{xcolor}
\usepackage{soul}
\usepackage{verbatim}
\usepackage{enumitem}
\usepackage{todonotes}
\usepackage{cancel}
\usepackage[toc,page]{appendix}
\usepackage[colorlinks,allcolors=black]{hyperref}
\usepackage{diagbox}

\begin{document}
\include{def-com}
\title{\boldmath Revealing the nature of hidden charm pentaquarks with machine learning}



\author {Zhenyu Zhang}
\address{Guangdong Provincial Key Laboratory of Nuclear Science, Institute of Quantum Matter, South China Normal University, Guangzhou 510006, China}
\affiliation{Guangdong-Hong Kong Joint Laboratory of Quantum Matter, Southern Nuclear Science Computing Center, South China Normal University, Guangzhou 510006, China}

\author{Jiahao Liu}
\address{Guangdong Provincial Key Laboratory of Nuclear Science, Institute of Quantum Matter, South China Normal University, Guangzhou 510006, China}
\affiliation{Guangdong-Hong Kong Joint Laboratory of Quantum Matter, Southern Nuclear Science Computing Center, South China Normal University, Guangzhou 510006, China}

\author{Jifeng Hu}\email{hujf@m.scnu.edu.cn}
\address{Guangdong Provincial Key Laboratory of Nuclear Science, Institute of Quantum Matter, South China Normal University, Guangzhou 510006, China}
\affiliation{Guangdong-Hong Kong Joint Laboratory of Quantum Matter, Southern Nuclear Science Computing Center, South China Normal University, Guangzhou 510006, China}

\author{Qian Wang}\email{qianwang@m.scnu.edu.cn}
\address{Guangdong Provincial Key Laboratory of Nuclear Science, Institute of Quantum Matter, South China Normal University, Guangzhou 510006, China}
\affiliation{Guangdong-Hong Kong Joint Laboratory of Quantum Matter, Southern Nuclear Science Computing Center, South China Normal University, Guangzhou 510006, China}

\author{Ulf-G. Mei{\ss}ner}\email{meissner@hiskp.uni-bonn.de}
\affiliation{Helmholtz-Institut f\"ur Strahlen- und Kernphysik and Bethe Center for Theoretical Physics, 
 Universit\"at Bonn, D-53115 Bonn, Germany}
\affiliation{Institute for Advanced Simulation, Institut f\"ur Kernphysik and J\"ulich Center for Hadron Physics,
Forschungszentrum J\"ulich, D-52425 J\"ilich, Germany}
\affiliation{Tbilisi State University, 0186 Tbilisi, Georgia}
 
\date{\today}

\begin{abstract}
We study the nature of the hidden charm pentaquarks, 
i.e. the $P_c(4312)$, $P_c(4440)$ and $P_c(4457)$, with a neural network approach in 
pionless effective field theory. In this framework, the normal $\chi^2$ fitting approach
cannot distinguish the quantum numbers of the $P_c(4440)$ and $P_c(4457)$.
In contrast to that, the neural network-based approach can discriminate them, which still cannot be seen as a proof of the spin of the states since pion exchange is not considered in the approach. 
In addition, we also illustrate the role of each experimental data bin of the invariant $J/\psi p$ mass distribution
on the underlying physics in both neural network and fitting methods.
Their similarities and differences demonstrate that neural network methods
can use data information more effectively and directly.
This study provides more insights about how the neural network-based approach 
predicts the nature of exotic states from the mass spectrum. 
\\ \par\
\end{abstract}
\maketitle

\section{Introduction}
\label{sec:introduction}

In the last two decades, we have witnessed the emergence of 
tens of candidates for the so-called exotic hadrons  which are beyond the 
conventional meson and baryon configurations, such as the 
hidden charm pentaquarks~\cite{LHCb:2015yax,LHCb:2016ztz,LHCb:2019kea},
the fully charmed tetraquark $X(6900)$~\cite{LHCb:2020bwg, CMS:2022yhl, ATLAS:2022hhx}, or the
doubly charmed tetraquark $T_{cc}^+$~\cite{LHCb:2021vvq,LHCb:2021auc}.
The first hidden charm pentaquarks were reported by the LHCb Collaboration
in 2015 in Ref.~\cite{LHCb:2015yax} by observing the $J/\psi p$ invariant mass via the process $\Lambda_b\to J/\psi p K^-$. 
Four years later, the  $P_c(4312)$, $P_c(4440)$ and $P_c(4457)$ were reported with an order of magnitude larger
luminosity in Ref.~\cite{LHCb:2019kea}.
Since then, many interpretations have been proposed for these states, e.g. compact multi-quark,
hadronic molecule, hybrid, and triangle singularity~\cite{Chen:2016spr,Chen:2016qju,Lebed:2016hpi,Guo:2017jvc,Dong:2017gaw,Albuquerque:2018jkn,Brambilla:2019esw,Liu:2019zoy,Guo:2019twa,Liu:2019zoy,JPAC:2021rxu,Zou:2021sha,Mai:2022eur}.
Traditionally, judging whether a model works or not
 is to compare it with the experimental data. 
However, such a top-down approach makes the conclusion model-dependent.
Physicists are thus trying to find bottom-up approaches~\cite{Ng:2021ibr}
to obtain a definitive conclusion. 
One direction is to use 
machine learning to explore hadron properties,
which benefits from the evolution of computing capabilities
during the last decades. For instance, 
genetic algorithms \cite{Ireland:2004kp,Fernandez-Ramirez:2008ixe,NNPDF:2014otw} and neural networks \cite{Forte:2002fg,Rojo:2004iq,Keeble:2019bkv,Adams:2020aax} can be utilized
to explore the nature of hadrons or the properties of nuclei. 
Machine learning has been widely used in nuclear 
physics~\cite{Niu:2018csp,Niu:2018trk,Ma:2020mbd,Kaspschak:2020ezh,Kaspschak:2021hbc,Bedaque:2021bja,Dong:2022wkd},
 high energy nuclear physics~\cite{Baldi:2016fql,Baldi:2014kfa,Boehnlein:2021eym}, experimental data analysis~\cite{Guest:2018yhq,WHITESON20091203}, and 
 theoretical physics~\cite{doi:10.1126/science.aag2302,doi:10.1126/science.abk3333},
even to discover the physical principles
 underneath the experimental data~\cite{Ng:2021ibr,Lu:2022joy,Zhang:2022uqk}. 
However, its application in hadron physics
is in the early stage.
A preliminary attempt was made in Refs.~\cite{Sombillo:2021rxv,Sombillo:2021yxe,Ng:2021ibr,Liu:2022uex} for the hadronic molecular picture~\cite{Guo:2017jvc},
 which is one of the natural explanations of near-threshold peaks. 
Refs.~\cite{Sombillo:2020ccg,Liu:2022uex}
 demonstrate that deep learning can be applied 
to classify poles and regress the model parameters in the one-channel coupling case.

Here, extending and improving the works~\cite{Sombillo:2021rxv,Sombillo:2021yxe}, 
we consider the multi-channel coupling case.
We take the hidden charm pentaquarks,  
i.e. $P_c(4312)$, $P_c(4440)$ and $P_c(4457)$~\cite{LHCb:2019kea}
as examples, to achieve the following goals:
\begin{itemize}
\item[(1)] In the $\Sigma_c^{(*)}\bar{D}^{(*)}$ hadronic molecular picture~\cite{Liu:2019tjn,Du:2019pij,Du:2021fmf},
both the $P_c(4440)$ and the $P_c(4457)$ are related to the 
$\Sigma_c\bar{D}^*$ threshold. Their spin-parity ($J^P$) could be
either $\frac 12^-,~\frac32^-$ (solution~A) or $\frac 32^-,~\frac 12^-$ (solution~B)~\cite{Liu:2019tjn,Du:2019pij,Du:2021fmf}, from the viewpoint of pionless effective field theory (EFT). In solution~A, the lower mass
hidden charm pentaquark has lower spin, and vice versa in solution~B.
Although the $\chi^2$ fitting cannot distinguish the two solutions,
we try to find out which solution is preferred in machine learning.

\item[(2)] In the traditional fitting approach, the physics is embedded
in the model parameters. One has to extract the model parameters
from the experimental data and further extract the physical quantities of interest.
We use a neural network (NN) based approach to extract the pertinent physical quantities 
directly and illustrate the role of each experimental data point (here the bins in the
mass distributions) in the physics.
\end{itemize}

We have used  one particular NN for the above goals and demonstrated that
the NN-based approach  consistently favors solution A over B
compared to the fitting approach, and clearly gives the properties of the poles in the multi-channel case.

\section{Model Description}
\label{sec:model}

The $P_c(4312)$ and  $P_c(4440)/P_c(4457)$ states are close to the $\Sigma_c\bar{D}$ and $\Sigma_c\bar{D}^*$ thresholds, respectively,
making them prime candidates for hadronic molecules. 
In the heavy quark limit, these hidden charm pentaquarks are related to the scattering between the $(\Sigma_c,\Sigma_c^*)$ and the $(\bar{D},\bar{D}^*)$ heavy quark spin doublets. Note that the heavy quark symmetry breaking effect is about $(m_\rho/m_D)^2\sim 25\%$ in the charm sector, with $m_\rho$ and $m_D$ the masses of the $\rho$ and $D$ mesons. Since both the $P_c(4440)$ and $P_c(4457)$ are related to the $\Sigma_c\bar{D}^*$ channel, their mass difference comes from the spin-spin interaction between the $\Sigma_c$ baryon and the $\bar{D}^*$ meson for a given total spin, which should vanish in the exact heavy quark limit. Thus in the heavy quark limit  the $P_c(4440)$ and $P_c(4457)$ should coincide with each other. Their hyperfine splitting stems from the heavy quark symmetry breaking effect. Consequently, we consider the heavy quark symmetry breaking effect from the mass difference in the same doublet.
By constructing the interactions between the two doublets, i.e. the $(\Sigma_c,\Sigma_c^*)$ and the $(\bar{D},\bar{D}^*)$ doublets, in the EFT
respecting the heavy quark spin symmetry (HQSS), Refs.~\cite{Du:2021fmf,Du:2019pij} extracted the pole
positions of the seven hidden charm pentaquarks by fitting the 
$J/\psi p$ invariant mass distribution for the process $\Lambda_b\to J/\psi p K^-$~\footnote{Note that there is additional dynamic $\Lambda_c\bar{D}^{(*)}$ channel in Ref.~\cite{Du:2021fmf} in comparison with Ref.~\cite{Du:2019pij}. However, the parameters of the $\Lambda_c\bar{D}^{(*)}$ channel are out-of-control due to the absence of the experimental data in the $\Lambda_c\bar{D}^*$ channel and the properties of the poles are driven by the $\Sigma_c^{(*)}\bar{D}^{(*)}$ channel. As the result, we only consider the elastic $\Sigma_c^{(*)}\bar{D}^{(*)}$ channel and the inelastic $\eta_c p$, $J/\psi p$ channels in our framework. }. 
Following the same procedure, we
consider the $\Sigma_c^{(*)}\bar{D}^{(*)}$ and $J/\psi p$, $\eta_c p$ channels
as elastic and inelastic channels, respectively. The potential quantum numbers of
hidden charm pentaquarks are $\frac 12^-,~\frac 32^-$ and $\frac 52^-$ which correspond
to the scattering of the $\Sigma_c\bar{D}-\Sigma_c\bar{D}^*-\Sigma_c^*\bar{D}^*$, 
$\Sigma_c^*\bar{D}-\Sigma_c\bar{D}^*-\Sigma_c^*\bar{D}^*$ and $\Sigma_c^*\bar{D}^*$,
respectively. In the heavy quark limit, their interaction only depends on the light degrees of freedom~\cite{Voloshin:2011qa}. 
To extract their interactions, one can expand the two-hadron basis in terms of the heavy-light basis $|H\otimes L\rangle$, with $H$ and $L$ the heavy and light degrees of freedom, respectively,
\begin{eqnarray}
\begin{pmatrix}
   |\Sigma_c\bar{D}\rangle \\[6pt]
   |\Sigma_c\bar{D}^*\rangle \\[6pt]
   |\Sigma_c^*\bar{D}^*\rangle
\end{pmatrix}_\frac{1}{2}=\begin{pmatrix}
   \frac{1}{2} & \frac{1}{2\sqrt{3}} & \sqrt{\frac{2}{3}}\\[6pt]
   \frac{1}{2\sqrt{3}} & \frac{5}{6} & -\frac{\sqrt{2}}{3}\\[6pt]
   \sqrt{\frac{2}{3}} & -\frac{\sqrt{2}}{3} & -\frac{1}{3}
\end{pmatrix}
\begin{pmatrix}
   |0\otimes\frac{1}{2}\rangle \\[6pt]
   |1\otimes\frac{1}{2}\rangle \\[6pt]
   |1\otimes\frac{3}{2}\rangle \\
\end{pmatrix},
\end{eqnarray}
\begin{eqnarray}
\begin{pmatrix}
   |\Sigma_c\bar{D}^*\rangle \\[6pt]
   |\Sigma_c^*\bar{D}\rangle \\[6pt]
   |\Sigma_c^*\bar{D}^*\rangle
\end{pmatrix}_\frac{3}{2}=\begin{pmatrix}
   \frac{1}{\sqrt{3}} & -\frac{1}{3} & \frac{\sqrt{5}}{3}\\[6pt]
   -\frac{1}{2} & \frac{1}{\sqrt{3}} & \frac{1}{2}\sqrt{\frac{5}{3}}\\[6pt]
   \frac{1}{2}\sqrt{\frac{5}{3}} & \frac{\sqrt{5}}{3} & -\frac{1}{6}
\end{pmatrix}
\begin{pmatrix}
   |0\otimes\frac{3}{2}\rangle \\[6pt]
   |1\otimes\frac{1}{2}\rangle \\[6pt]
   |1\otimes\frac{3}{2}\rangle \\
\end{pmatrix},
\end{eqnarray}
\begin{eqnarray}
    |\Sigma_c^*\bar{D}^*\rangle_\frac{5}{2}=|1\otimes\frac{3}{2}\rangle.
\end{eqnarray}

Here, the subscripts $\frac 12$, $\frac 32$ and $\frac 52$ give the total spin of the two-hadron system. In the HQSS, as the interaction only depends on the light degrees of freedom~\cite{Voloshin:2011qa}, the scattering amplitude is described by the two parameters~\cite{Du:2021fmf} 
\begin{eqnarray}
C_{\frac 12}&\equiv& \langle H\otimes \frac 12\left|\hat{\mathcal{H}}\right|H\otimes \frac 12\rangle,\\
C_{\frac 32}&\equiv& \langle H\otimes \frac 32\left|\hat{\mathcal{H}}\right|H\otimes \frac 32\rangle,
\end{eqnarray}
with the subscripts corresponding to the spin of the light degrees of freedom.
The coupling for the $S$-wave and $D$-wave inelastic channels,
i.e. the $J/\psi p$ and $\eta_c p$ channels, 
are described by two parameters
\begin{eqnarray}
g_S&\equiv& \langle 1\otimes \frac 12 \left|\hat{\mathcal{H}}\right|J/\psi p\rangle_S= \langle 0\otimes \frac 12 \left|\hat{\mathcal{H}}\right|\eta_c p\rangle_S,\\
g_Dk^2&\equiv& \langle 1\otimes \frac32 \left|\hat{\mathcal{H}}\right|J/\psi p\rangle_D= \langle 0\otimes \frac32 \left|\hat{\mathcal{H}}\right|\eta_c p\rangle_D,
\end{eqnarray}
respectively. The weak bare production amplitude can also be parameterized in terms of the $|H\otimes L\rangle$ basis
\begin{eqnarray}
\mathcal{F}_i^J\equiv\langle\Lambda_b\left|\hat{\mathcal{H}}_W\right|(H\otimes L)_i^J K^-\rangle,
\label{eq:barepro}
\end{eqnarray}
with $(H\otimes L)_i^J$ the $i$th state in the $|H\otimes L\rangle$ basis for spin $J$. With this definition, the bare decay amplitudes read
\begin{eqnarray}
    P^\frac{1}{2}=\begin{pmatrix}
    \frac{1}{2}\mathcal{F}_1^\frac{1}{2}+\frac{1}{2\sqrt{3}}\mathcal{F}_2^\frac{1}{2}+\sqrt{\frac{2}{3}}\mathcal{F}_3^\frac{1}{2}\\[6pt]
    \frac{1}{2\sqrt{3}}\mathcal{F}_1^\frac{1}{2}+\frac{5}{6}\mathcal{F}_2^\frac{1}{2}-\frac{\sqrt{2}}{3}\mathcal{F}_3^\frac{1}{2}\\[6pt]
    \sqrt{\frac{2}{3}}\mathcal{F}_1^\frac{1}{2}-\frac{\sqrt{2}}{3}\mathcal{F}_2^\frac{1}{2}-\frac{1}{3}\mathcal{F}_3^\frac{1}{2}
    \end{pmatrix},
\end{eqnarray}
\begin{eqnarray}
    P^\frac{3}{2}=\begin{pmatrix}
    \frac{1}{\sqrt{3}}\mathcal{F}_1^\frac{3}{2}-\frac{1}{3}\mathcal{F}_2^\frac{3}{2}+\frac{\sqrt{5}}{3}\mathcal{F}_3^\frac{3}{2}\\[6pt]
    -\frac{1}{2}\mathcal{F}_1^\frac{3}{2}+\frac{1}{\sqrt{3}}\mathcal{F}_2^\frac{3}{2}+\frac{1}{2}\sqrt{\frac{5}{3}}\mathcal{F}_3^\frac{3}{2}\\[6pt]
    \frac{1}{2}\sqrt{\frac{5}{3}}\mathcal{F}_1^\frac{3}{2}+\frac{\sqrt{5}}{3}\mathcal{F}_2^\frac{3}{2}-\frac{1}{6}\mathcal{F}_3^\frac{3}{2}
    \end{pmatrix},
\end{eqnarray}
\begin{eqnarray}
P^\frac{5}{2}=\mathcal{F}_1^\frac{5}{2},
\end{eqnarray}
with the upper index $J=\frac 12, \frac 32, \frac 52$ for the total spin of the two-hadron system.
 Putting pieces together,
the inelastic amplitude of the $J/\psi p$ for the decay process $\Lambda_b \to J/\psi p K^-$ can be expressed as~\cite{Du:2021fmf,Du:2019pij} 
\begin{align}
    U^J_i(E,k)=-\sum_{\beta}\int\frac{d^3\bf{q}}{(2\pi)^3}\mathcal{V}^J_{i\beta}(k)G_{\beta}(E,q)U^J_{\beta}(E,q).
\label{eq:amplitude_1}
\end{align}
\begin{figure}[htbp]
    \centering
    \includegraphics[width=0.48\textwidth]{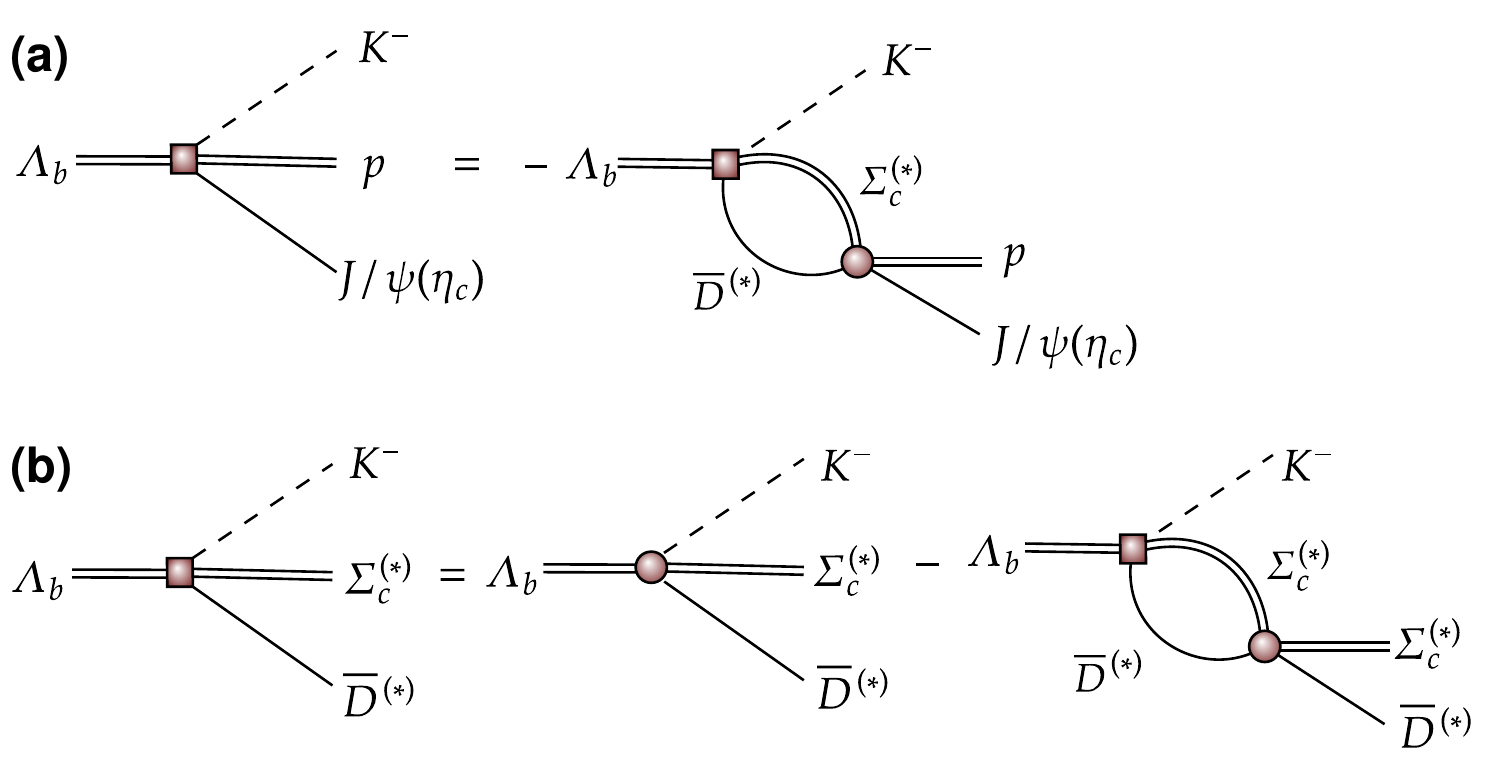}
    \captionsetup{justification=raggedright}
    \caption{(Color online) Graphical representation for the decay amplitudes $\Lambda_b\to K^-J/\psi (\eta_c)p$ (a) and $\Lambda_b\to K^-\Sigma_c^{(*)}\bar{D}^{(*)}$ (b), with the filled squares representing their physical decay amplitudes. The physical decay amplitude $\Lambda_b\to K^-\Sigma_c^{(*)}\bar{D}^{(*)}$ is obtained by solving the Lippmann-Schwinger equation (LSE) Eq.\eqref{eq:amplitude_2}.
    The first filled circle of figure (b) stands for the bare decay amplitude of the $\Lambda_b\to K^-\Sigma_c^{(*)}\bar{D}^{(*)}$ process. 
    The other filled circles represent the transitions 
    among the elastic $\Sigma_c^{(*)}\bar{D}^{(*)}$ and the inelastic channels $J/\psi p$ and $\eta_c p$.}
    \label{fig:Fey1}
\end{figure}
The corresponding diagrams are shown in Fig.~\ref{fig:Fey1}a. 
Here, $U^J_i$ is the $i$th $J/\psi p$ inelastic channel amplitude with the quantum number $J$.
$\mathcal{V}^J_{i\beta}$ is the transition vertex bet\-ween the $\beta$th elastic and the $i$th
inelastic channel, 
and described by the coupling constants $g_S$ for the $S$-wave and $g_D$  for the $D$-wave. 
\begin{eqnarray}
G_\beta(E,q)=\frac{2\mu_\beta}{q^2-p_\beta^2-i\epsilon},\quad p_\beta^2\equiv 2\mu_\beta (E-m_\mathrm{th}^\beta),
\end{eqnarray}
is the two-body propagator of the $\beta$th elastic channel~\cite{Du:2021fmf,Du:2019pij}, with $\mu_\beta$ and $m_\mathrm{th}^\beta$ the reduced mass and threshold of the $\beta$th channel.
The physical production amplitude of the $\beta$th elastic channel $U^J
_{\beta}$ for total spin $J$ is obtained from the Lippmann-Schwinger (LSE)
\begin{align}
    &U^J_{\alpha}(E,p)\notag\\
    &=P^J_{\alpha}-\sum_{\beta}\int\frac{d^3\bf{q}}{(2\pi)^3}V^J_{\alpha\beta}(E,p,q)G_{\beta}(E,q)U^J_{\beta}(E,q),
\label{eq:amplitude_2}
\end{align}
with $P^J_{\alpha}$, constructed by seven parameters $\mathcal{F}^J_n$ defined as in Eq.~\eqref{eq:barepro},
the production amplitude for the $\alpha$th elastic channel. 
The corresponding diagrams are shown in Fig.~\ref{fig:Fey1}b.

Note that the above integral equations can be reduced to algebric equations in the framework of a pionless EFT. For a general introduction to this type of EFT, see~\cite{Meissner:2022odx}.
$V^J$ is the effective potential
\begin{eqnarray}
    V^J=V_C+V^J_{\mathrm{in}}(E),
\end{eqnarray}
which contains 
both the scattering between the elastic channels $V_C$ described by
two parameters $C_{1/2}$, $C_{3/2}$ and that $V^J_{\mathrm{in}}(E)$ between the elastic and inelastic channels described by the two parameters $g_S$, $g_D$. In total, eleven parameters are combined in a vector
\begin{eqnarray}
 \mathcal{P} &=&  \bigl(g_S, g_D, C_{\frac 12}, C_{\frac 32},\nonumber \\ 
  &&  \mathcal{F}^{\frac{1}{2}}_1,\mathcal{F}^{\frac{1}{2}}_2,\mathcal{F}^{\frac{1}{2}}_3,\mathcal{F}^{\frac{3}{2}}_1,\mathcal{F}^{\frac{3}{2}}_2,\mathcal{F}^{\frac{3}{2}}_3,\mathcal{F}^{\frac{5}{2}}_1\bigr).
\label{eq:para vector}
\end{eqnarray} 

To describe the $J/\psi p$ invariant mass distribution,
a composite model is constructed via a probability distribution function (PDF) as:
\begin{align} 
    {\rm PDF}(E;\mathcal{P})\notag
    =\alpha \sum_J\int |U^J|^2 \mathrm{p.s.}(E) G(E'-E)dE' \\ 
     + \quad (1-\alpha) {\rm Chebyshev_6}(E)~,
    \label{eq:PDF}
\end{align}
with $U^J$ the $\Lambda_b\to J/\psi p K^-$ amplitude~\eqref{eq:amplitude_2} 
for total spin $J=\frac 12,~\frac{3}{2},~\frac 52$ and $\mathrm{p.s.}(E)$
is the phase space of the corresponding process.
 A Gaussian function $G(E'-E)$ which represents the experimental detector resolution is convoluted with the physical invariant mass distribution.
 Here, we take the energy resolution as a constant of $\sigma \sim 2.3$ MeV~\cite{LHCb:2019kea} 
without considering its energy dependence. Further,
 ${\rm Chebyshev_6}$ is the sixth-order Chebyshev polynomial which represents the background distribution, 
 with $1-\alpha$ its fraction, $\alpha \in [0,1]$.
The background fraction in the data is determined to be (96.0$\pm$0.8)\% as discussed in the Supplementary materials.
The coefficients ($c_0,...,c_6$) for the background component 
are obtained by fitting the Chebyshev polynomials to data~\cite{Brun:1997pa}, as listed in the Supplementary materials.
Note that the area enclosed by the signal (background) distribution, i.e. the remnant of the first term
after removing the factor $\alpha$ (the second term after removing the factor $1-\alpha$ ) in Eq.~\eqref{eq:PDF} is normalized  to one. 

\section{States and Labels}
\label{sec:labels}
\begin{figure}[ht]
    \centering
    \includegraphics[width=0.48\textwidth]{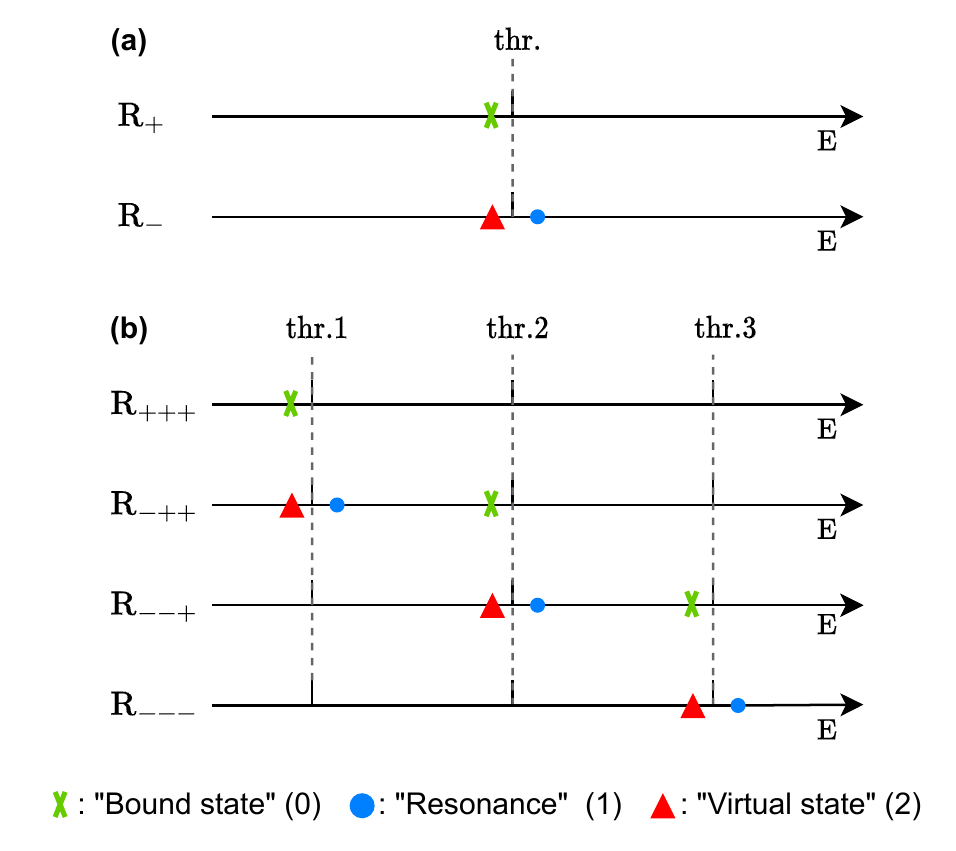}
    \captionsetup{justification=raggedright}
    \caption{(Color online) The definition of the pole for the one coupled channel case, 
    i.e. the $J^P=\frac 52^-$ state (a) and the three coupled channel case, i.e. the $J^P=\frac 12^-$ and $\frac 32^-$ states (b).  
 The crossed, circle and triangle points stand for ``Bound state", ``Resonance" and ``Virtual state", respectively.}
    \label{fig:Riemann_surface}
\end{figure}

The hidden charm pentaquarks are described as generated from the scattering of the $(\Sigma_c,\Sigma_c^*)$ and $(\bar{D},\bar{D}^*)$ doublets. 
 As the thresholds of the inelastic channels $J/\psi p$ and $\eta_c p$ are far away from those of elastic channels, their effects on   the classification of the poles relevant to the physical observables are marginal.
 Thus, we consider the classification of poles by considering the elastic channels. In this case,
the $\frac 12^-$ and $\frac 32^-$ states are given by a three-channel case with
$2^3=8$ Riemann surfaces denoted as $\mathrm{R}_{\pm\pm\pm}$, see~\cite{Mai:2022eur} for a pictorial.
Here, the ``+" and ``$-$" signs in the $i$th position
mean the physical and unphysical sheet of the $i$th channel, respectively.
The physical observables, i.e. the $J/\psi p$ invariant mass distribution in our case, are real, reflecting the amplitude in the real axis. Those are related to the poles on the complex plane. We consider the poles most relevant to the physical observables, i.e. the poles on the physical sheet $\mathrm{R}_{+++}$ and those close-by ones. As we aim at extracting the most important information of those poles,
it is sufficient to work with leading order contact interactions.
In this case, the poles accounting for the near threshold structures
can be classified as in Fig.~\ref{fig:Riemann_surface}. 
For the classification~\footnote{For the classification of poles, strictly, bound states do not decay to any final particles, which means them stable. However, in the standard model, there are only 11 stable particles, i.e. photon, the three generations of neutrino and antineutrino, electron, positron, proton, and antiproton. Other particles can decay into final states with lower masses. In our case, bound state, virtual state, and resonance are defined based on the most coupled elastic channels.} of the states, we start from the one-channel case, i.e. $J^P=\frac{5}{2}^-$ state related to the $\Sigma_c^*\bar{D}^*$ channel. The ``bound state", ``resonance", and ``virtual state" are defined as a pole on the  physical sheet below threshold, a pole on the unphysical sheet above the threshold, and a pole on the unphysical sheet below the threshold, respectively (Fig.~\ref{fig:Riemann_surface}a). Those poles are labeled as $0$, $1$, $2$ in order. 
For the three-channel case (Fig.~\ref{fig:Riemann_surface}b), i.e. $J^P=\frac 12^-$ and $J^P=\frac 32^-$ channels, the poles are defined according to the most strongly coupled channel. The case with three poles on the $\mathrm{R}_{+++}$, $\mathrm{R}_{-++}$ and $\mathrm{R}_{--+}$ sheets slightly below the first, second, and third thresholds, respectively, is defined as a ``bound state" which is labeled as $0$. The case with three poles on the $\mathrm{R}_{-++}$, $\mathrm{R}_{--+}$ and $\mathrm{R}_{---}$ sheets slightly above the first, second, and third thresholds, respectively, is defined as a ``resonance" which is labeled as $1$. The case with three poles on the $\mathrm{R}_{-++}$, $\mathrm{R}_{--+}$ and $\mathrm{R}_{---}$ sheets slightly below the first, second, and third thresholds, respectively, is defined as a ``virtual state" which is labeled as $2$. Note that these three states are defined according to their most strongly coupled channel, which the word ``slightly" means. For instance, on the sheet $\mathrm{R}_{-++}$, both the ``resonance" (blue circle point) and ``bound state" (green crossed point) locate between the first and the second thresholds. 

However, the former one strongly couples to the first channel and is located close to
the first threshold, defined as a resonance of the first channel. The latter one strongly couples to the
second channel and is located  close to the second threshold, defined as a bound state of the second channel. 
The situation for the other states is similar.

We use a four-digit label $``olll"$, with $l =0,1,2$ and $o=0,1$, to denote the situation of the poles. The last three labels are for the nature of the poles in the $J^P=\frac12^-,~\frac 32^-$ and $\frac 52^-$ channels in order. The first label ``$o$", which we put in front,
is used  for the mass order. 
That is due to the indeterminacy of the quantum numbers of the $P_c(4440)$ and $P_c(4457)$.  
As they are close to the $\Sigma_c\bar{D}^*$ threshold, 
the $S$-wave interaction can give both $J^P=\frac 12^-$ and $\frac 32^-$
states. In the limit of HQSS, there are two solutions, i.e. Solution~A and Solution~B defined in Refs.~\cite{Liu:2019tjn,Du:2019pij,Du:2021fmf}, as discussed before.
In Solution~A, the higher spin state has larger mass, i.e. $J_{P_{c}(4440)}=\frac 12$ and  $J_{P_{c}(4457)}=\frac 32$. In Solution~B, the situation is reversed, i.e. $J_{P_{c}(4457)}=\frac 12$ and  $J_{P_{c}(4440)}=\frac 32$.
To distinguish these two scenarios, another label $o=1$ and $o=0$  for 
Solution~A and Solution~B, respectively, is required.
In total, we have a four-digit label ``$olll$" to denote
the situation of the poles. Taking the ``1012" case for example,
it means that the poles for $J^P=\frac 12^-, \frac 32^-$ and $\frac 52^-$ channels are ``bound state", ``resonance" and ``virtual state", respectively. The first label $1$ means Solution~A, i.e. $J_{P_c(4440)}=\frac 12,~J_{P_c(4457)}=\frac 32$.
Note that we can also have cases  with the 
poles different from the three situations discussed above, 
e.g. the poles are far away from the thresholds.
However, their probabilities are almost zero and are thus neglected.  

\section{Monte Carlo simulation}
\label{sec:MC_simulation}
\begin{figure}[!h]
    \centering
    \includegraphics[width=0.48\textwidth]{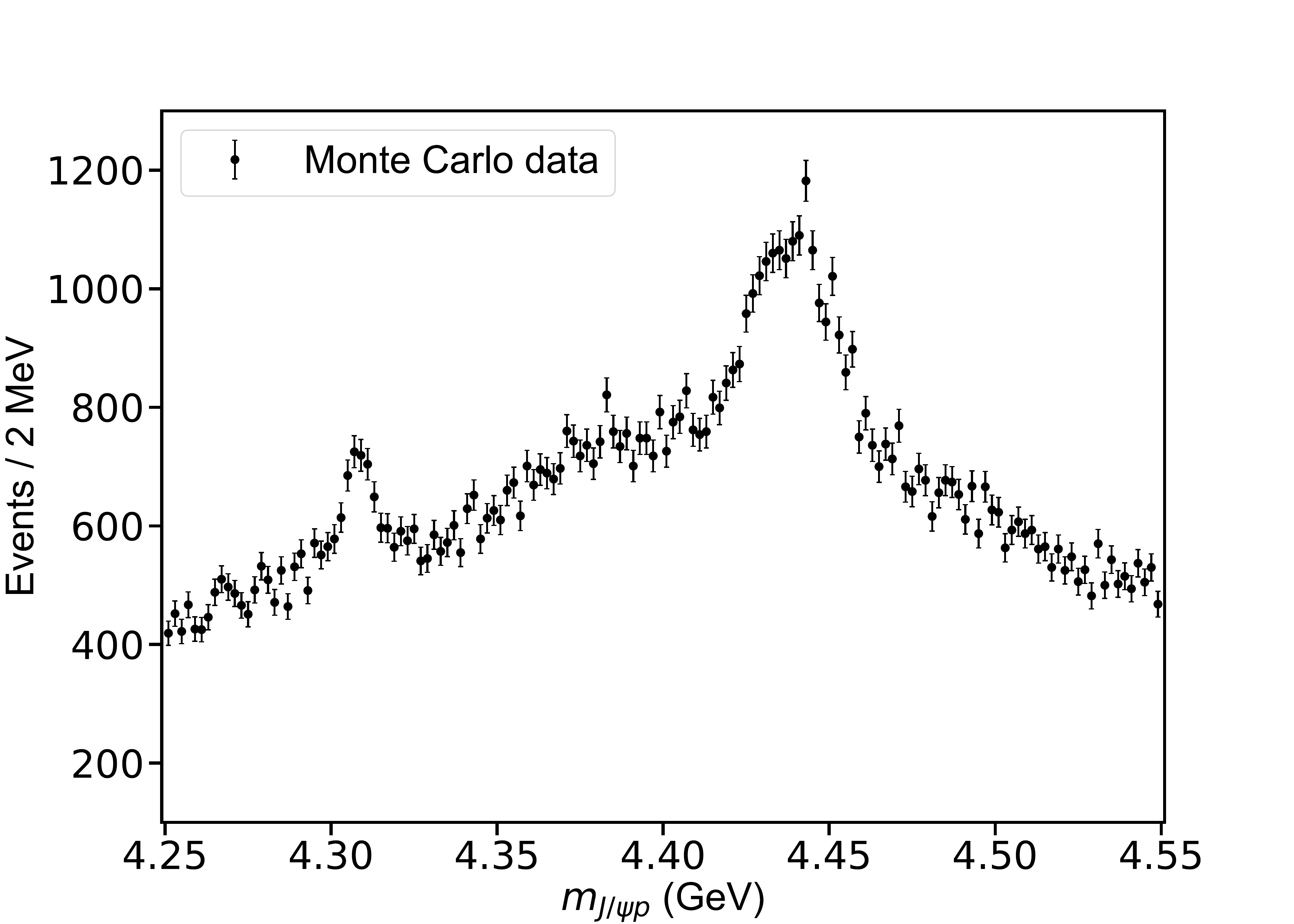}
    \captionsetup{justification=raggedright}
    \caption{(Color online) An example of the simulated invariant mass distribution $\mathcal{M}(J/\psi p)$.}
    \label{fig:invmass}
\end{figure}

Based on the PDF given in Eq.~\eqref{eq:PDF}, 
240184 samples corresponding to physical states, i.e. with poles not too far away from the real axis in the
considered energy range,  are selected among 1.85 million samples uniformly generated in the space of $\mathcal{P}$,
 distributed in the mass window from 4.25~$\mathrm{GeV}$ to 4.55~$\mathrm{GeV}$.
These samples are represented as a set: $\{\mathcal{H}_j, \mathcal{L}_j\}$ where the  $j$ index refers to a given sample. A histogram {$\mathcal{H}$} with 150 bins (see Fig.~\ref{fig:invmass}) denotes the invariant mass spectrum of  the $P_c$ states (background included) and the label $\mathcal{L}$ indicates the state label 
defined in Sec.~\ref{sec:labels}.
\noindent The values of $\mathcal{P}$ are uniformly sampled in the following ranges,
\begin{alignat}{3}
&g_S\in [0,10]\ {\rm GeV}^{-2},     &\quad \mathcal{F}^{\frac{1}{2}}_3&\in [-3600,-3300],  \notag\\
&g_D\in [0.5,1.5]\times g_S,        &\quad \mathcal{F}^{\frac{3}{2}}_1&\in [-3900,-3600],  \notag\\ 
&C_{1/2}\in [-20,0] {\rm GeV}^{-2}, &\quad  \mathcal{F}^{\frac{3}{2}}_2&\in [-1900,-1600],  \label{eq:para}\\
&C_{3/2}\in [0.5,1.5]\times C_{1/2},&\quad  \mathcal{F}^{\frac{3}{2}}_3&\in [-4800,-4500],  \notag\\
&\mathcal{F}^{\frac{1}{2}}_1\in [0,300],&\quad\mathcal{F}^{\frac{5}{2}}_1&\in [600,900].  \notag\\
&\mathcal{F}^{\frac{1}{2}}_2\in [700,1000],\notag
\end{alignat}
The ranges are empirically taken to produce a mass spectrum similar to the experimental data.
Note that $g_S$ and $g_D$ as well as $C_{1/2}$ and $C_{3/2}$  are strongly correlated, so the second
of these coefficients is related by a factor in the range  $[0.5, 1.5]$ to the corresponding first one. 
Note that these strong correlations were found in Refs.~\cite{Du:2019pij,Du:2021fmf}, and are also
found if one uses a much larger set of samples to train the NNs.  
The Monte Carlo (MC) production is performed with the open source software ROOT~\cite{Brun:1997pa} and GSL~\cite{Galassi2011GNU}, as categorized in Tab.~\ref{tab:label0}.
Considering different background fractions, the set of MC samples produced at the background fraction $1-\alpha=$ 90\% is denoted as $\{\mathcal{S}^{90}\}$, so are the other sets.

\begin{table}[ht]
    \centering
    \captionsetup{justification=raggedright}
    \caption{ Distribution of parameters.
    The state label of the second column is defined as in Fig.~\ref{fig:Riemann_surface}.
    The first label $``0"$ means $J_{P_c(4440)}=\frac 32$ and $~J_{P_c(4457)}=\frac 12$.
    The first label $``1"$ means $J_{P_c(4440)}=\frac 12$ and $~J_{P_c(4457)}=\frac 32$.
    The last column represents the number of samples for this label.}
    \begin{tabular}{ccc}
    \hline
    Mass relation label & State label &   Number of samples\\
    \hline
     0&000   &46,951\\
     1&000   &4283\\
     1&001   &1260\\
     1&002   &4360\\
     0&100   &3740\\
     0&110   &4320\\
     0&111   &7520\\
     1&111   &360\\
     0&200   &9590\\
     1&200   &280\\
     1&210   &3980\\
     1&211   &2690\\
     1&220   &50,240\\
     1&221   &50,512\\
     1&222   &50,098\\
    \hline
    \end{tabular}  \label{tab:label0}
\end{table}

\section{Training and Verification}
\label{sec:TaV}
The NN is implemented with an infrastructure of ResNet~\cite{He2016DeepRL} as discussed in the Supplementary materials, and solved with the Adam~\cite{2014Adam} optimizer parametrized by the cross-entropy loss function. In order to balance the impact of unequal sample sizes, the cross-entropy loss function is weighted by the reciprocal of the number of samples. To avoid the network falling into a local minimum solution during the training process, the weights of the neurons are randomly initialized with a normal distribution centered at 0 and with a width of 0.01, and the biases of the neurons are set to zero.
A reasonable solution can be obtained after training 500 epochs, using an
initial learning rate value of 0.001. 
\begin{figure*}[htbp]
\centering
\includegraphics[width=0.48\textwidth]{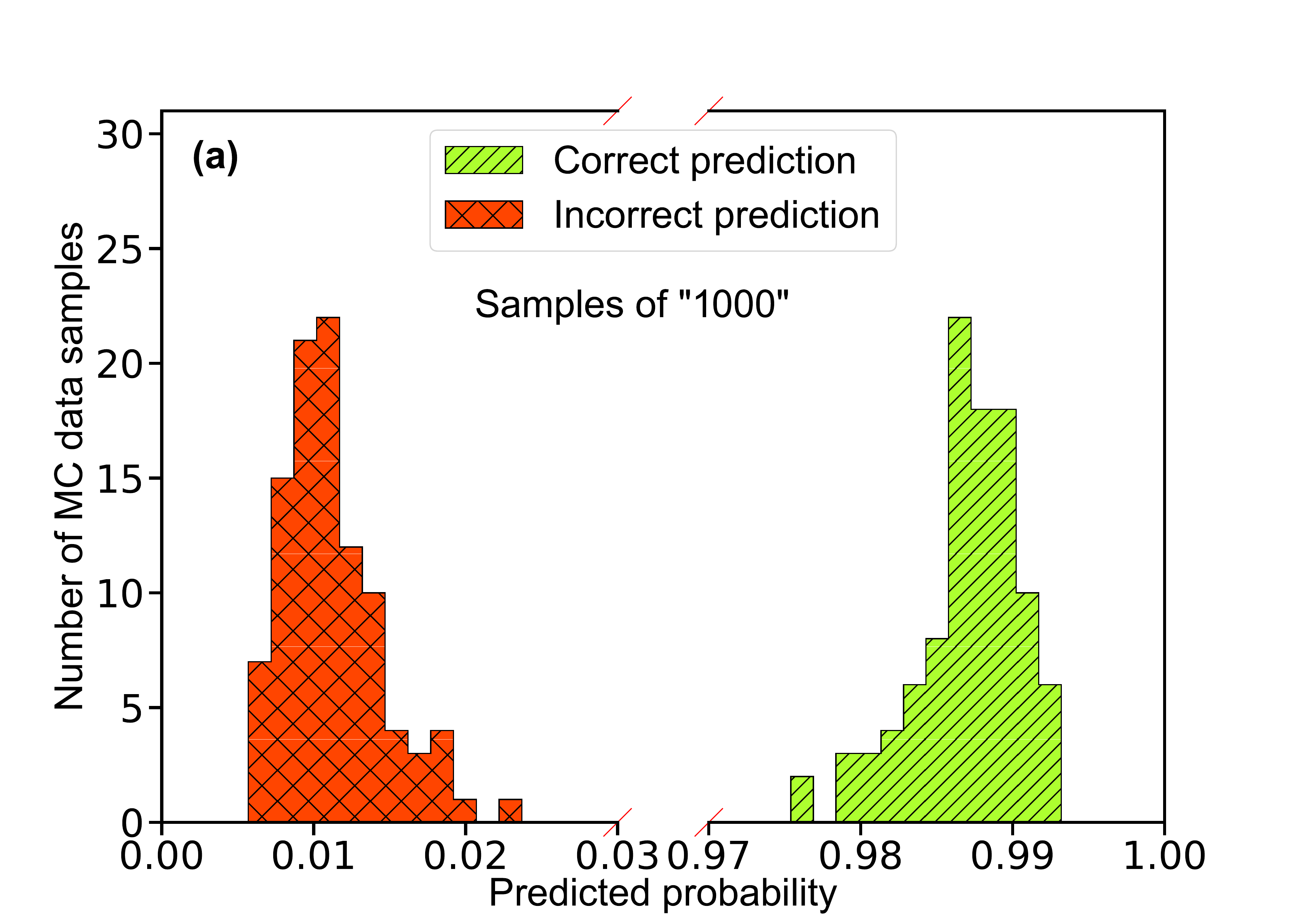}
\includegraphics[width=0.48\textwidth]{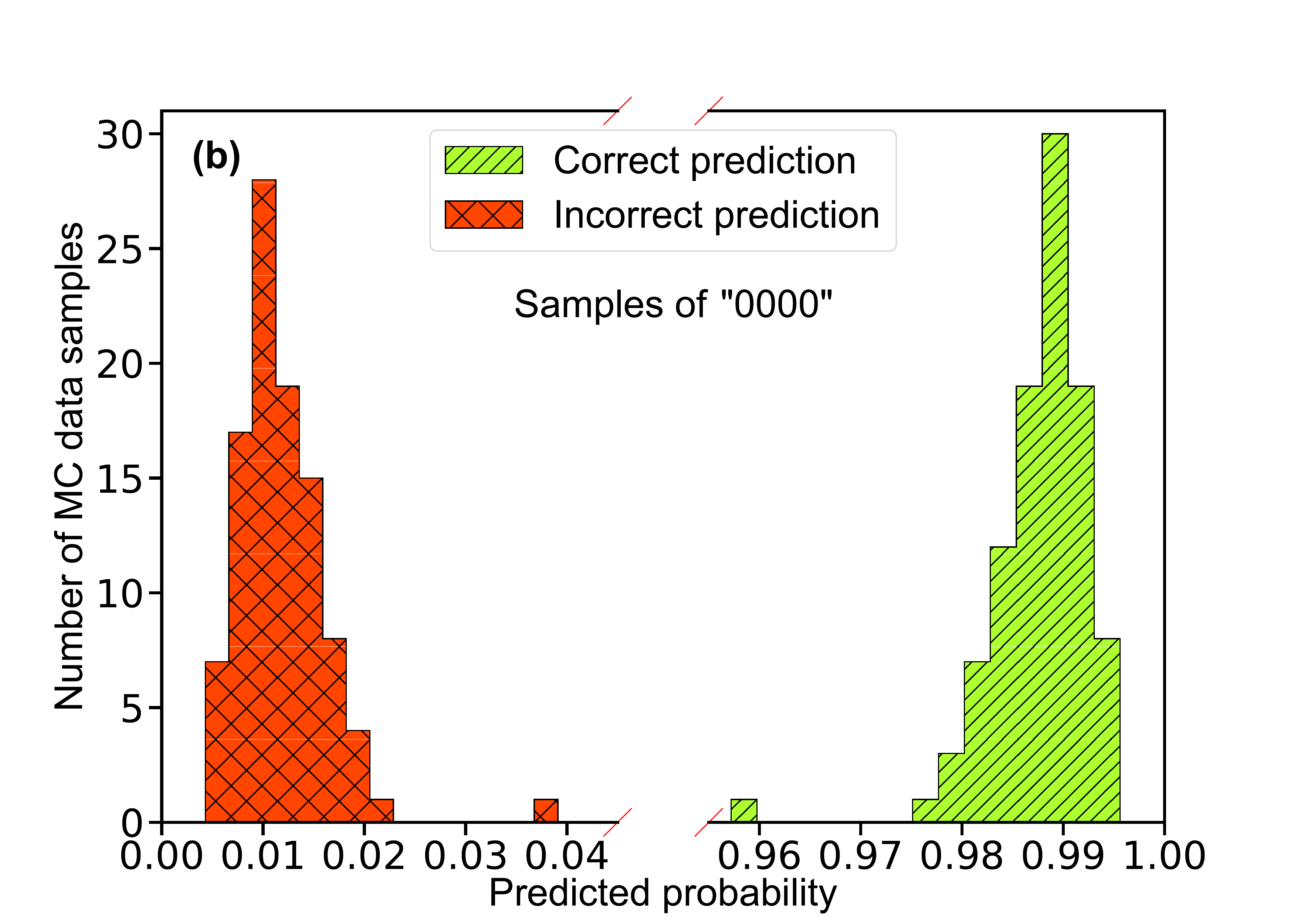}
\includegraphics[width=0.48\textwidth]{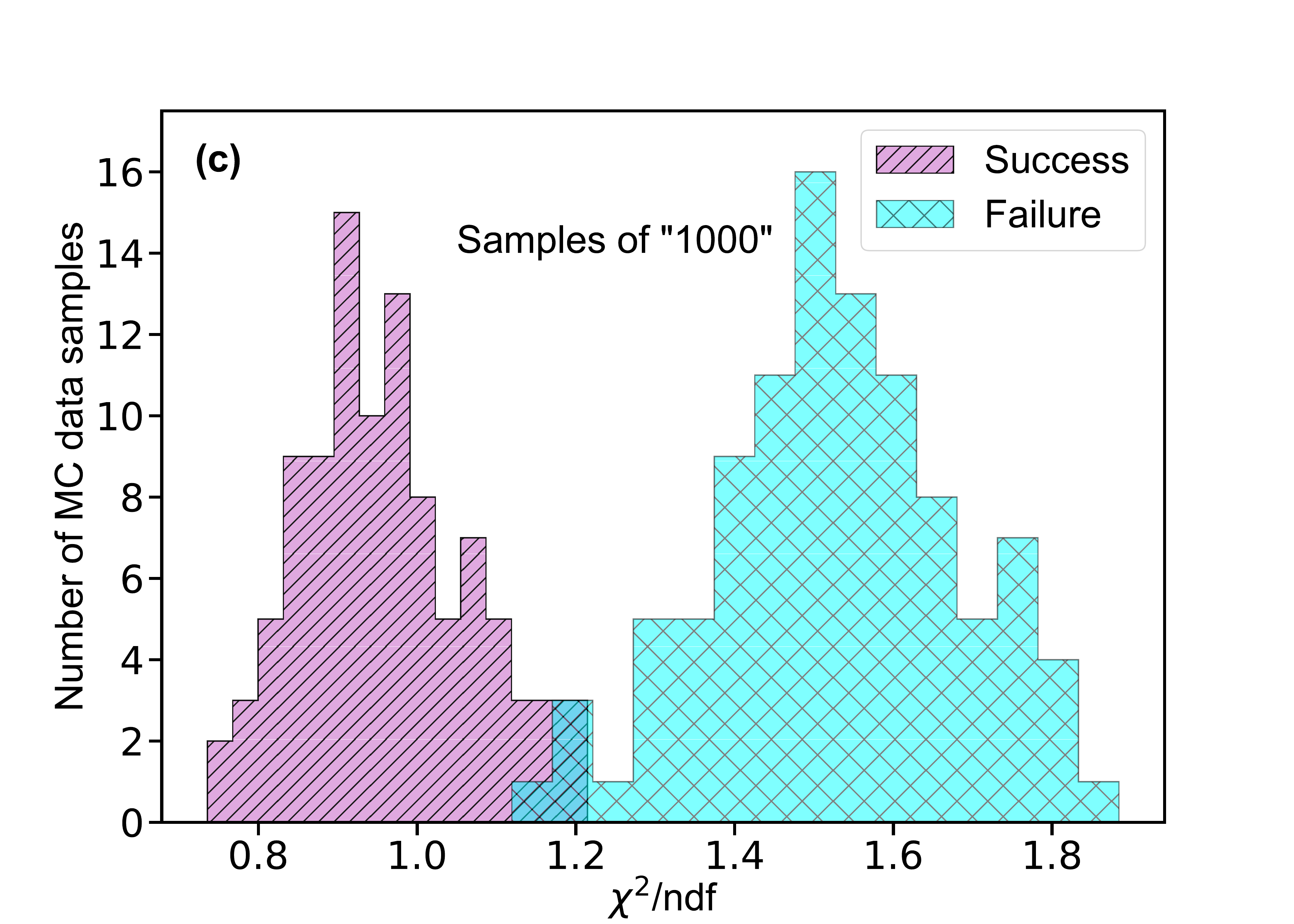}
\includegraphics[width=0.48\textwidth]{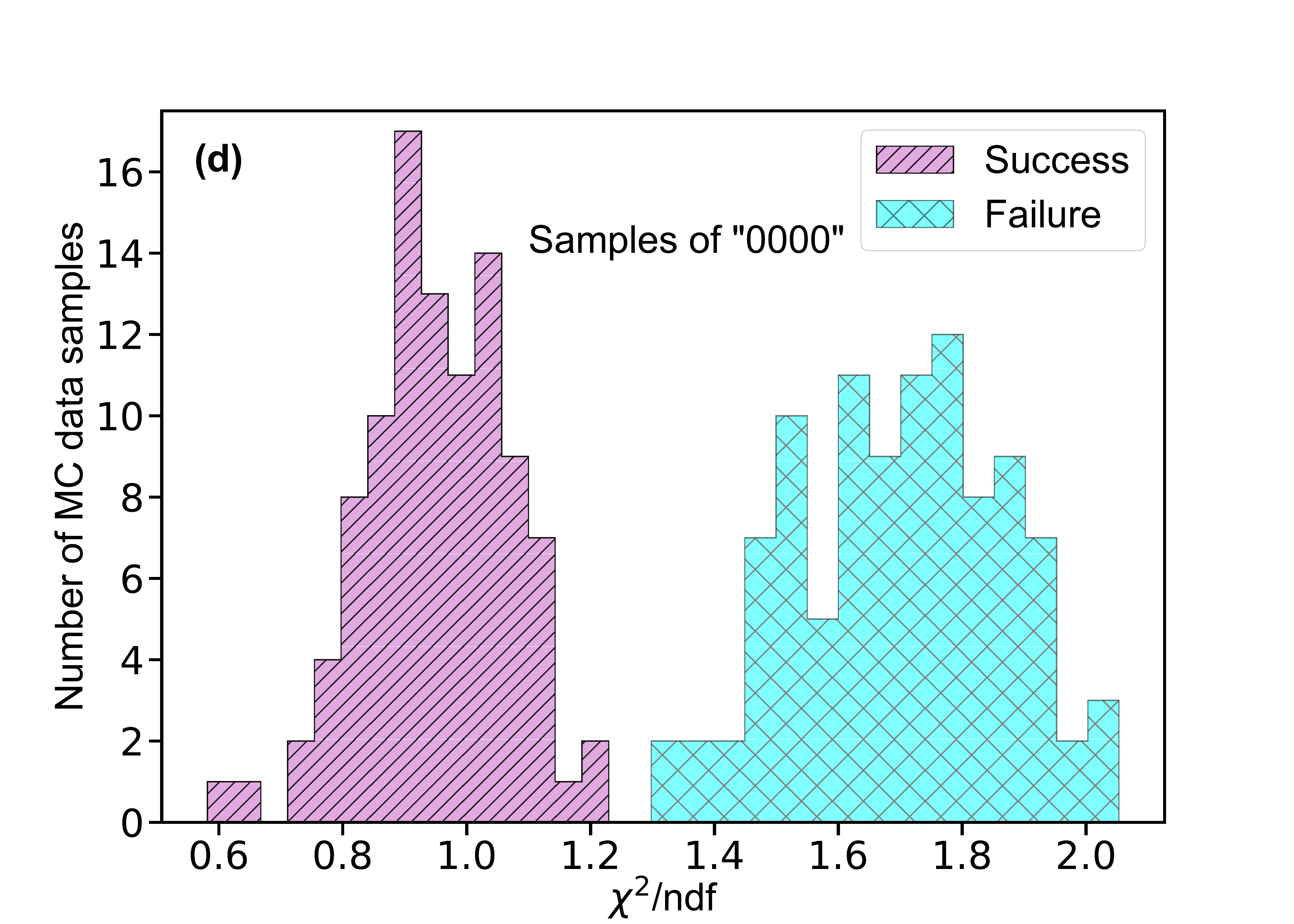}
\captionsetup{justification=raggedright}
\caption{(Color online) The predicted probability distributions for samples corresponding to the ``1000" label (a) and the ``0000" label (b) in the neural network. The normalized $\chi^2$/ndf distributions for samples corresponding to the ``1000" label (c), samples corresponding to the ``0000" label (d) in the normal fitting approach.}
\label{fig:network_MCdata}
\end{figure*}
Here the learning rate of the next interval is reduced to half of the previous one,
controlled by the stepLR method. The interval is set to 100 epochs in our case. Note that 70\% of MC simulation samples are used for training and the rest 30\% for a benchmark. To avoid overfitting in the training, dropout layers are introduced in the network during the training. The dropout probability is chosen to 0.3 to optimise performance. In case of overfitting, the obtained NN works well on training datasets while probably does not on testing datasets.
Thus the loss function calculated with training datasets should obviously differ from the one calculated with testing datasets. 
However, Fig.~9 in the Supplementary materials indicates that there is no obvious difference. The training details are summarized in the Supplementary materials. 
In short, the NNs successfully retrieve the state label
with an accuracy (standard deviation) of 75.91(1.18)\%, 73.14(1.05)\%, 65.25(1.80)\% and 54.35(2.32)\% for MC simulation samples 
 $\{\mathcal{S}^{90}\}$, $\{\mathcal{S}^{92}\}$, $\{\mathcal{S}^{94}\}$ and $\{\mathcal{S}^{96}\}$.
The prediction accuracy decreases as the background increases, as expected.

Turning to the mass relation, 
we verified 100 $\{\mathcal{S}^{90}\}$ samples corresponding to the label ``1xxx", and another 100 samples corresponding to the label ``0xxx".
Fig.~\ref{fig:network_MCdata}a/b
shows the predicted probability distribution for the label ``1xxx/0xxx" samples,
in which all labels are successfully retrieved.
In other words, the NN can accurately predict the mass relationship. 
For comparison, we also perform an analysis with the $\chi^2$/ndf-fitting method.
Fig.~\ref{fig:network_MCdata}c/d shows the normalized $\chi^2$/ndf distributions
corresponding to the label ``1xxx/0xxx".
A 3\% misidentification is observed for the label ``1xxx" samples.

\section{Application to Data}
\label{sec:appdata}
We now use the well-trained NNs to analyse the experimental data. The  results are listed in Tab.~\ref{tab:Output_state_Simplify}.
To reduce the systematic uncertainties arising from the NNs, we trained five and ten NN models, which have an identical structure under different initialization, 
for each group of samples with different background fractions.
The probabilities of the sum of all the other labels are 
smaller than 1\%. The labels with the top three probabilities
are ``1000", ``1001" and ``1002", with the sum of them larger than $99\%$. The standard deviation decreases with the increasing number of NN models, as expected. It means that our NNs can well determine the first three labels and favors Solution~A,   
i.e. $J_{P_c(4440)}=\frac 12$ and $J_{P_c(4457)}=\frac 32$, 
as the first label is $1$.  
\begin{table*}[htbp]
    \centering
    \captionsetup{justification=raggedright}
    \caption{Predicted probability (standard deviation) for the  15 states corresponding to the different labels ${\mathcal L}$. 
 The label ``100X" means the sum of ``1000", ``1001", and ``1002".
    The probability of the sum of other labels is listed in the last column.}
    \begin{tabular}{ccccccc}
    \hline
     \multirow{2}{*}{Models} & \multicolumn{6}{c}{Label}\\
     \cline{2-7}
     &0000&1000&1001&1002&100X&Others\\
    \hline
         \multicolumn{7}{c}{Prediction(\%) of NN trained with $\{S^{90}\}$ samples.} \\
    \hline
    NN 1 &0.69&89.13&1.42&8.75&99.30&0.01\\
    NN 2 &0.03&5.83&38.47&55.30&99.60&0.37\\
    NN 3 &0.03&5.39&15.79&78.41&99.59&0.11\\
    NN 4 &0.01&1.9&27.01&70.95&99.86&0.13\\
    NN 5 &2.40&94.45&0.15&2.99&97.59&0.01\\
    5 NNs average  &0.63(1.03)&39.34&16.57&43.28&99.19(0.91)&0.13(0.15)\\
    10 NNs average   &0.36(0.74)&21.16&20.69&57.62&99.47(0.68)&0.12(0.13)\\
    \hline
        \multicolumn{7}{c}{Prediction(\%) of NN trained with $\{S^{92}\}$ samples.} \\
    \hline
    NN 1 &0.00&0.15&5.37&94.47&99.99&0.00\\
    NN 2 &0.00&0.07&4.11&95.81&99.99&0.00\\
    NN 3 &0.00&0.78&13.57&85.61&99.96&0.03\\
    NN 4 &0.00&0.81&19.02&80.16&99.99&0.00\\
    NN 5 &0.14&15.13&16.91&67.80&99.84&0.00\\
    5 NNs average  &0.03(0.06)&3.39&11.80&84.77&99.95(0.06)&0.01(0.01)\\
    10 NNs average  &0.01(0.04)&1.78&9.50&88.70&99.97(0.04)&0.00(0.01)\\
    \hline
     \multicolumn{7}{c}{Prediction(\%) of NN trained with $\{S^{94}\}$ samples.} \\
    \hline
    NN 1 &0.00&1.20&2.26&96.53&99.99&0.00\\
    NN 2 &0.02&2.23&11.51&86.24&99.98&0.00\\
    NN 3 &0.00&0.26&10.26&89.47&99.99&0.00\\
    NN 4 &0.15&24.05&48.79&26.89&99.73&0.13\\
    NN 5 &0.02&4.12&72.61&23.23&99.96&0.02\\
    5 NNs average  &0.04(0.06)&6.37&29.09&64.47&99.93(0.11)&0.03(0.06)\\
    10 NNs average  &0.02(0.05)&4.25&26.00&69.70&99.95(0.08)&0.02(0.04)\\
    \hline
        \multicolumn{7}{c}{Prediction(\%) of NN trained with $\{S^{96}\}$ samples.} \\
    \hline
    NN 1 &0.00&1.29&19.79&78.92&100.00&0.00\\
    NN 2 &0.00&3.38&25.64&70.97&99.99&0.01\\
    NN 3 &0.00&6.57&32.20&61.23&100.00&0.00\\
    NN 4 &0.60&55.77&21.65&21.96&99.38&0.00\\
    NN 5 &0.00&0.27&3.52&96.20&99.99&0.00\\
    5 NNs average  &0.12(0.27)&13.46&20.56&65.86&99.87(0.28)&0.00(0.01)\\
    10 NNs average  &0.11(0.21)&13.13&24.97&61.79&99.89(0.22)&0.00(0.01)\\
    \hline
    \end{tabular}  \label{tab:Output_state_Simplify}
\end{table*}
That shows one of the advantages of machine learning comparing to the normal fitting approach. However, this conclusion is only based on the pionless EFT. One cannot obtain a solid conclusion about the hadronic molecules without the one-pion-exchange (OPE) potential. Thus it makes little sense to compare our results with that of Refs.~\cite{Du:2021fmf,Du:2019pij}. However,
the hidden charm pentaquarks in pionless EFT are sufficient to illustrate the difference between machine learning and the normal fitting approach, as well as the potential merit of machine learning.
The second label $0$ and the third label $0$
mean that the poles for the $J^P=\frac 12^-$ and $J^P=\frac 32^-$ channels behave
as ``bound states", which is different from the virtual state conclusion for the $P_c(4312)$ in Ref.~\cite{Ng:2021ibr}
also using machine learning. The pole situation for the $J^P=\frac 52^-$
is undetermined, i.e. all of the three labels $0$, $1$, $2$ appearing
for the fourth label, as the structure around the $\Sigma_c^*\bar{D}^*$
is not significant. 
To resolve this issue, precise data in this energy region would be needed.
 In the normal fitting approach, the $J$=1/2, 3/2 and 5/2 channels are described by the same set of parameters in the heavy quark limit. As the result, once the set of parameters is determined, all the poles of the three channels are determined as well. However, in our NN approach, we extract the types of poles without extracting the intermediate parameters directly from the experimental data. This makes the poles of the $J$=1/2, 3/2 well determined, but the $J$=5/2 channel undetermined. 
Of importance is to understand how the NN learns information from the input mass spectrum $\mathcal{H}$. The SHapley Additive exPlanation (SHAP) value is widely used as a feature importance metric for well-established models in machine learning. One of the prevailing methods for estimating NN model features is the DeepLIFT algorithm based on DeepExplainer. A positive (negative) SHAP value indicates that
a given data point is pushing the NN classification in favor of (against) a given class. A large absolute SHAP value implies a large impact of a given mass bin on the
classification. This method requires a certain amount of dataset as a reference for evaluating network features. We choose 3000 \{$\mathcal{S}^{90}$\} samples in the dataset as a reference to test the 
SHAP values of both 1000 \{$\mathcal{S}^{90}$\} samples and the experimental data on our network. The upper panel of Fig.~\ref{fig:shap_value} shows the SHAP values of experimental data corresponding to 5 labels, i.e. ``0000", ``1001", ``0100", ``1000" and ``1002". The lower panel of Fig.~\ref{fig:shap_value} shows the SHAP values of MC samples corresponding to the ``1200" label, with others presented in the Supplementary materials.  
Both  panels of Fig.~\ref{fig:shap_value} indicate that data points around the peaks in the mass spectrum have a greater impact.
\begin{figure}[htbp]
    \centering
    \includegraphics[width=0.48\textwidth]{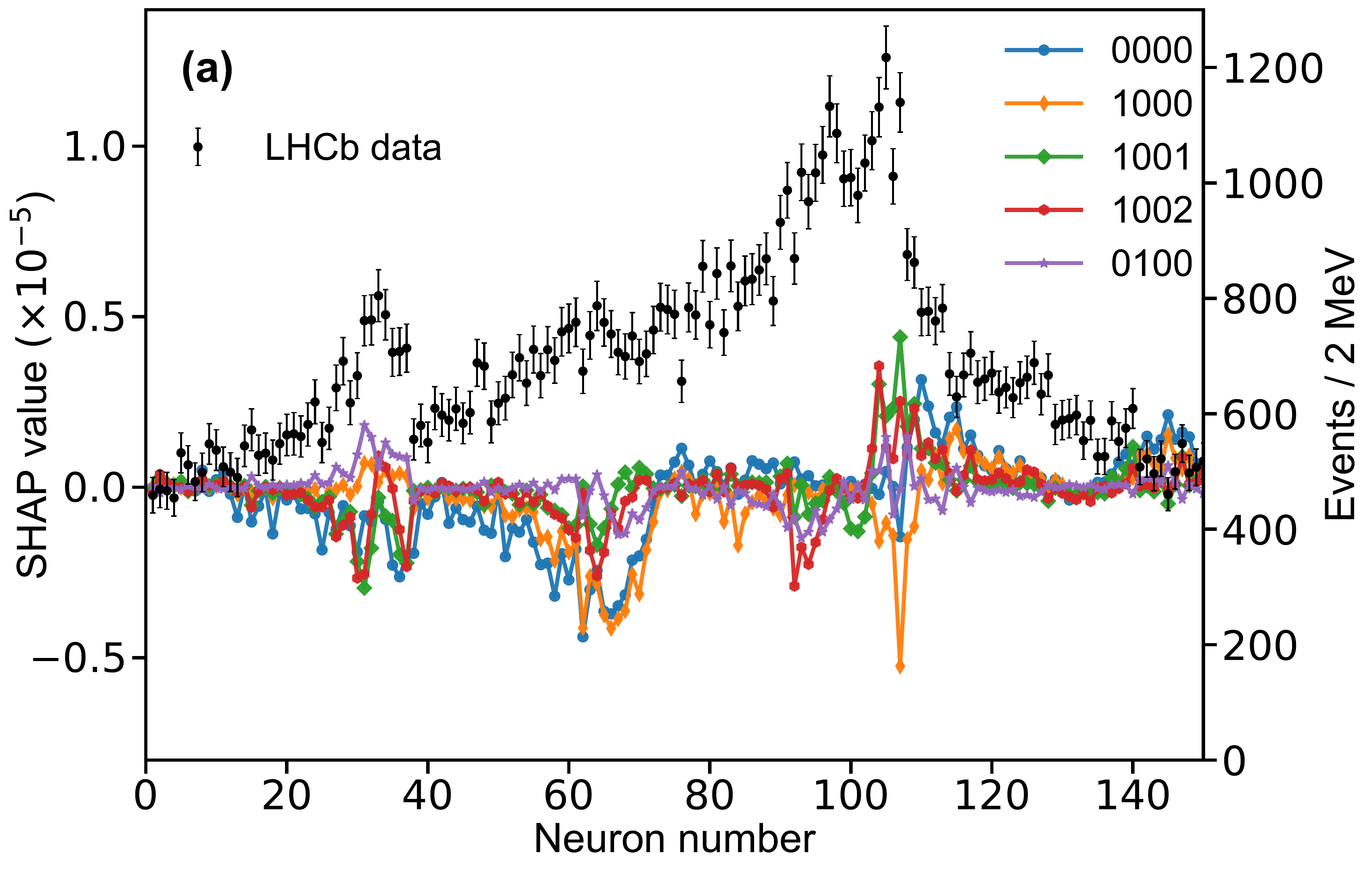}
        \includegraphics[width=0.48\textwidth]{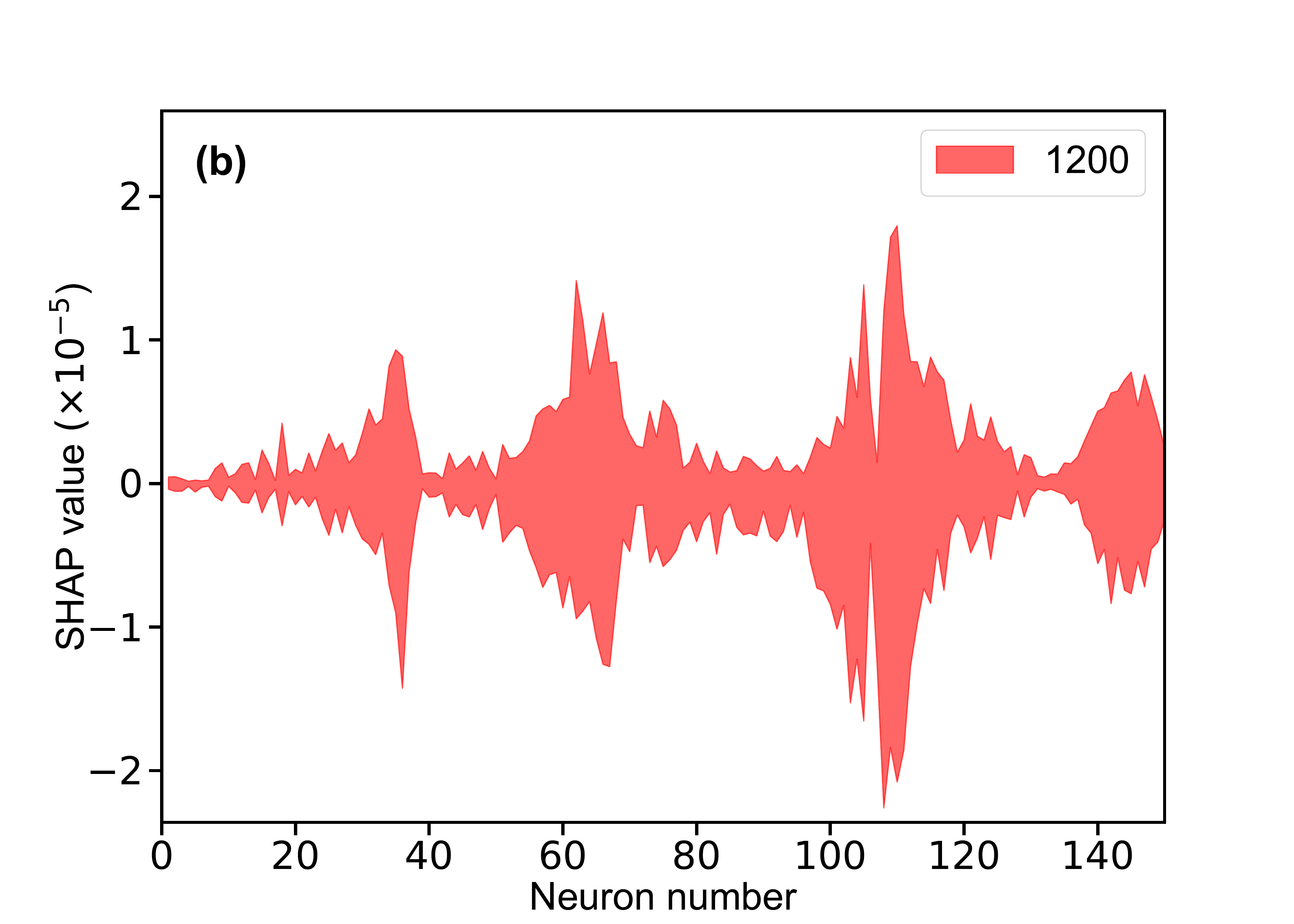}
        \captionsetup{justification=raggedright}
        \caption{(Color online) The distributions of SHAP values for the experimental data represented by the NN (a) and 1000 $\{\mathcal{S}^{90}\}$ samples in the neurons of the ``1200" label (b). The points in (a) are the 150 experimental data. Different color lines
     correspond to five different labels. The analogous distributions to (b) for the samples of other labels can be found in the Supplementary materials.}
    \label{fig:shap_value}
\end{figure}
\begin{figure}[htbp]
    \centering
    \includegraphics[width=0.48\textwidth]{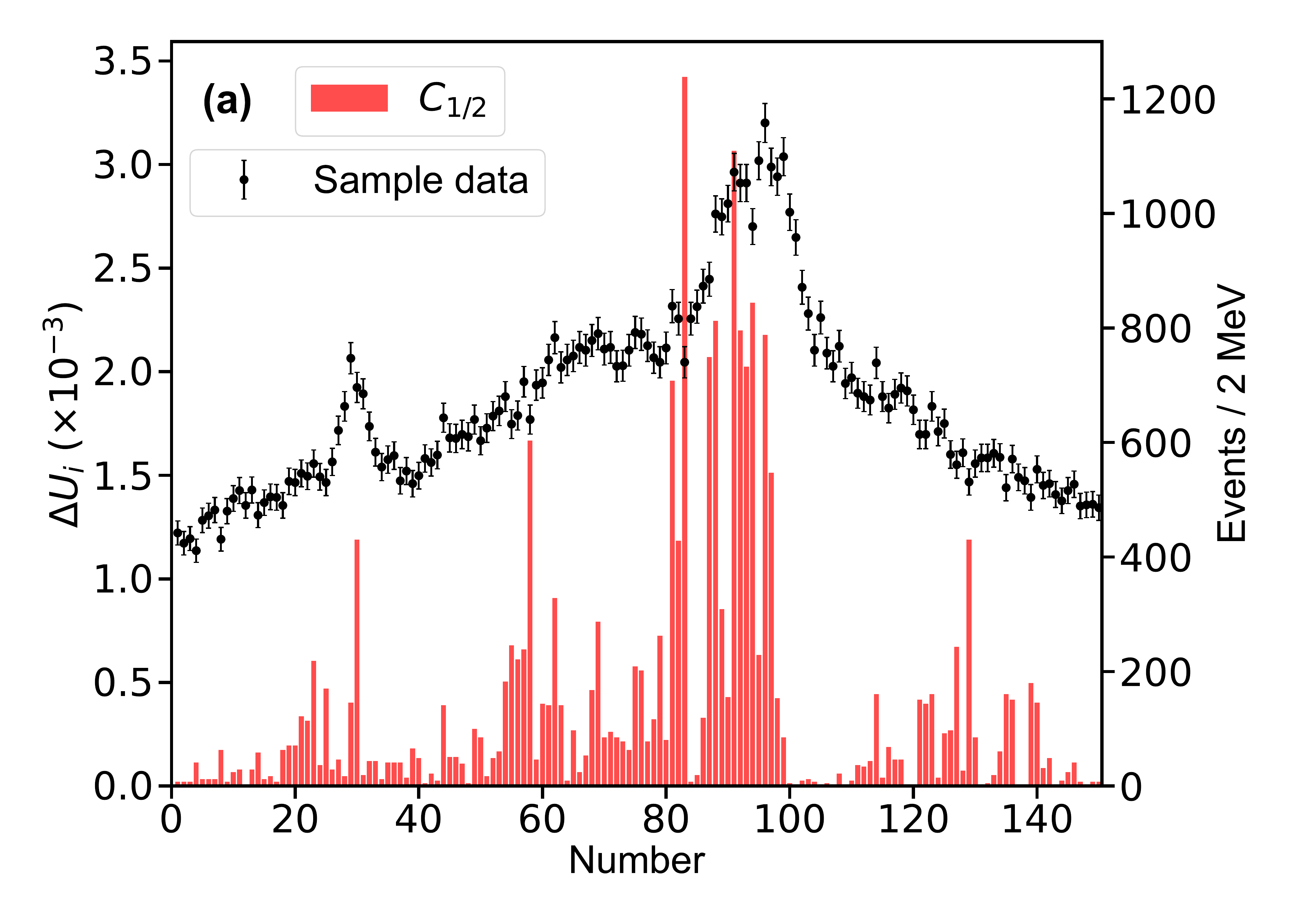}
    \includegraphics[width=0.48\textwidth]{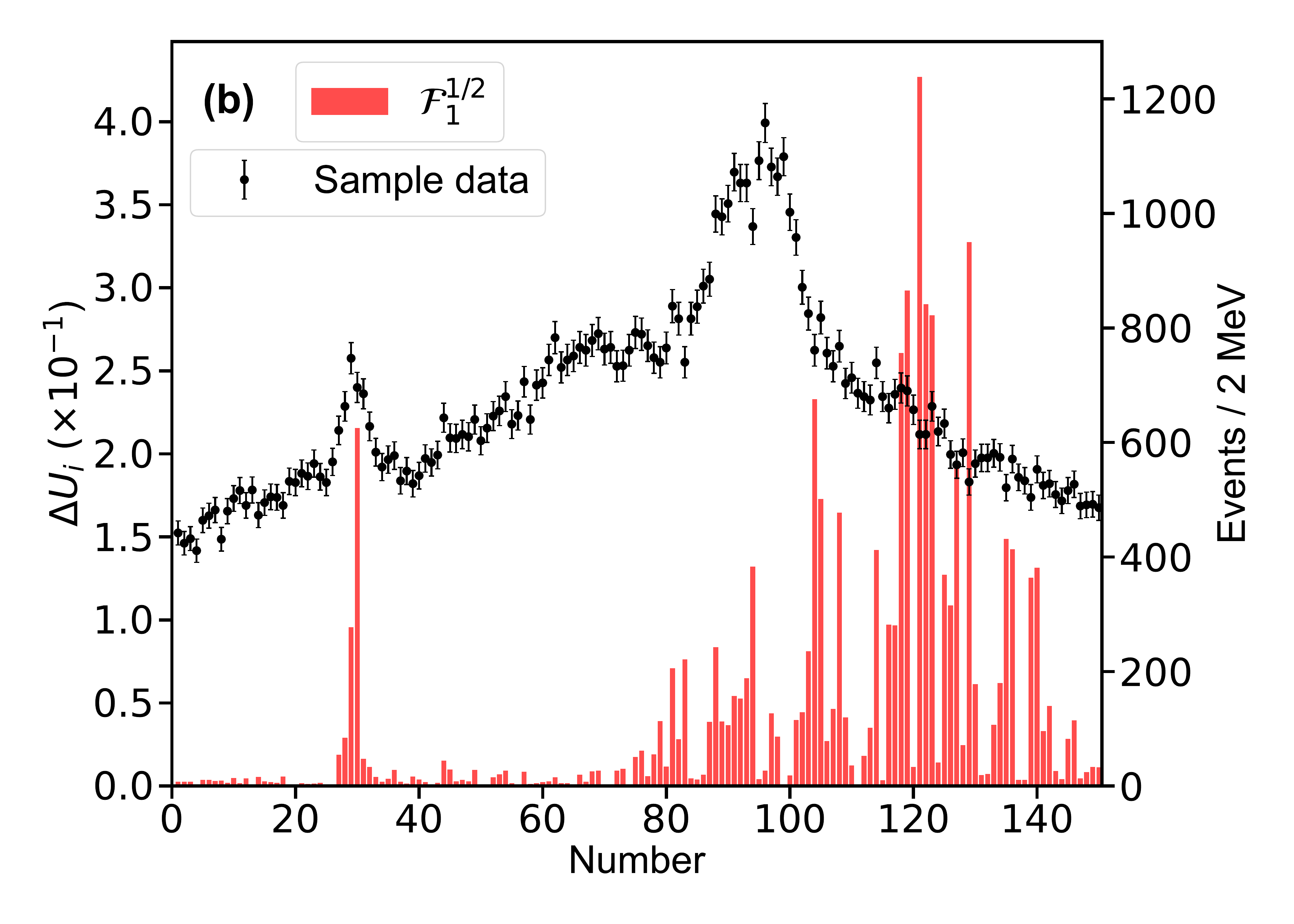}
    \captionsetup{justification=raggedright}
    \caption{(Color online) The $\Delta U_i$ distributions 
    of the parameters $C_{1/2}$ (a) and $\mathcal{F}^{1/2}_1$ (b) for the 150 sample data points.}
    \label{fig:fitresult_11x}
\end{figure}
A comparison is performed by checking the traditional $\chi^2$/ndf-fitting approach,
which minimizes the
\begin{equation}
 \chi^2\equiv\sum_k^{N} \left(\frac{N(k;\mathcal{P})-N^d(k)}{\sigma_{N^d}(k)}\right)^2
\end{equation}
between the theoretical prediction $N(k;\mathcal{P})$ and experimental data $N^d(k)$ by summing over all bins of $\mathcal{H}$,
where $\mathcal{P}$ represents the model parameters to be determined and $k$ is the bin index.
We check how the $j$th bin of $\mathcal{H}$ impacts on the parameters $\mathcal{P}$ and 
the pole positions.
\begin{figure*}[hbtp]
    \centering
    \includegraphics[width=0.32\textwidth]{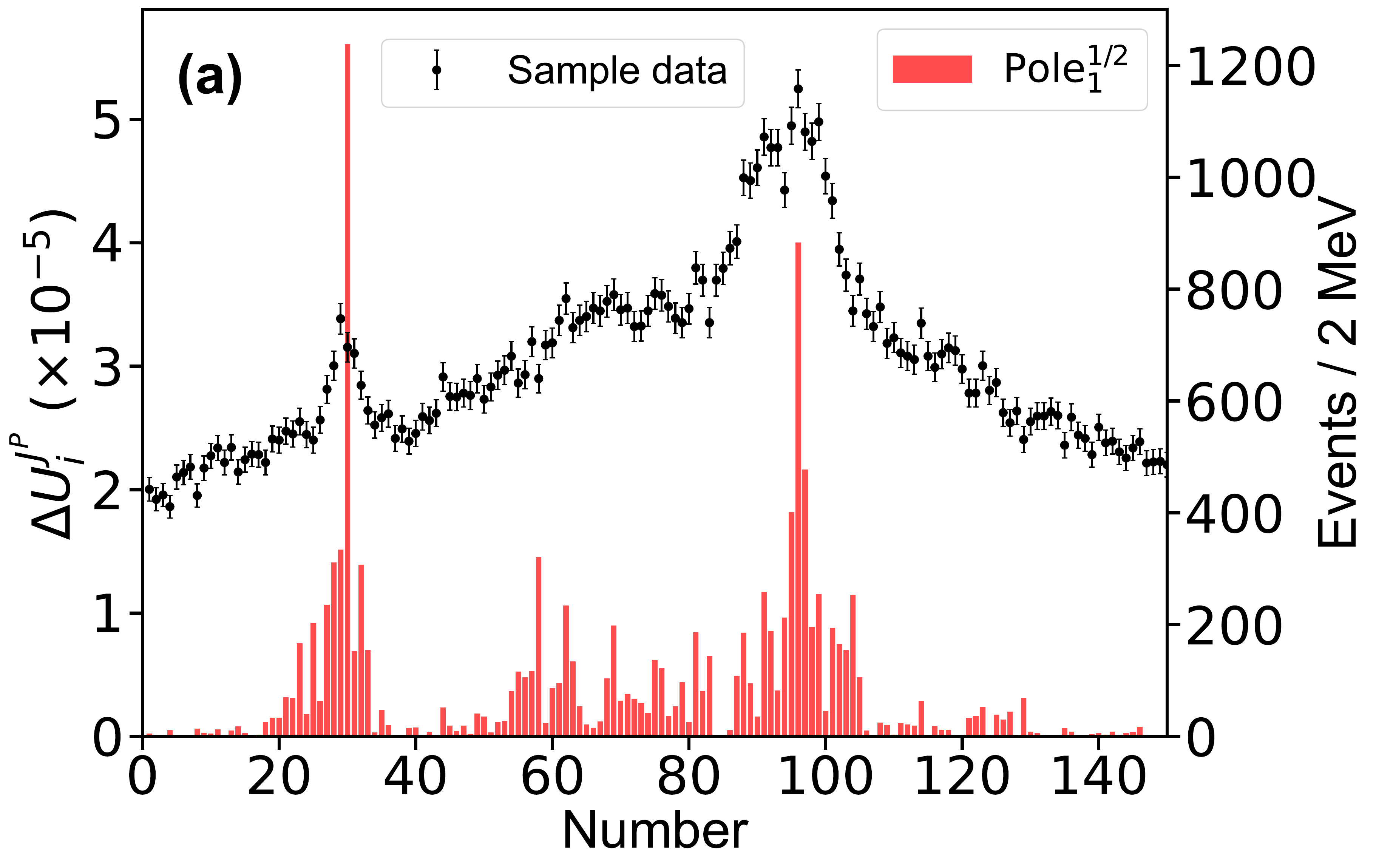}
    \includegraphics[width=0.32\textwidth]{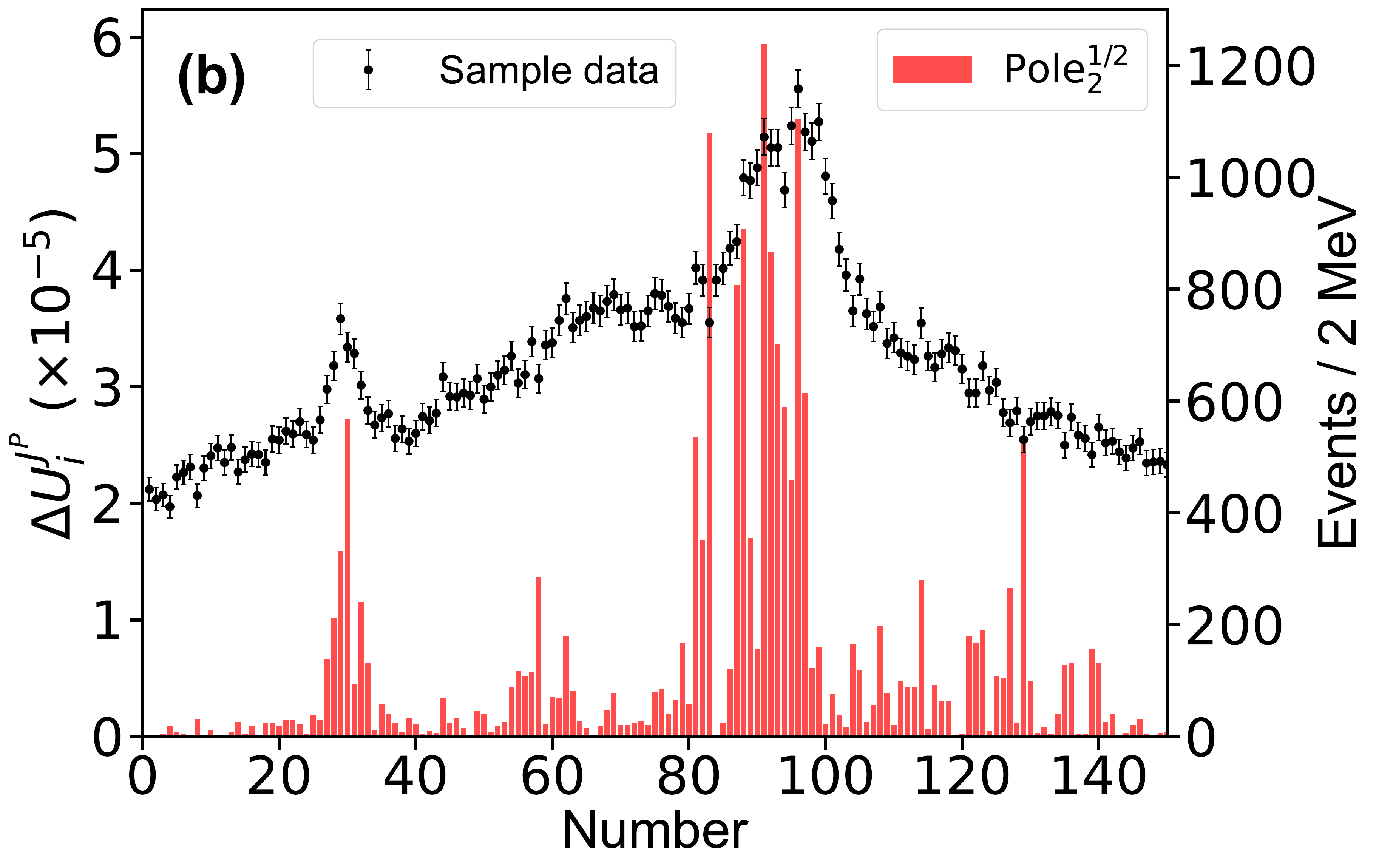}
    \includegraphics[width=0.32\textwidth]{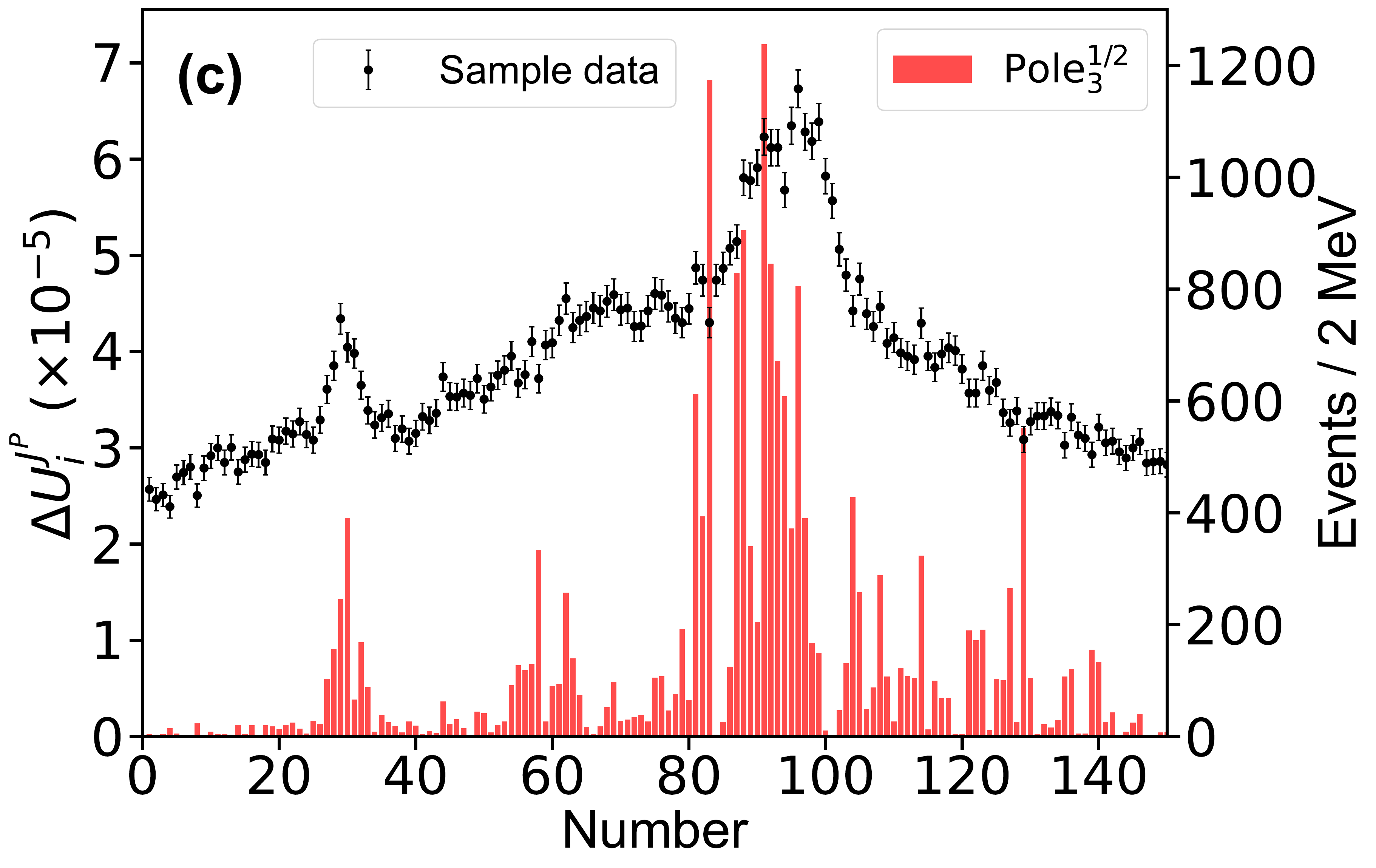}
    \includegraphics[width=0.32\textwidth]{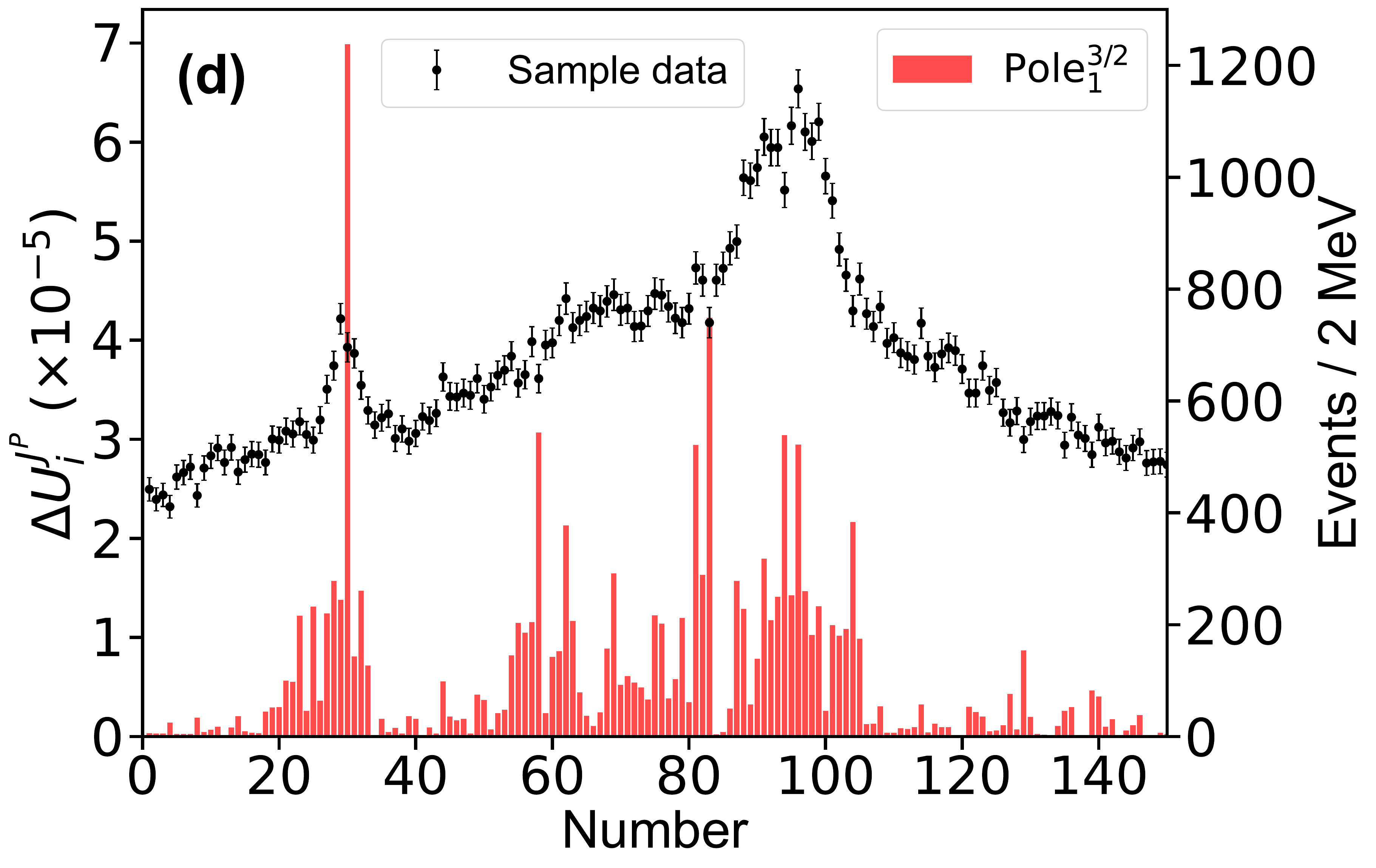}
    \includegraphics[width=0.32\textwidth]{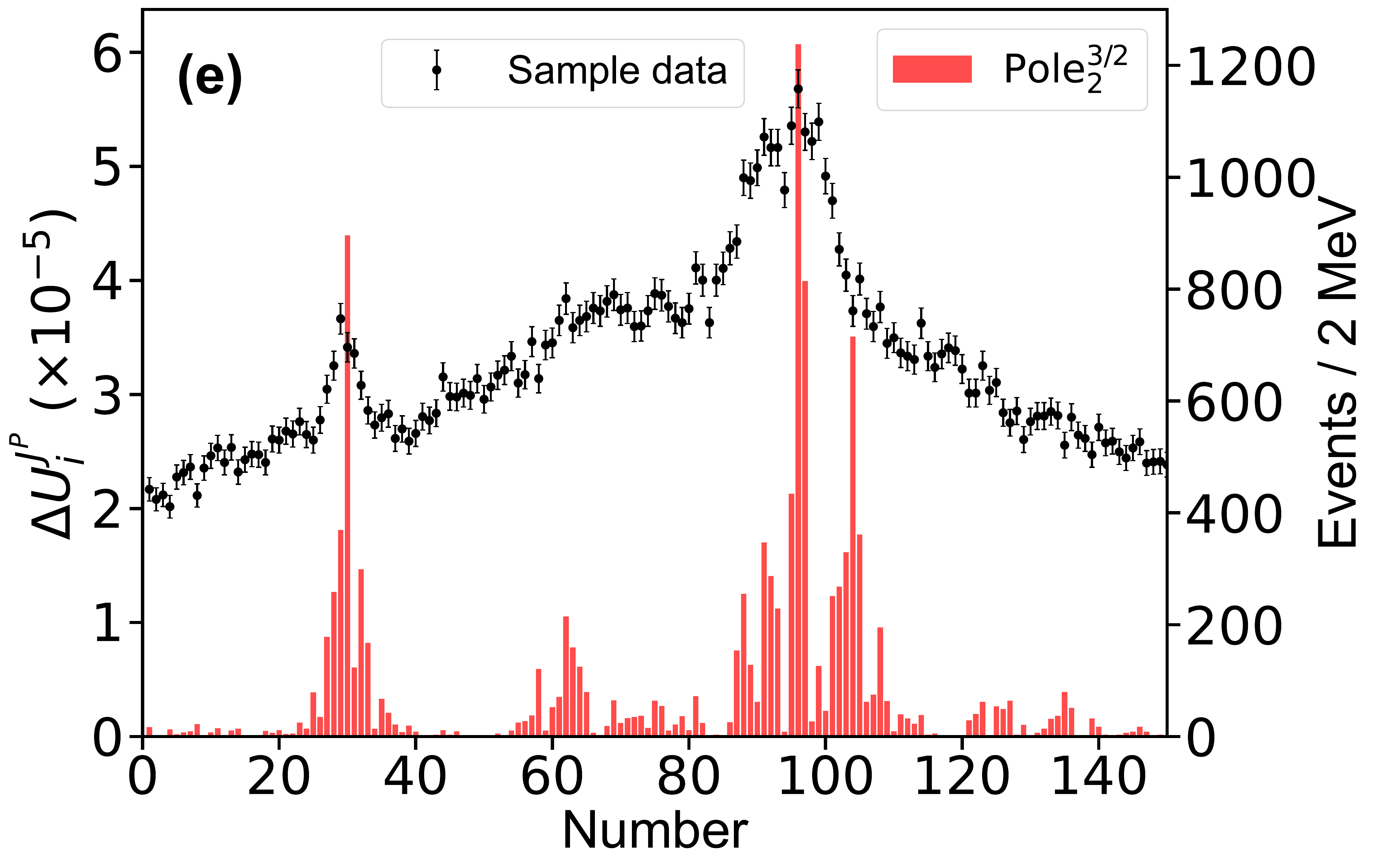}
    \includegraphics[width=0.32\textwidth]{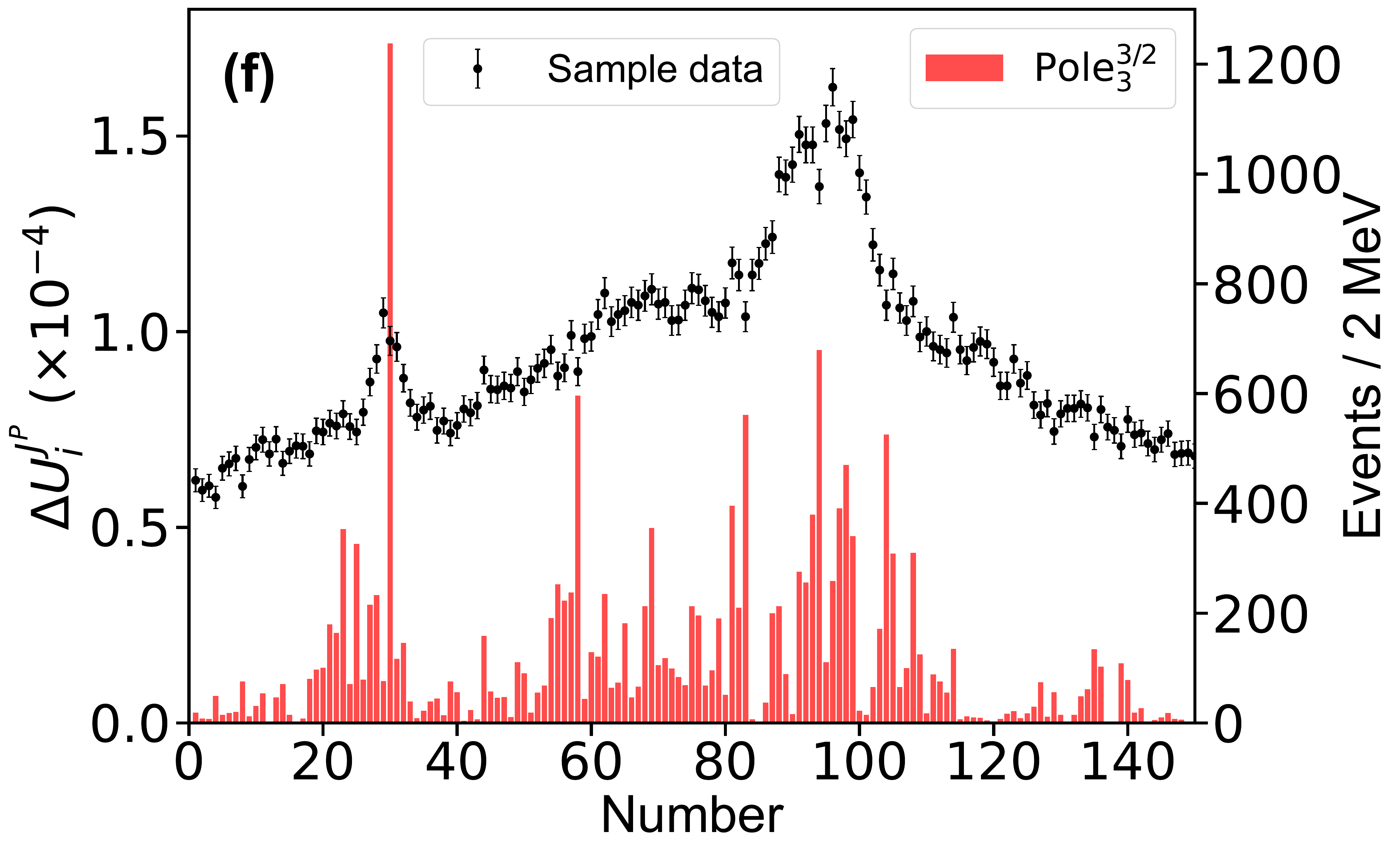}
    \includegraphics[width=0.32\textwidth]{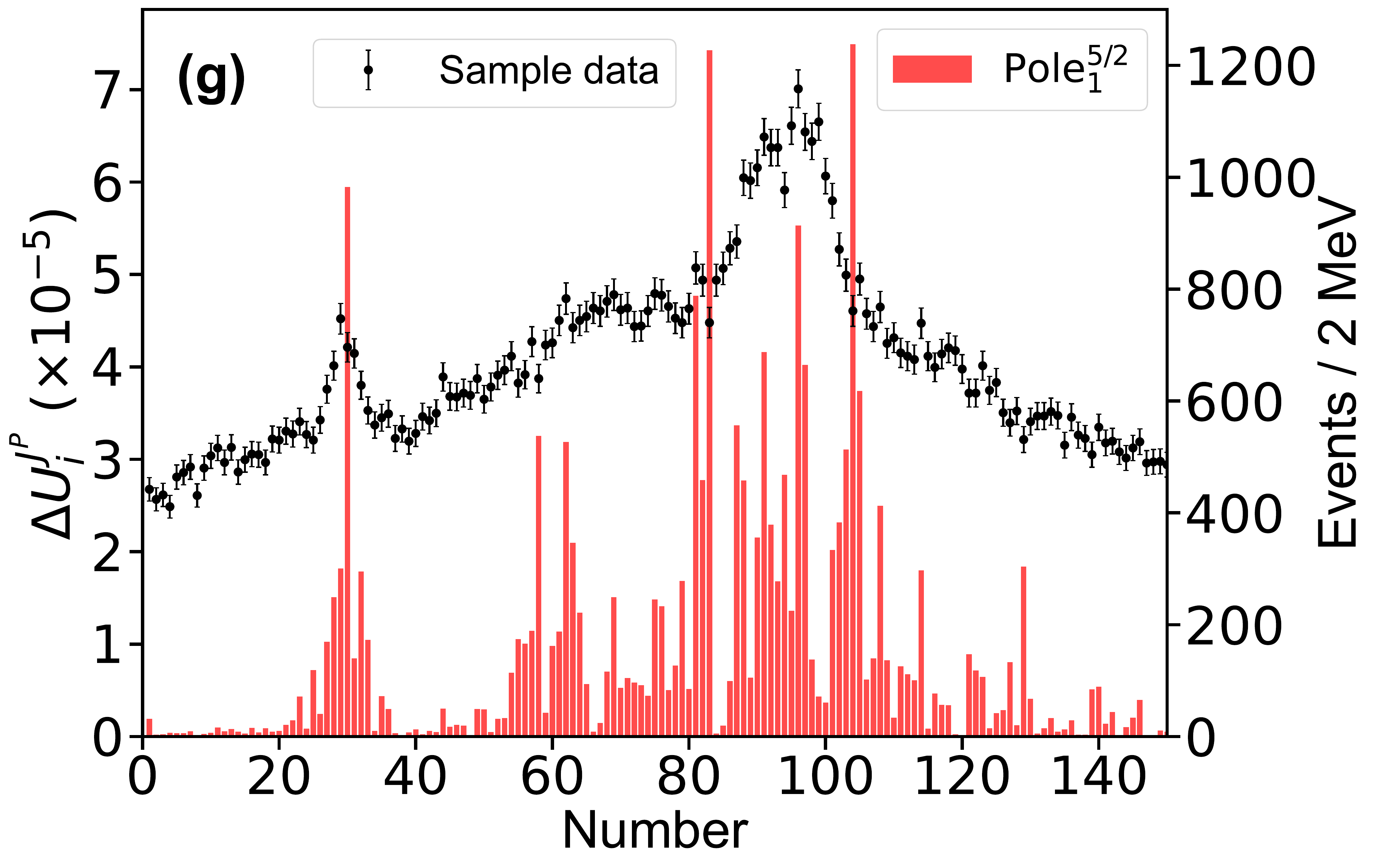}
    \captionsetup{justification=raggedright}
    \caption{(Color online) The $\Delta U_i^{J^P}$ distributions for the seven pole positions of the $J^{P}=\frac{1}{2}^-$
    channel (a), (b), (c), the $J^{P}=\frac{3}{2}^-$
    channel (d), (e), (f) and the $J^{P}=\frac{5}{2}^-$
    channel (g). The poles in the $J^P=\frac{1}{2}^-,\frac{3}{2}^-$ channels are
    denoted as $\mathrm{Pole}_i^{J}$, with 
    $i=1,2,3$ from lower to higher energy.
    The pole in the $J^P=\frac{5}{2}^-$ channel is
    denoted as $\mathrm{Pole}_1^{5/2}$.
    The data is a simulated sample for Solution A. 
    The distributions do not consider the correlation among the parameters.}
    \label{fig:fitpoleresult_12}
\end{figure*}
 
For this purpose, we define the uncertainty $\Delta U_i \equiv |\frac{\mathcal{P}_i(\mathrm{on})-\mathcal{P}_i(\mathrm{off})}{\mathcal{P}_i(\mathrm{on})}|$ for the $i$th parameter of $\mathcal{P}$, which is determined with the central values obtained by switching on/off 
the $j$th bin of $\mathcal{H}$ in the fitting.
The bigger uncertainty observed by excluding a bin means a larger impact power of the bin on a parameter.
Fig.~\ref{fig:fitresult_11x} illustrates the $\Delta U_i$ distributions for two of  the eleven parameters.
The distributions for other parameters can be found in the Supplementary materials.
The bins near the peaks in the mass spectrum do not show stronger constraints on the uncertainties of the parameters $C_{1/2}$ and $\mathcal{F}^{1/2}_1$ than the others. That is because that the model parameters are correlated to each other and do not reflect the underlying physics directly. 
In addition, to check the impact of the $j$th experimental point 
on the pole positions of the $J^P$ channel, the uncertainty 
\begin{equation}
\Delta U_i^{J^P} \equiv \left|\frac{\mathrm{Re}[\mathrm{Pole}_i](\mathrm{on})-\mathrm{Re}[\mathrm{Pole}_i]
(\mathrm{off})}{\mathrm{Re}[\mathrm{Pole}_i](\mathrm{on})}\right|
\end{equation}
for the real part of the $i$th pole position is defined by switching on/off 
the $j$th bin of $\mathcal{H}$ in the fitting. A similar behavior can be found for the imaginary part of the poles, see Supplementary materials.
Fig.~\ref{fig:fitpoleresult_12} illustrates
the $\Delta U_i^{J^P}$ distributions for the $J^P=\frac{1}{2}^-, \frac{3}{2}^-$ and $\frac{5}{2}^-$ channels,
respectively, without considering the correlation among the parameters. 
Although those values are of the order $10^{-5}$, 
one can still see that the experimental data around
the $\Sigma_c\bar{D}$, $\Sigma_c\bar{D}^*$, and $\Sigma_c^*\bar{D}$
thresholds are more important than the others.  
The reason why the data around the $\Sigma_c^*\bar{D}^*$ threshold is not 
important is that the small production amplitude of the $\Sigma_c^*\bar{D}^*$ channel
and the experimental data have little constraint about the corresponding poles~\cite{Du:2019pij,Du:2021fmf}. 
One can also obtain the same conclusion from Fig.~\ref{fig:fitpoleresult_12}g,
where the experimental data around the $J^P=\frac 52^-$ dynamic channel $\Sigma_c^*\bar{D}^*$
do not show any significance. Due to the coupled-channel effect,
the experimental data around the coupled channels still have strong constraints
on the physics, e.g. the seven poles, around the other coupled channels.
Taking the first pole of the $J^P=\frac 12^-$ channel as an example, e.g.
Fig.~\ref{fig:fitpoleresult_12}a, the data around the $\Sigma_c\bar{D}^*$ threshold are
as significant as those around the $\Sigma_c\bar{D}$ threshold.
As the sample data of Fig.~\ref{fig:fitpoleresult_12} corresponds to Solution~A, the data around $P_c(4440)$ ($P_c(4457)$)
are more important than those around the $P_c(4457)$ ($P_c(4440)$)
in Fig.~\ref{fig:fitpoleresult_12}b (Fig.~\ref{fig:fitpoleresult_12}e) as expected.

\vspace{0.2cm}
\section{Conclusion}
\label{sec:con}

We have investigated the nature of the famous 
hidden charm pentaquarks with a NN-based approach in a pionless EFT,
which strongly favors $J_{P_{c}(4440)}=\frac 12$ and  $J_{P_{c}(4457)}=\frac 32$,
i.e. solution~A in Refs~\cite{Liu:2019tjn,Du:2019pij,Du:2021fmf,Wang:2019ato,Yang:2011wz}.
From this NN,
we find that solution~A is systematically preferred over  solution~B.
This conclusion is based on the pionless EFT, 
which is used to illustrate the difference between the machine learning and the normal fitting approach.
Furthermore, we also performed checks on both 
the NN-based approach and the $\chi^2$/ndf-fitting approach.
Our conclusion is that both approaches work well on the MC simulation samples.
In the NN-based approach, the role of each data bin on the underlying physics
is well reflected by the SHAP value. For the $\chi^2$/ndf-fitting approach, 
such a direct relation does not exist. This further explains why the two solutions
can be better distinguished in the NN-based approach than in the $\chi^2$/ndf-fitting approach.
At the same time, the effect of the background fraction on the accuracy of the network
 is accurately obtained, which provides a paradigm for NNs to study exotic
  hadrons by first extracting the background fraction from the data, and then analyzing
   the physics of the possible states from the data with the network trained from
    the corresponding background fraction data.
This study provides more insights about how the NN-based approach 
predicts the nature of exotic states from the mass spectrum. 

\vspace{0.2cm}
{\bf \color{gray}Acknowledgements:}~~We are grateful to Meng-Lin Du, Feng Kun Guo and Christoph Hanhart for the helpful discussion. 
This work is partly supported by 
the National Natural Science Foundation of China with Grant No.~12035007,
Guangdong Provincial funding with Grant No.~2019QN01X172,
Guangdong Major Project of Basic and Applied Basic Research No.~2020B0301030008.
Q.W. and U.G.M. are also supported by the NSFC and the Deutsche Forschungsgemeinschaft (DFG, German
Research Foundation) through the funds provided to the Sino-German Collaborative
Research Center TRR110 ``Symmetries and the Emergence of Structure in QCD"
(NSFC Grant No. 12070131001, DFG Project-ID 196253076-TRR 110).
The work of U.G.M. is further supported by the
Chinese Academy of Sciences (CAS) President's International Fellowship
Initiative (PIFI) (Grant No. 2018DM0034) and Volkswagen Stiftung
(Grant No. 93562).

\vspace{0.2cm}
{\bf \color{gray}Author contributions:}~~Zhenyu Zhang and Jiahao Liu did the calculations. Jifeng Hu, Qian Wang and Ulf-G. Mei{\ss}ner focused on how to use machine learning to solve the problems in hadron physics. All the authors made substantial contributions to the physical and technical discussions as well as the editing of the manuscript. All the authors have read and approved the final version of the manuscript.



\onecolumngrid
\newpage
\appendix
\clearpage
{\large \bf Supplemental Material for ``Revealing the nature of hidden charm pentaquarks with machine learning'' }

\section{Fitting to the  Background Distributions}

\begin{figure}[ht]
    \centering
    \includegraphics[width=0.5\textwidth]{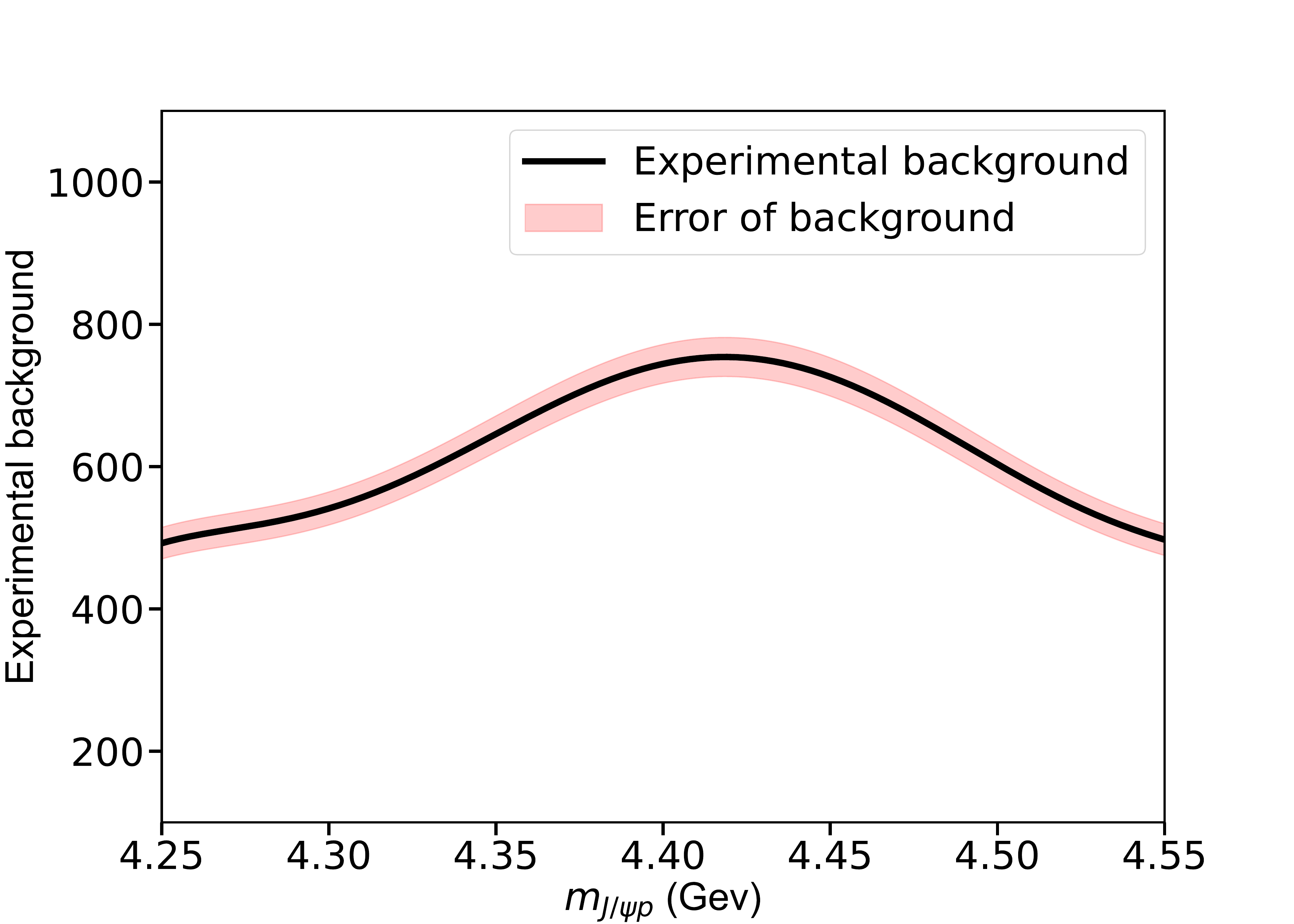}
    \caption{The experimental background distribution of  the invariant mass $\mathcal{M}(J/\psi p)$.}
    \label{fig:Background_data}
\end{figure}

The background, see Fig.~\ref{fig:Background_data}, is described by sixth-order Chebyshev polynomials~\eqref{eq:Chebyshev},
with the coefficients listed in  Tab.~\ref{tab:coefficients}, by
fitting the experimental background. The template recursive functions~\eqref{eq:recursive functions} for
defining the  Chebyshev polynomials is,
\begin{align}
    &T_0(x)=1,\notag\\
    &T_1(x)=x,\notag\\
    &T_2(x)=2x^2-1,\label{eq:recursive functions}\\
    &...\notag\\
    &T_N(x)=2xT_{N-1}(x)-T_{N-2}(x).\notag
\end{align}
This allows for the  implementation of the  Chebyshev polynomials as
\begin{align}
    &\mathrm{Chebyshev}_0(x)=c_0,\notag\\
    &\mathrm{Chebyshev}_1(x)=c_0+c_1x,\notag\\
    &\mathrm{Chebyshev}_2(x)=c_2T_2(x)+\mathrm{Chebyshev}_1(x),\label{eq:Chebyshev}\\
    &...\notag\\
    &\mathrm{Chebyshev}_6(x)=c_6T_6(x)+\mathrm{Chebyshev}_5(x),\notag
\end{align}
where the $c_0,c_1,...$ are coefficients, as given in Tab.~\ref{tab:coefficients} for the problem at hand.

\begin{table}[ht]
    \centering
    \caption{The coefficients of the sixth-order Chebyshev polynomials.}
    \begin{tabular}{ccc}
    \hline
     Coefficients &  Value & Error \\
    \hline
     $c_0$   &67.96 &$\pm$0.18\\
     $c_1$   &4.13  &$\pm$0.18\\
     $c_2$   &$-$16.60&$\pm$0.15\\
     $c_3$   &$-$2.55 &$\pm$0.15\\
     $c_4$   &3.27  &$\pm$0.14\\
     $c_5$   &2.88  &$\pm$0.14\\
     $c_6$   &$-$1.66 &$\pm$0.14\\
    \hline
    \end{tabular}  \label{tab:coefficients}
\end{table}

\section{Determination of the Background Fraction in Data}

\begin{figure}[ht]
    \centering
    \includegraphics[width=0.48\textwidth]{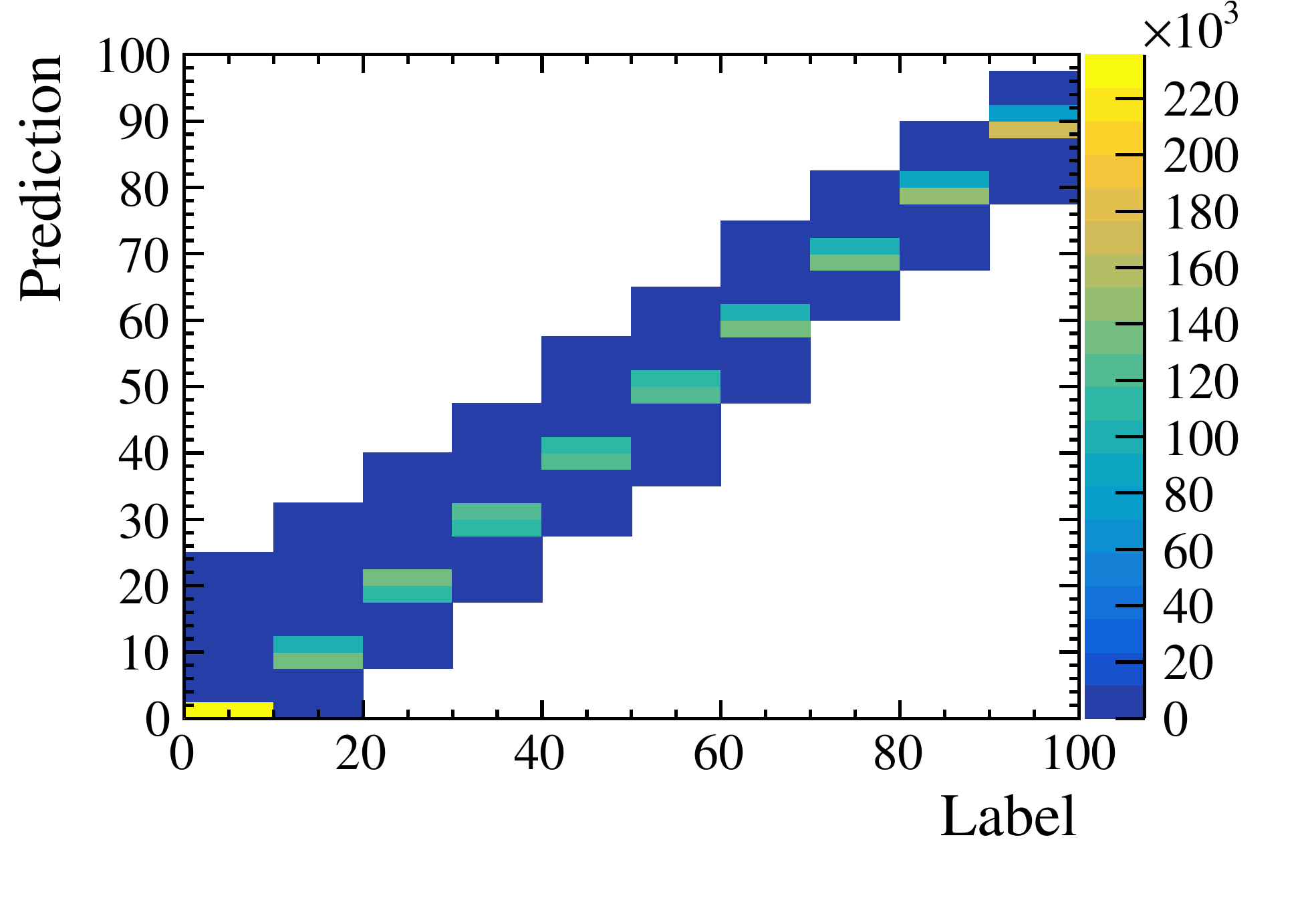}
        \includegraphics[width=0.48\textwidth]{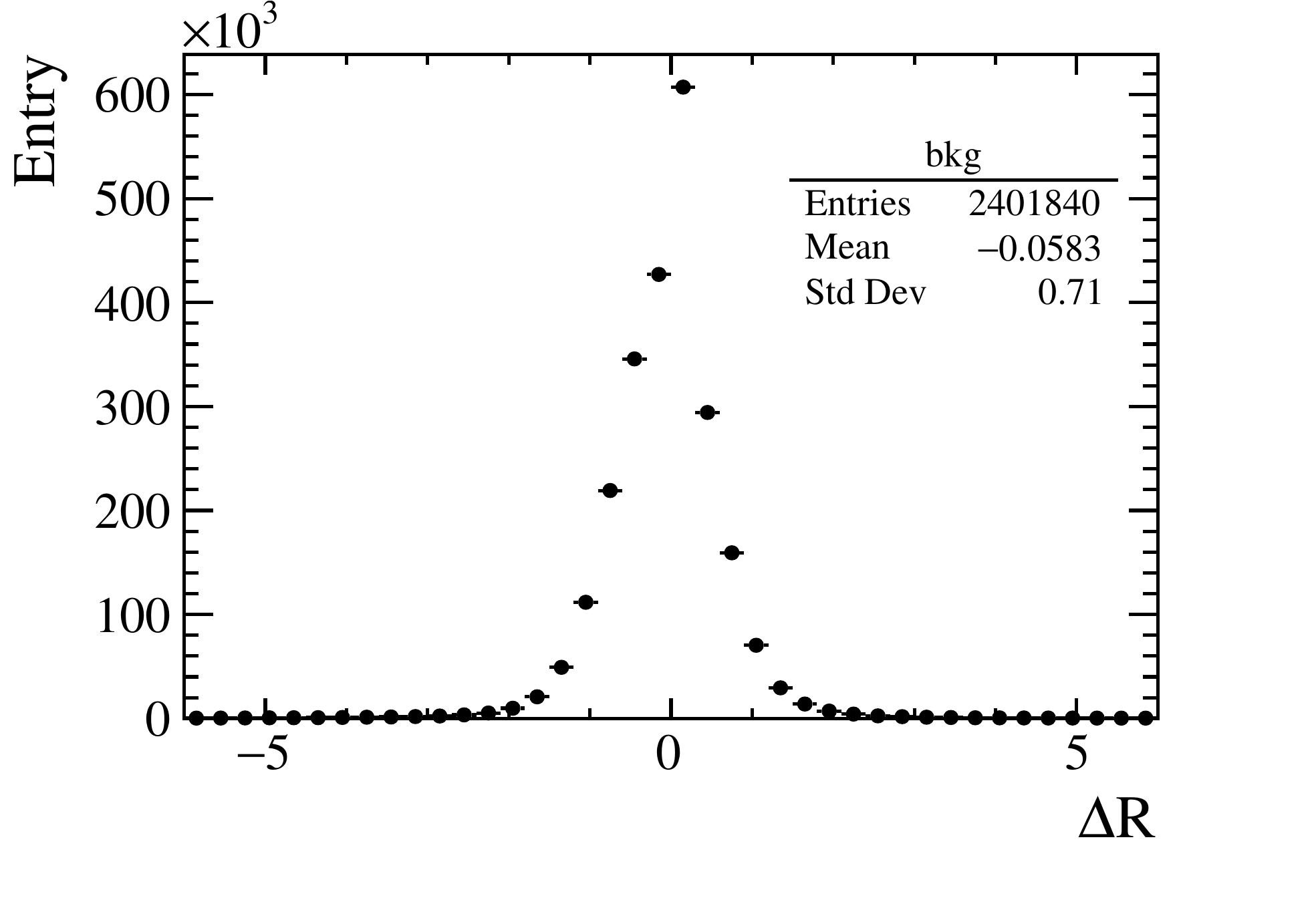}
    \caption{(left) Predicted background fractions versus the background values for samples: $\{\mathcal{S}^{0}...\mathcal{S}^{90}\}$, (right) the difference distributions between predicted background fractions and ground-truth values.}
    \label{fig:AbsErrorDistribution}
\end{figure}

We produce a group of samples for different  background  fractions
from 0\% to 90\% every 10\%, as well as  (92\%, 94\%, 96\%).
We trained a ResNet-based NN which employs the MSELoss function to measure the Euclidean distance between the predicted background fractions and the truth values. 
The NN successfully retrieves the background fraction, as illustrated in Fig.~\ref{fig:AbsErrorDistribution} (left).
The right plot shows the difference between the predicated values and the background values.
The bias and uncertainty values are -0.0583 and 0.71, respectively. We then use this NN to determine the background fraction for experimental data.
The background fraction for data is determined to be $(96.02\pm0.76$)\% in case of training the NN with samples $\{\mathcal{S}^{50}...\mathcal{S}^{90}, \mathcal{S}^{92},\mathcal{S}^{94},\mathcal{S}^{96}\}$,
and is determined to be $(95.72\pm0.72)$\% in case of training the NN with samples  $\{\mathcal{S}^{0}...\mathcal{S}^{90}, \mathcal{S}^{92},\mathcal{S}^{94},\mathcal{S}^{96}\}$.
The background fraction is determined to be $(95.79\pm0.76)$\% predicted with a different training. 
In short, the background fraction is taken as  their average value $(96.0\pm0.8)\%$.

\section{The impact of the energy bins on the model parameters}
\label{supp_weights_onpars}
\hypertarget{item:C}{}

\begin{figure}[ht]
    \centering
    \includegraphics[width=0.32\textwidth]{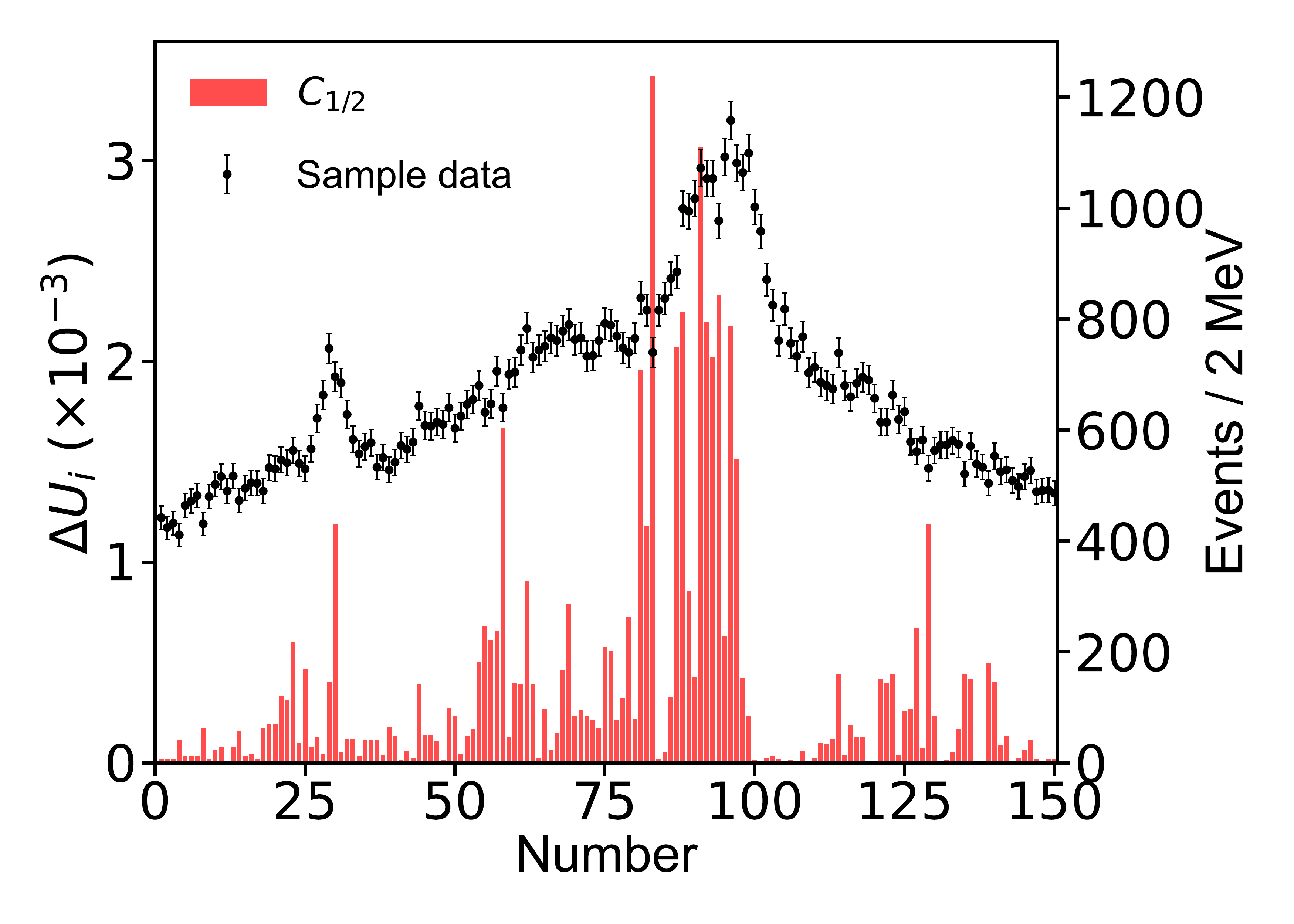}
    \includegraphics[width=0.32\textwidth]{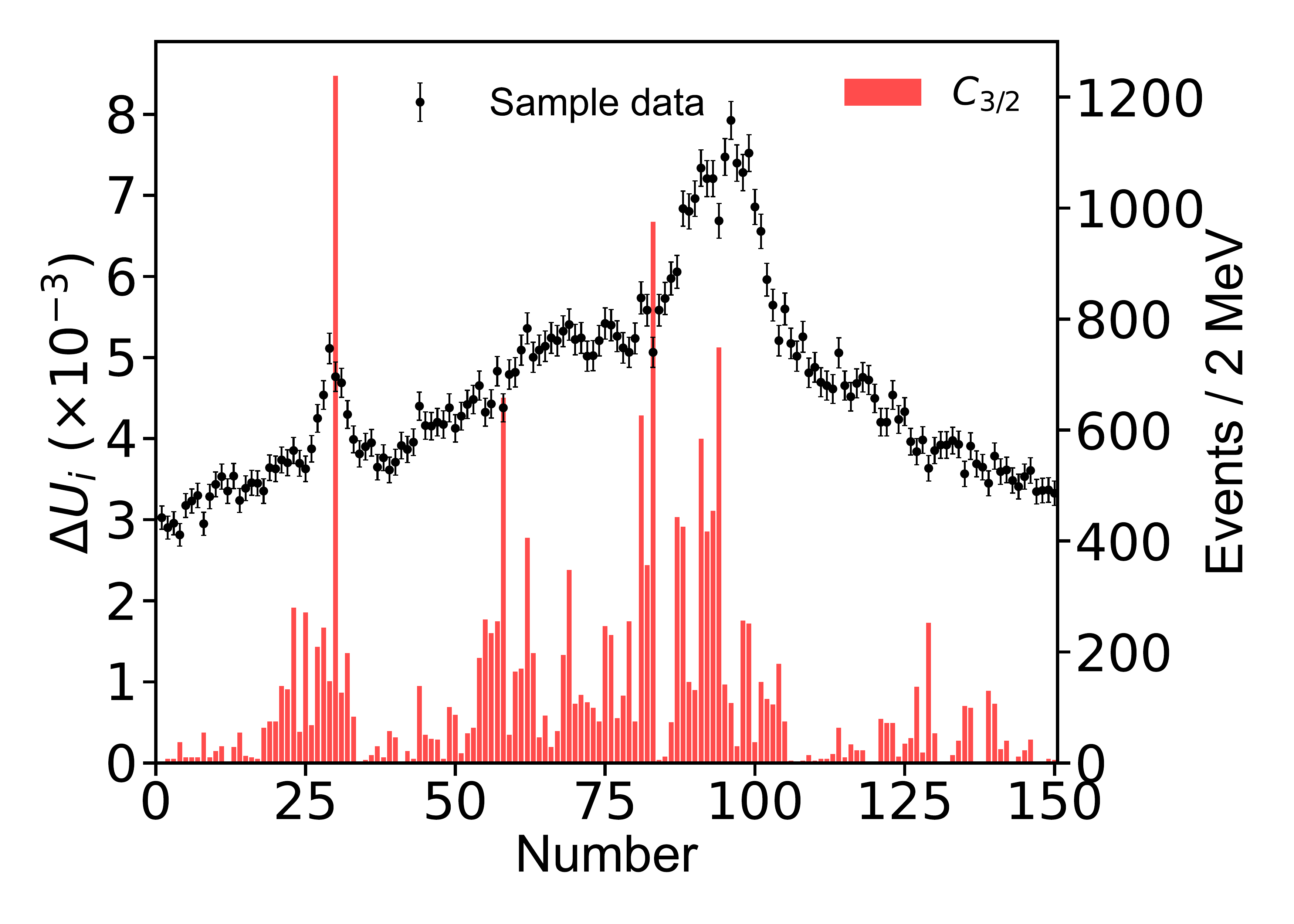}
    \includegraphics[width=0.32\textwidth]{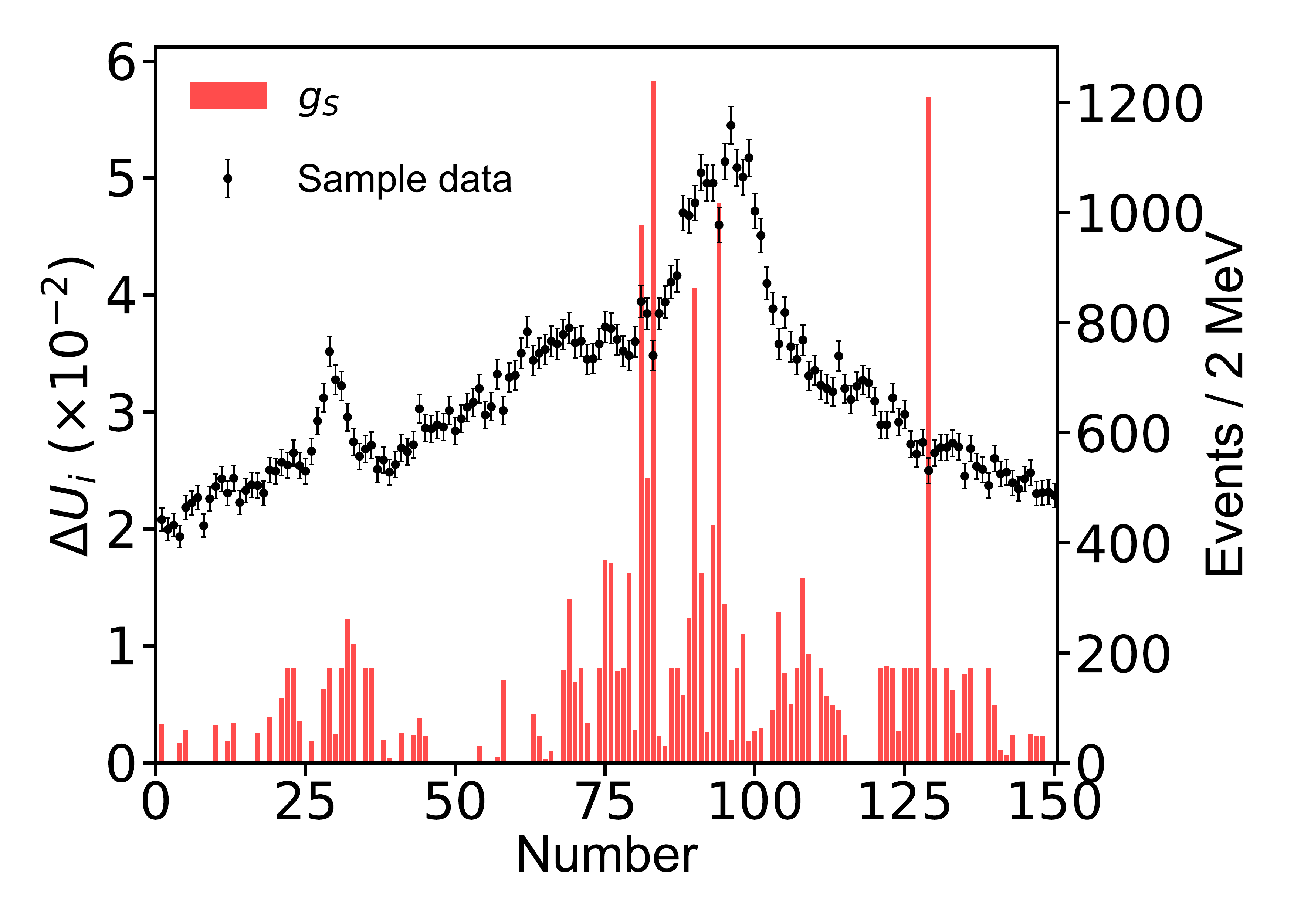}
    \includegraphics[width=0.32\textwidth]{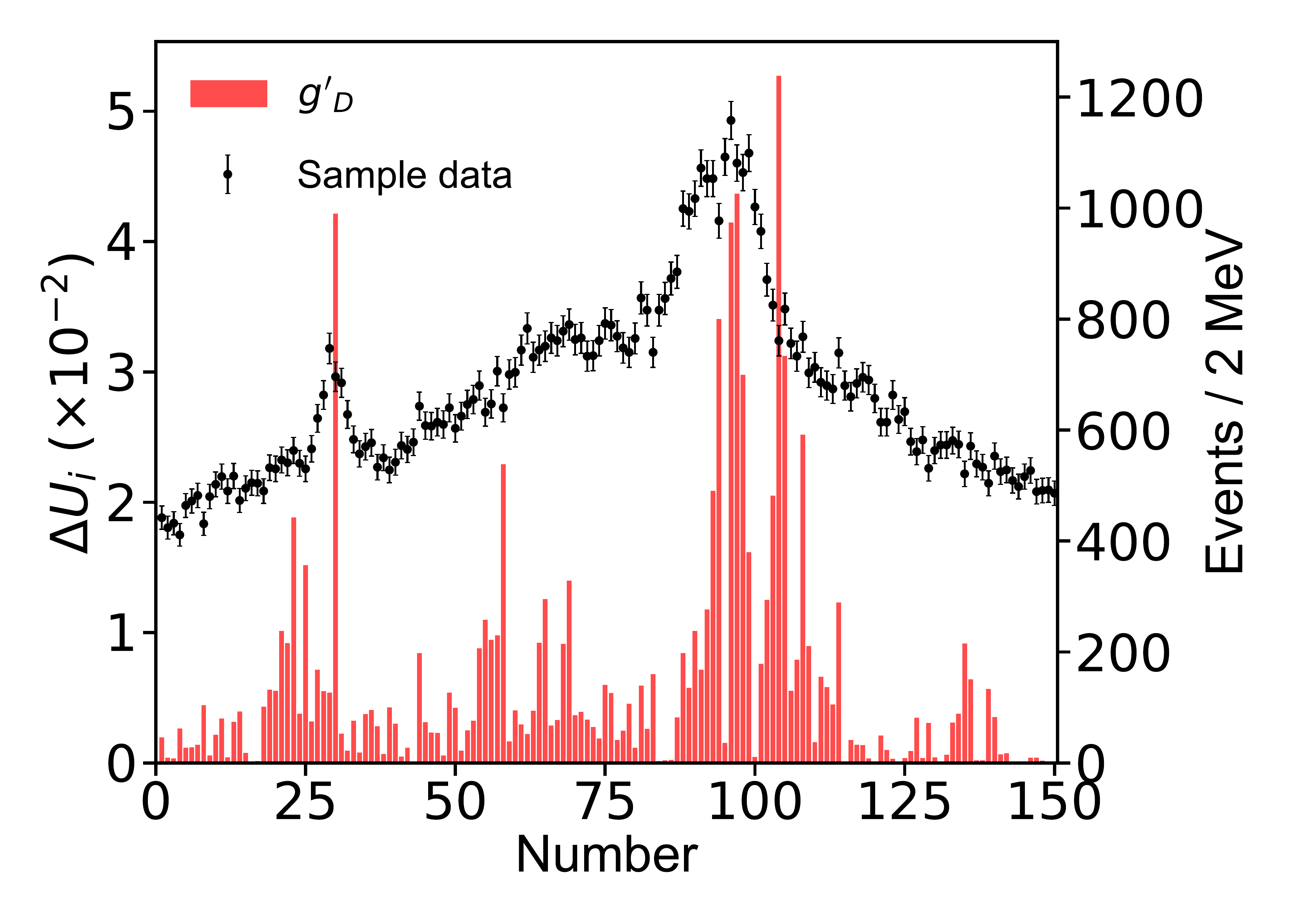}
    \includegraphics[width=0.32\textwidth]{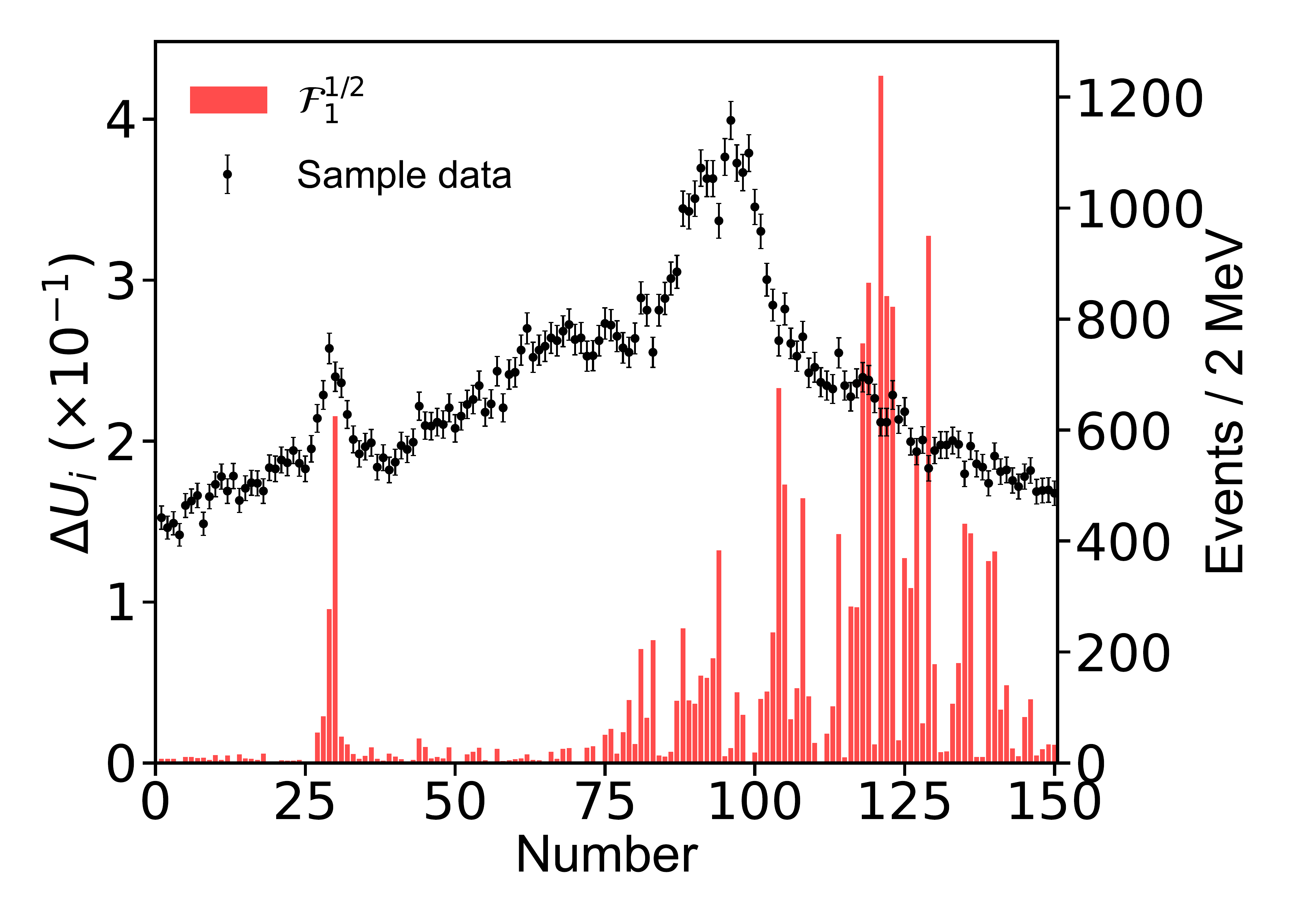}
    \includegraphics[width=0.32\textwidth]{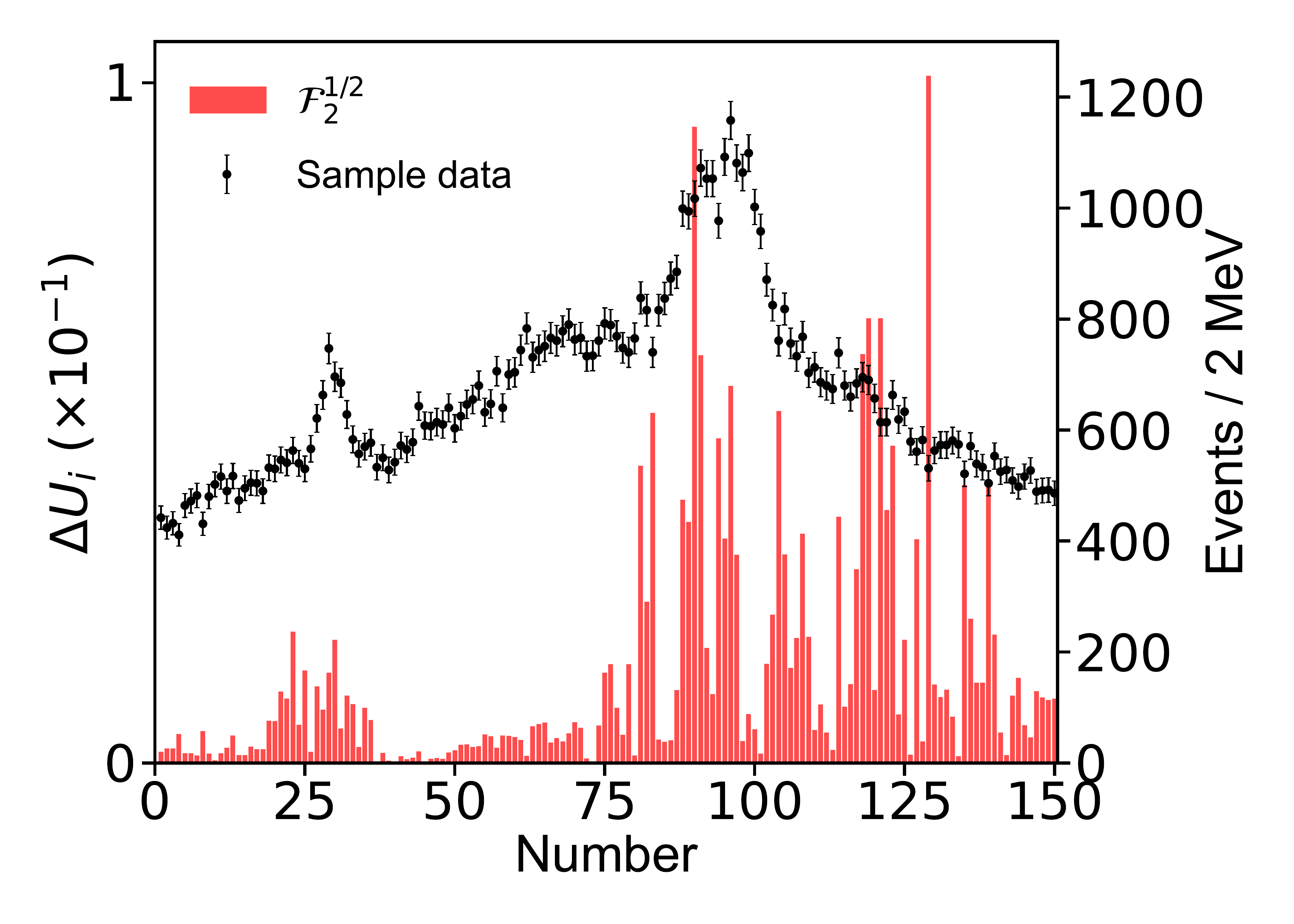}
    \includegraphics[width=0.32\textwidth]{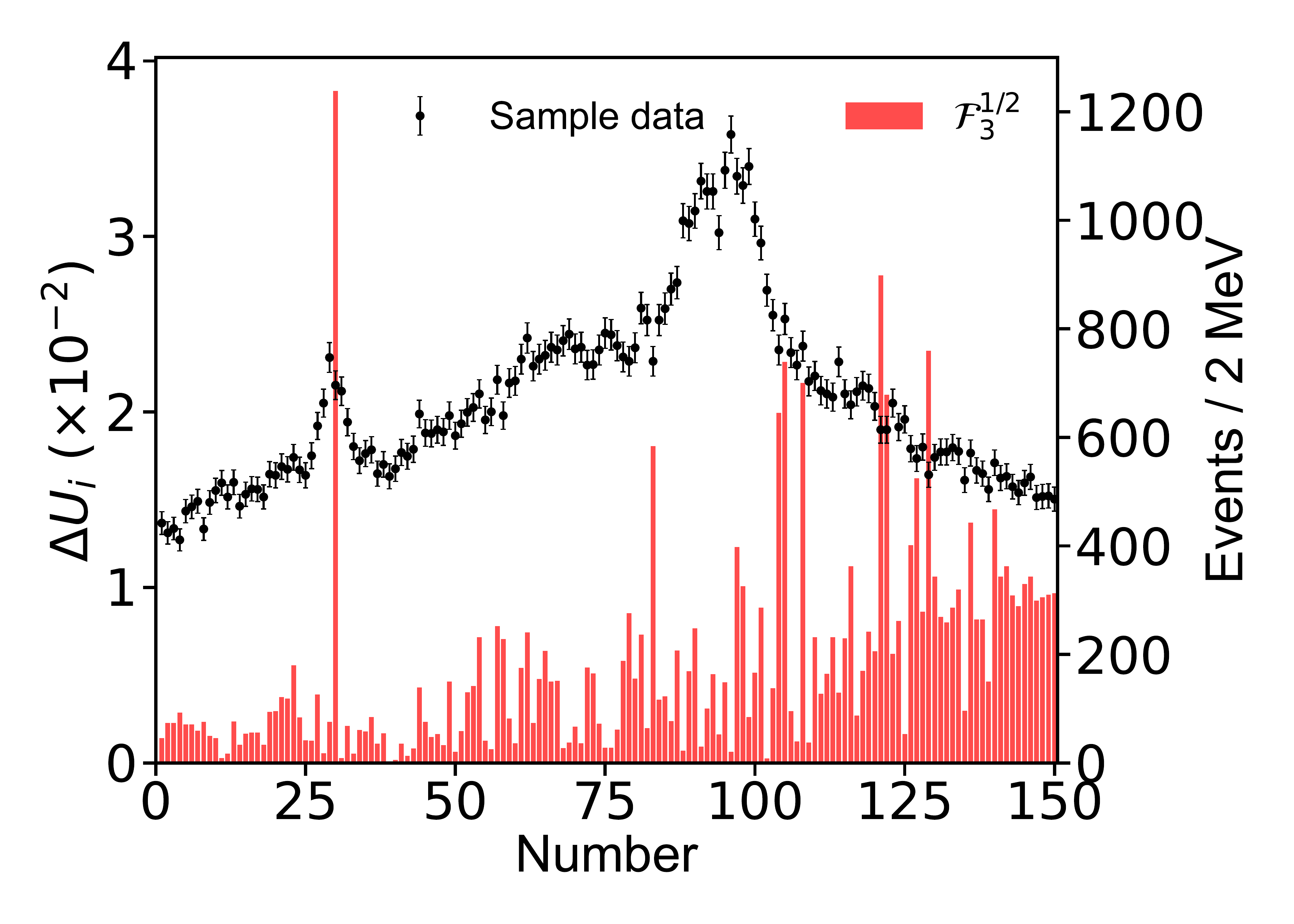}
    \includegraphics[width=0.32\textwidth]{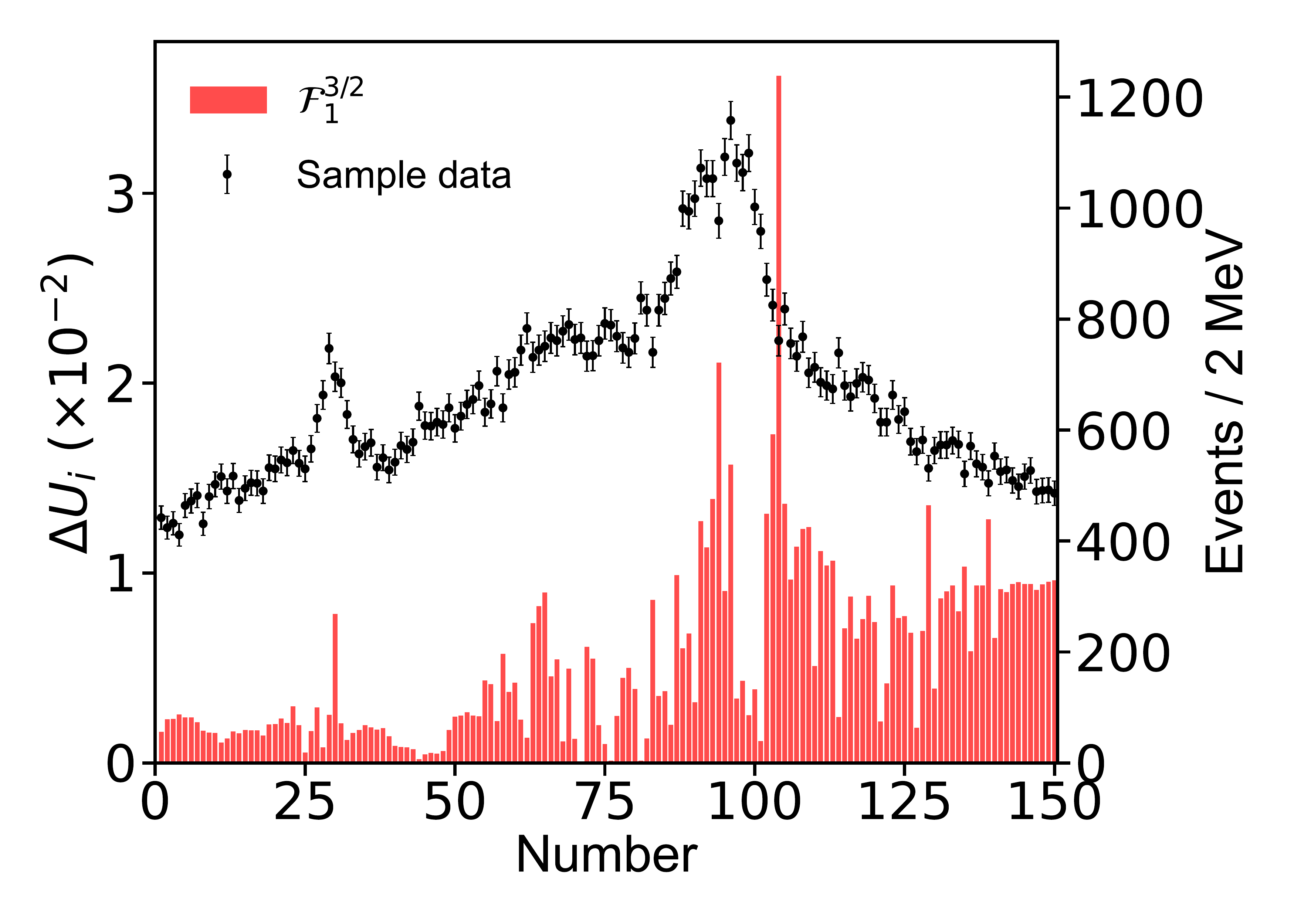}
    \includegraphics[width=0.32\textwidth]{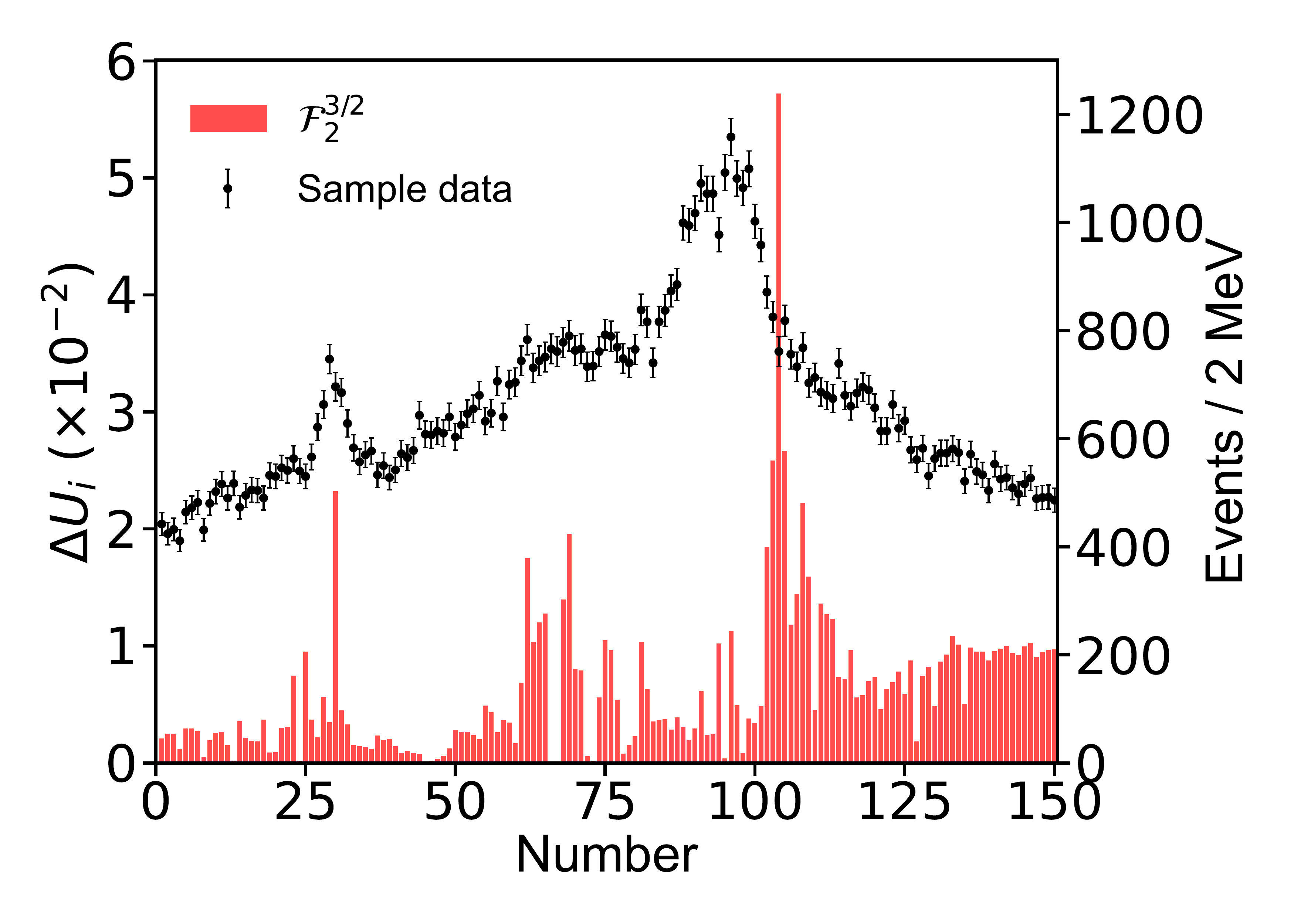}
    \includegraphics[width=0.32\textwidth]{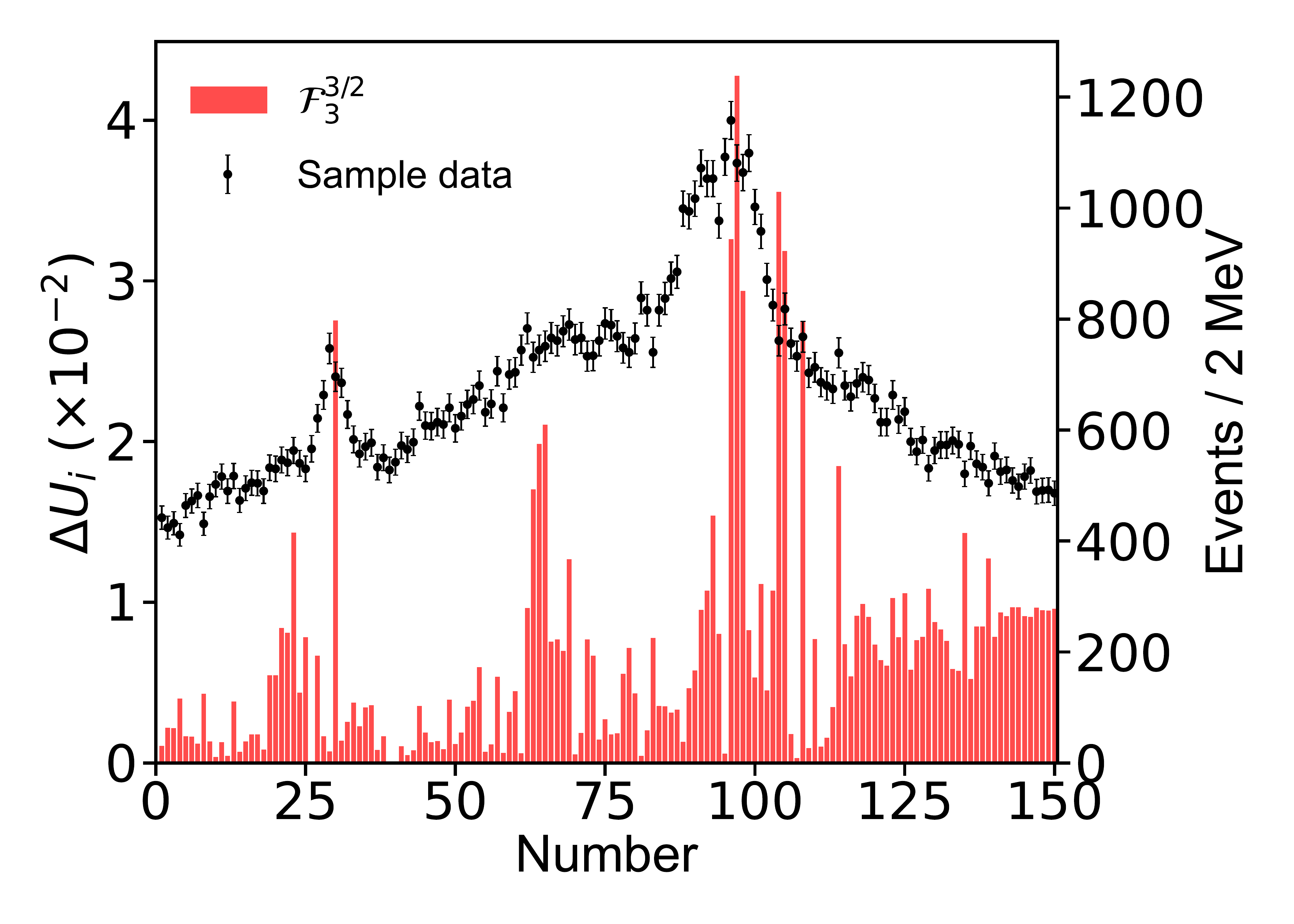}
    \includegraphics[width=0.32\textwidth]{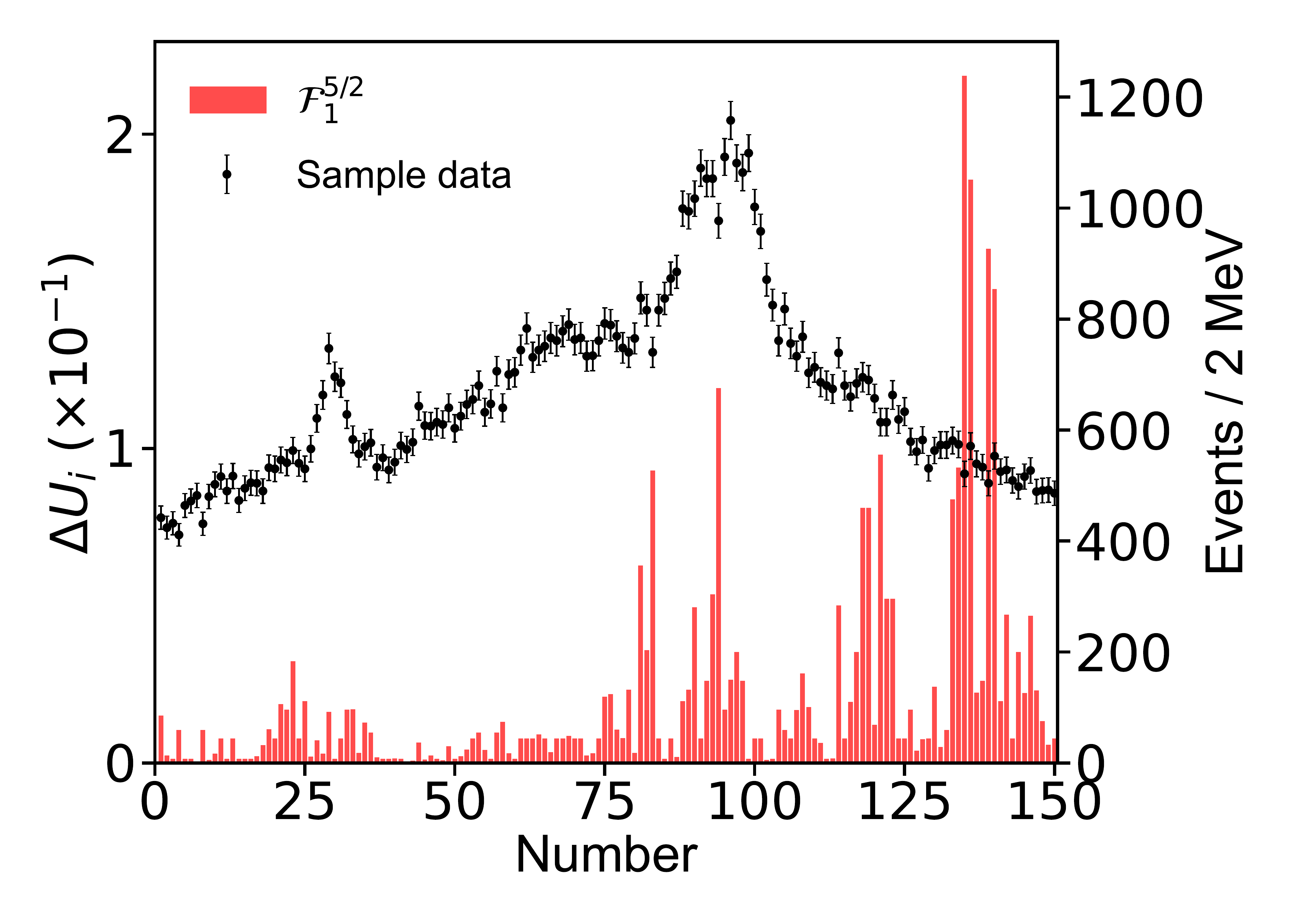}
    \caption{The $\Delta U_i$ distributions for each parameter (from left to right and top to bottom): $C_{1/2}$, $C_{3/2}$, $g_S$, $g'_D$, $\mathcal{F}^{1/2}_1$, $\mathcal{F}^{1/2}_2$, $\mathcal{F}^{1/2}_3$, $\mathcal{F}^{3/2}_1$, $\mathcal{F}^{3/2}_2$, $\mathcal{F}^{3/2}_3$, $\mathcal{F}^{5/2}_1$.}
    \label{fig:fitresult_11}
\end{figure}

We define the uncertainty
\begin{equation}
\Delta U_i \equiv \left|\frac{\mathcal{P}_i(\mathrm{on})-\mathcal{P}_i(\mathrm{off})}{\mathcal{P}_i(\mathrm{on})}\right|
\end{equation}
for the $i$th parameter of $\mathcal{P}$, which is determined with the central values obtained by switching on/off 
the $j$th bin of $\mathcal{H}$ in the fitting. 
We use $\Delta U_i$ to measure the influence of each data point on a parameter.
Figure~\ref{fig:fitresult_11} illustrates $\Delta U_i$ distributions w.r.t eleven parameters.
The larger $\Delta U_i$ points correspond to data points deviating strongly from the interpolation of
its neighbour data points. 

\clearpage
\section{The impact of the energy bins on the imaginary part of the pole positions.}
\label{supp_weights}
\hypertarget{item:C}{}

\begin{figure*}[ht]
    \centering
    \includegraphics[width=0.32\textwidth]{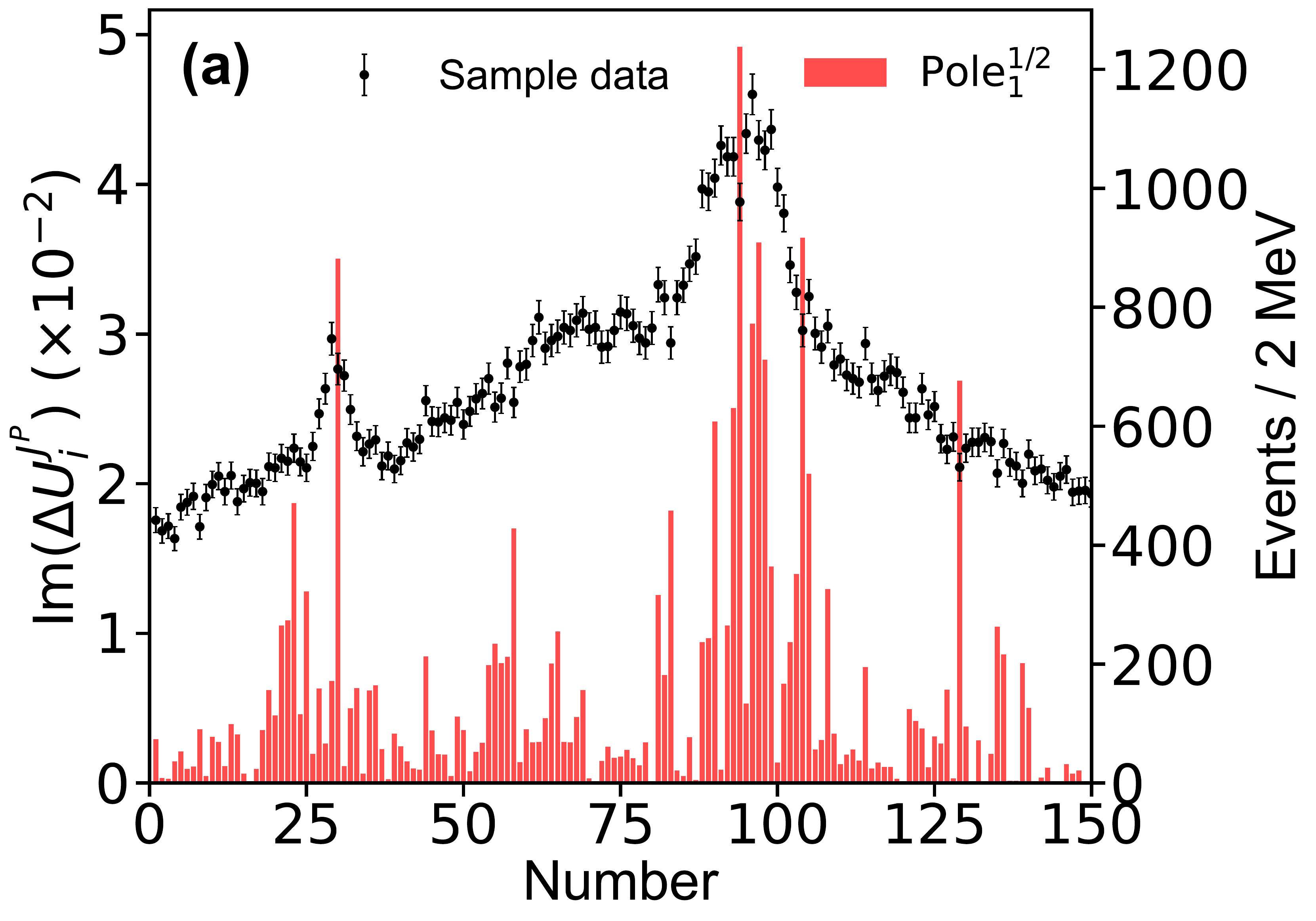}
    \includegraphics[width=0.32\textwidth]{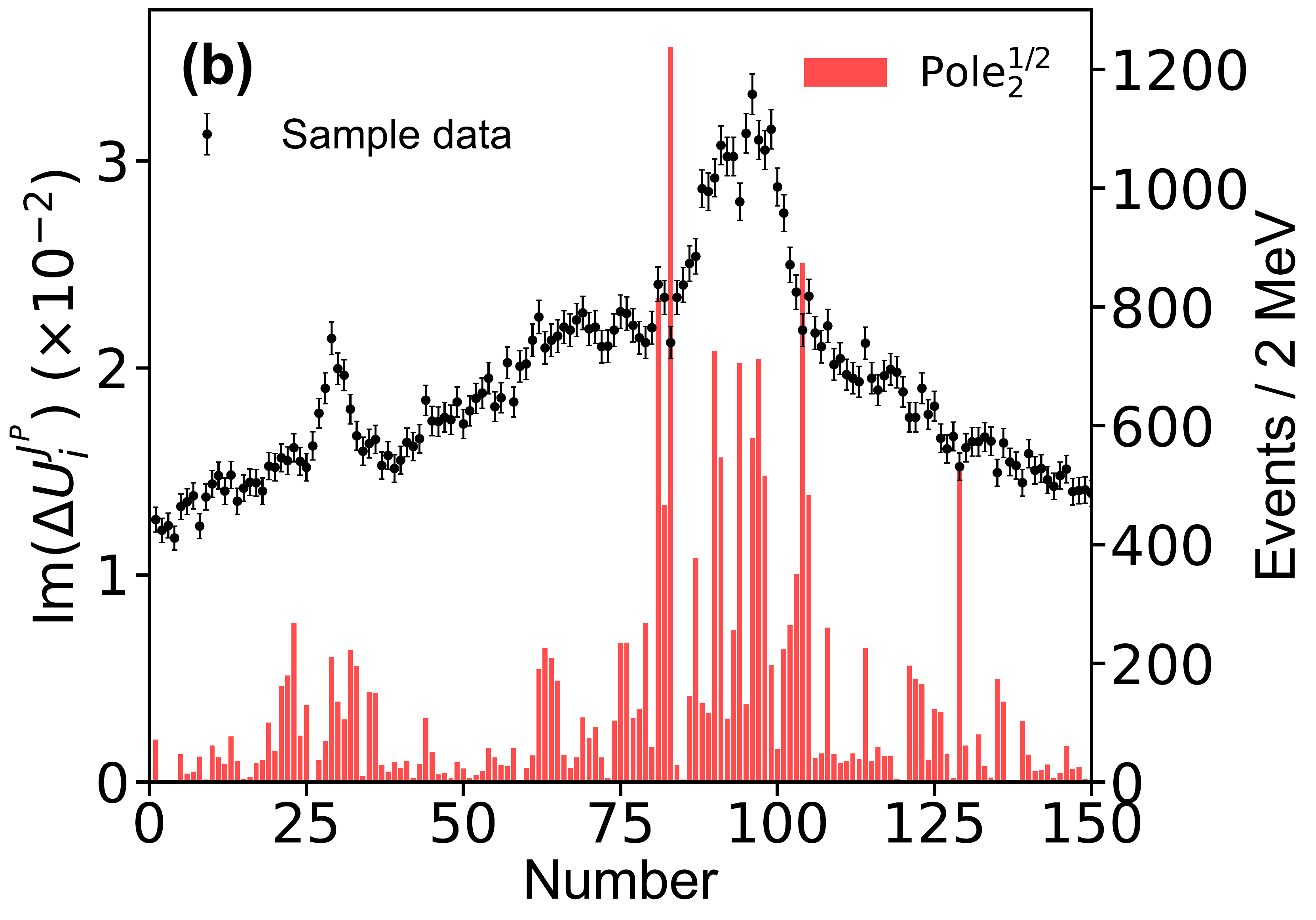}
    \includegraphics[width=0.32\textwidth]{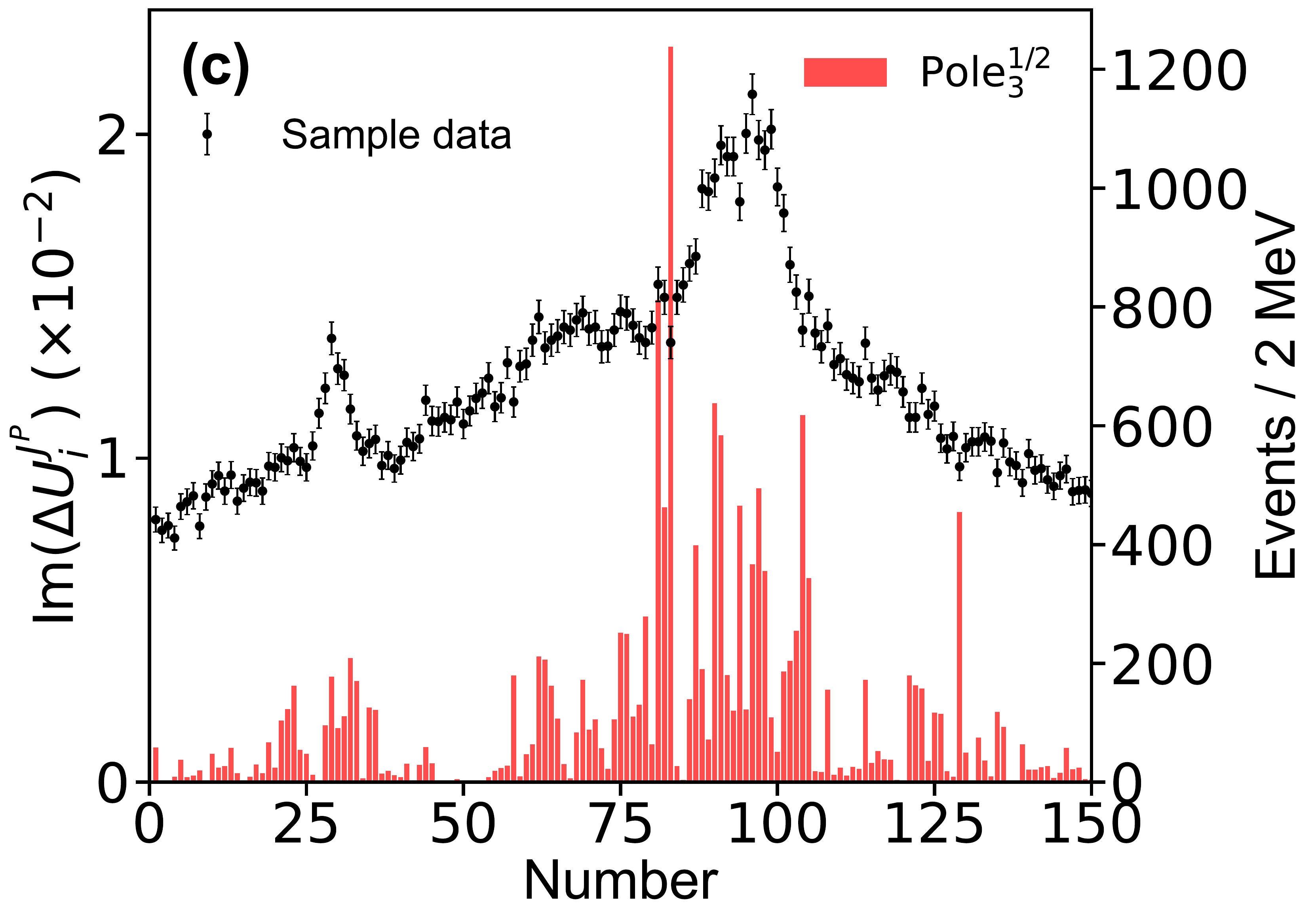}
    \caption{The $\mathrm{Im}(\Delta U_i^{J^P})$ distributions for the three pole positions of the $J^{P}=\frac{1}{2}^-$
    channel. (a), (b), (c) are for the poles from lower to higher energy. The data is a simulated sample for Solution A. 
    The distributions do not consider the correlation among the parameters.}
    \label{fig:fitpoleIm_result_12}
\end{figure*}

\begin{figure*}[ht]
    \centering
    \includegraphics[width=0.32\textwidth]{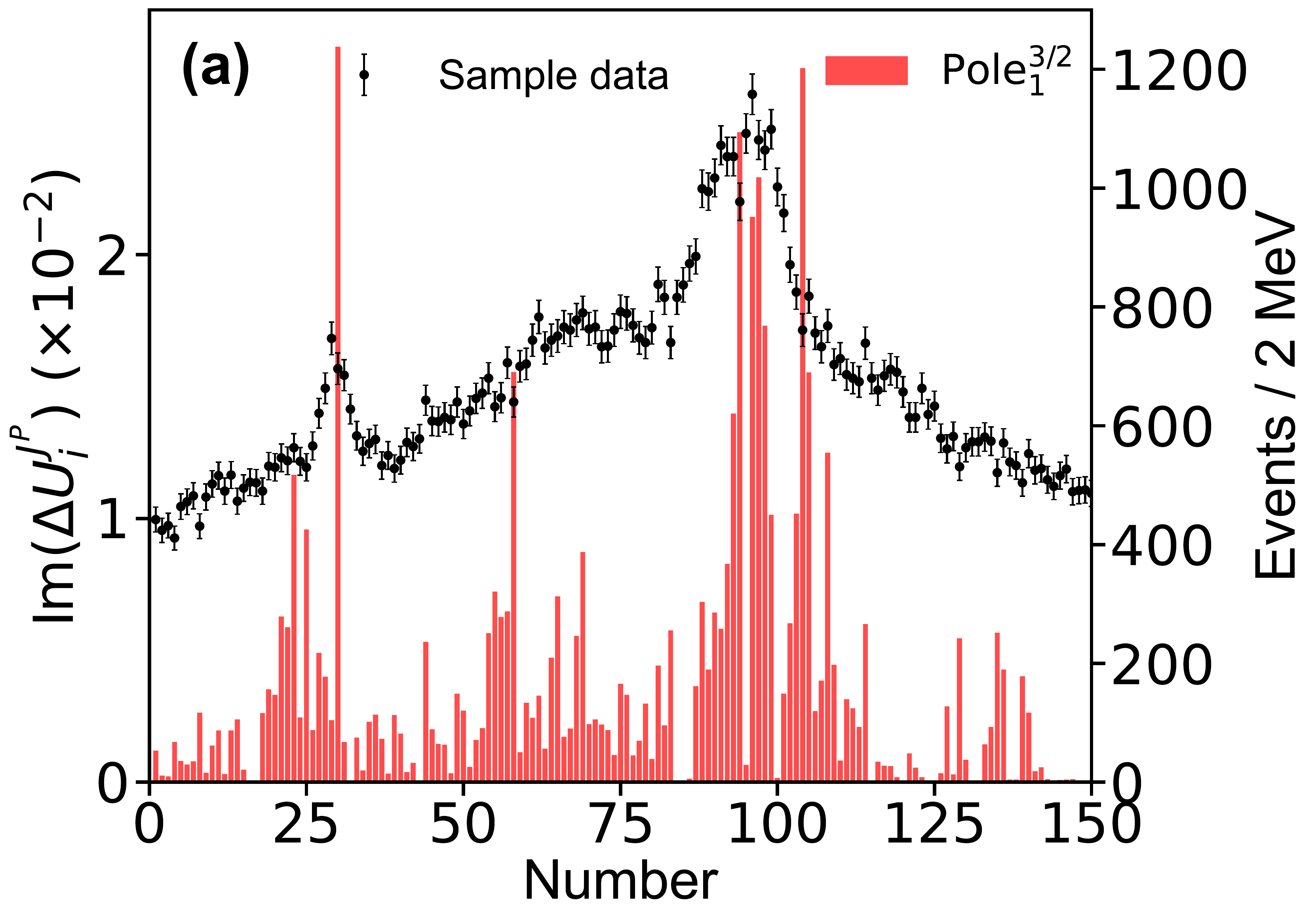}
    \includegraphics[width=0.32\textwidth]{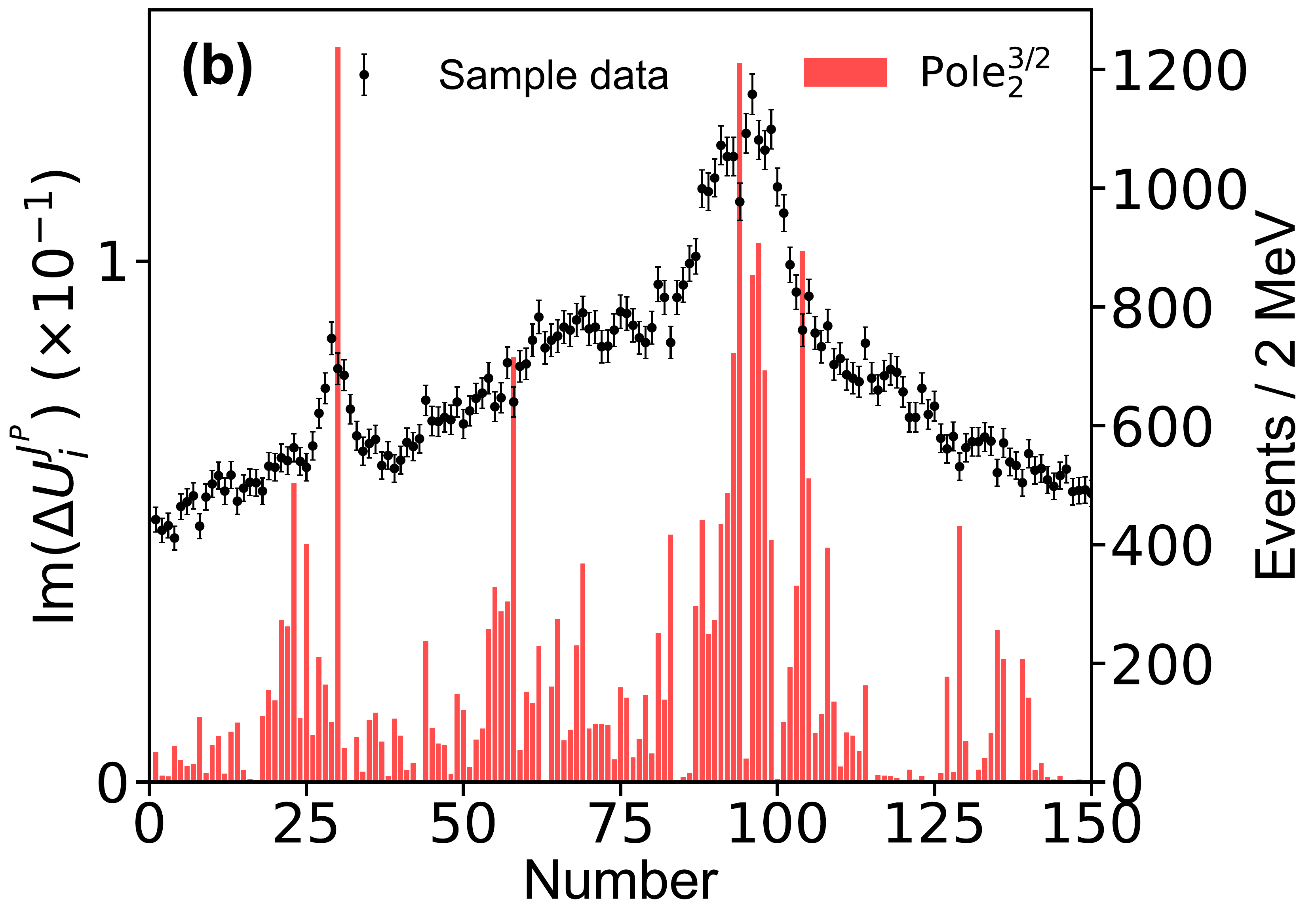}
    \includegraphics[width=0.32\textwidth]{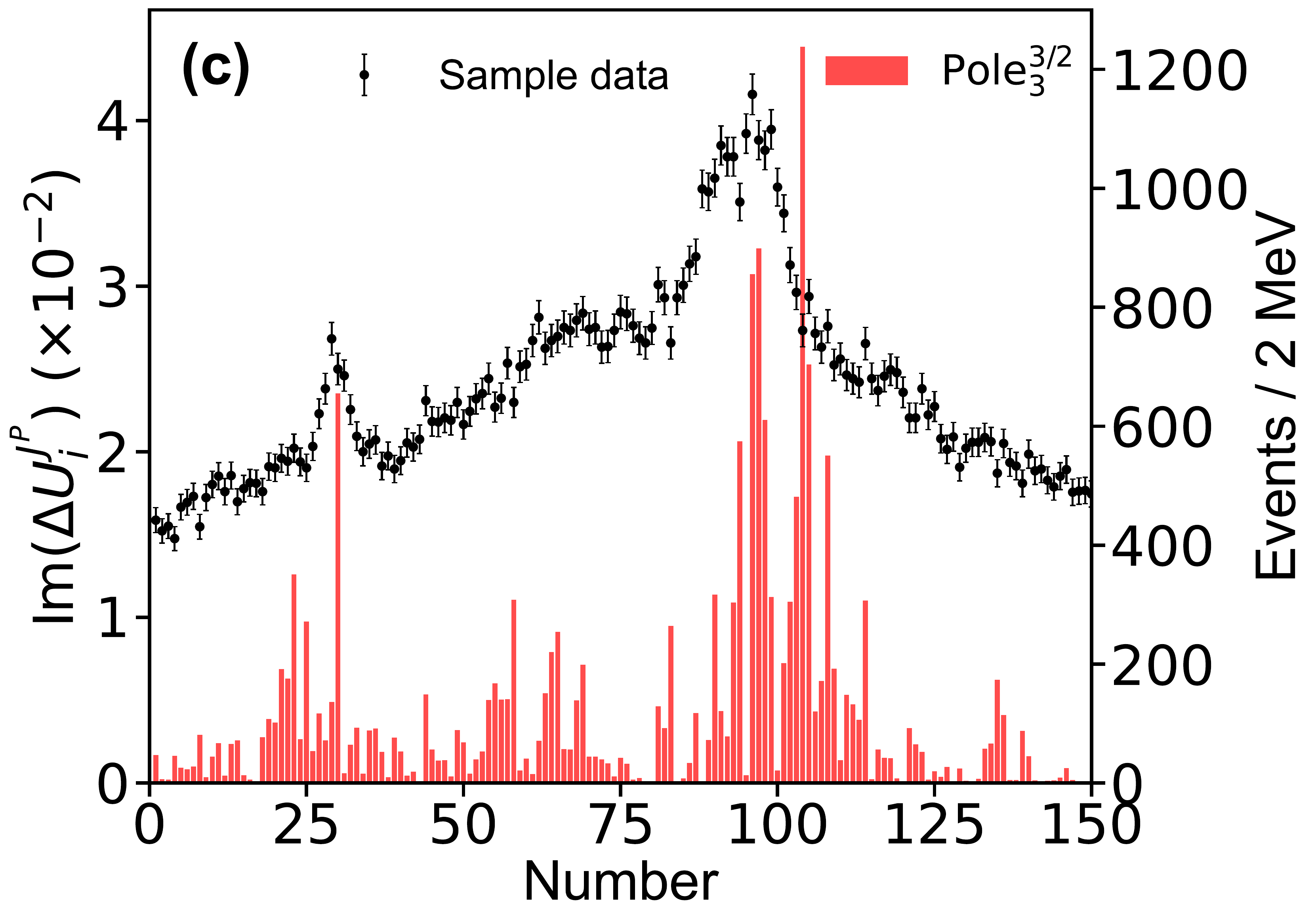}
    \caption{Analogous plot as Fig.~\ref{fig:fitpoleIm_result_12} but for $J^P=\frac 32^-$.}
    \label{fig:fitpoleIm_result_32}
\end{figure*}

\begin{figure}[ht]
    \centering
    \includegraphics[width=0.32\textwidth]{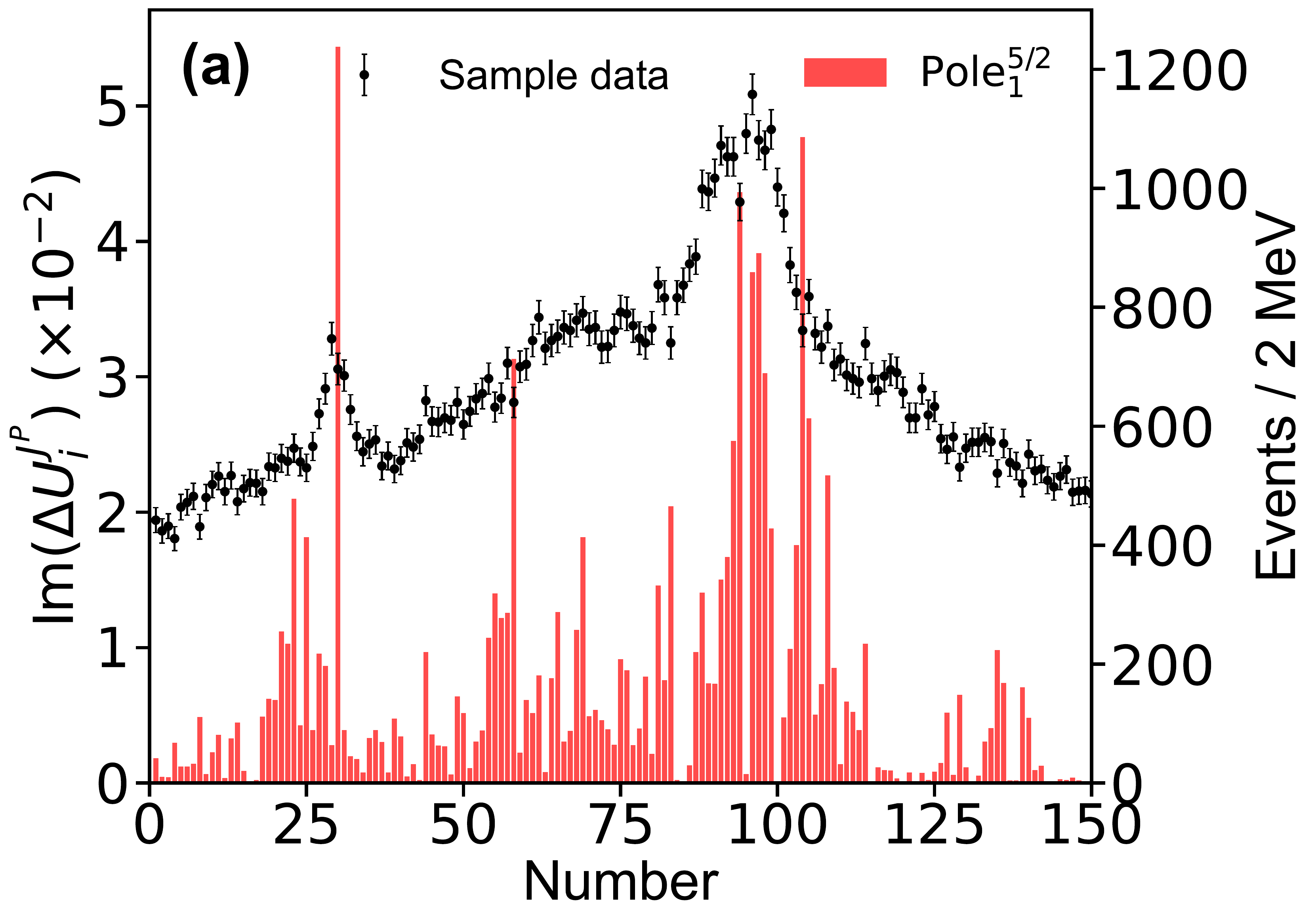}
    \caption{Analogous plot as Fig.~\ref{fig:fitpoleIm_result_12} but for $J^P=\frac 52^-$.}
    \label{fig:fitpoleIm_result_52}
    \vspace{-0.2cm}
\end{figure}

We define the uncertainty 
\begin{equation}
\mathrm{Im}(\Delta U_i^{J^P}) \equiv \left|\frac{\mathrm{Im}[\mathrm{Pole}_i](\mathrm{on})-\mathrm{Im}[\mathrm{Pole}_i]
(\mathrm{off})}{\mathrm{Im}[\mathrm{Pole}_i](\mathrm{on})}\right|
\end{equation}
for the imaginary part of the $i$th pole position by switching on/off 
the $j$th bin of $\mathcal{H}$ in the fitting. 
Figs.~\ref{fig:fitpoleIm_result_12}, ~\ref{fig:fitpoleIm_result_32},~\ref{fig:fitpoleIm_result_52} illustrates the $\mathrm{Im}(\Delta U_i^{J^P})$ distributions for the $J^P=\frac{1}{2}^-, \frac{3}{2}^-, \frac{5}{2}^-$ channels,
respectively, without considering the correlation among the parameters.

\clearpage
\section{Detailed Predictions}
To reduce the systematic uncertainties arising from the NNs,
we trained several NN models with identical structure but different initializations. 
In particular, for the background fractions $\{S^{90}\}$, $\{S^{92}\}$, $\{S^{94}\}$, and $\{S^{96}\}$,
we train ten NNs, respectively. 
 Their predicted results are presented
in Tab.~\ref{tab:Output_state} with maximum probabilities in boldface. 
As shown in the table, the first labels of all the maximum  probabilities are ``1", 
which indicates that all the NNs favor solution A, i.e. $J_{P_c(4440)}=\frac 12$ and $J_{P_c(4457)}=\frac 32$. 
As these 40 NNs
can distinguish solution~A from solution~B.
\begin{table}[ht]
    \centering
    \caption{Predicted probability of 15 state labels. The bold numbers in each line are the maximum probability.}
    \begin{tabular}{l|ccccccccccccccc}
    \hline
     \multirow{2}{*}{\diagbox{NN(Accuracy)}{Output(\%)}} & \multicolumn{15}{c}{Label}\\
     \cline{2-16}
     &0000&1000&1001&1002&0100&0110&0111&1111&0200&1200&1210&1211&1220&1221&1222 \\
    \hline
         \multicolumn{16}{c}{Prediction of NN trained with samples at a background level 90\%.} \\
    \hline
    NN 1 (77.29\%)&0.69&\bf{89.13}&1.42&8.75&0.00&0.00&0.00&0.00&0.00&0.00&0.00&0.00&0.00&0.01&0.00\\
    NN 2 (74.64\%)&0.03&5.83&38.47&\bf{55.30}&0.00&0.03&0.08&0.00&0.07&0.04&0.02&0.00&0.00&0.09&0.04\\
    NN 3 (74.97\%)&0.03&5.39&15.79&\bf{78.41}&0.00&0.02&0.03&0.00&0.00&0.00&0.00&0.00&0.00&0.03&0.03\\
    NN 4 (75.94\%)&0.01&1.90&27.01&\bf{70.95}&0.00&0.01&0.01&0.00&0.00&0.01&0.00&0.00&0.01&0.03&0.06\\
    NN 5 (74.07\%)&2.40&\bf{94.45}&0.15&2.99&0.00&0.00&0.00&0.00&0.00&0.00&0.00&0.00&0.00&0.01&0.00\\
    NN 6 (77.56\%)&0.05&2.76&31.64&\bf{65.36}&0.00&0.01&0.03&0.00&0.00&0.02&0.00&0.00&0.01&0.02&0.08\\
    NN 7 (76.79\%)&0.11&4.16&35.17&\bf{60.23}&0.00&0.04&0.06&0.00&0.01&0.04&0.00&0.00&0.00&0.04&0.13\\
    NN 8 (76.78\%)&0.18&4.23&16.10&\bf{79.44}&0.00&0.00&0.00&0.00&0.00&0.00&0.00&0.00&0.00&0.02&0.02\\
    NN 9 (75.22\%)&0.11&2.54&18.26&\bf{79.02}&0.00&0.00&0.02&0.00&0.00&0.00&0.00&0.00&0.00&0.03&0.01\\
    NN 10 (75.87\%)&0.00&1.24&22.85&\bf{75.77}&0.00&0.00&0.02&0.00&0.00&0.00&0.00&0.00&0.00&0.00&0.09\\
    \hline
     \multicolumn{16}{c}{Prediction of NN trained with samples at a background level 92\%.} \\
    \hline
    NN 1 (72.94\%)&0.00&0.15&5.37&\bf{94.47}&0.00&0.00&0.00&0.00&0.00&0.00&0.00&0.00&0.00&0.00&0.00\\
    NN 2 (72.68\%)&0.00&0.07&4.11&\bf{95.81}&0.00&0.00&0.00&0.00&0.00&0.00&0.00&0.00&0.00&0.00&0.00\\
    NN 3 (74.58\%)&0.00&0.78&13.57&\bf{85.61}&0.00&0.00&0.01&0.00&0.00&0.00&0.00&0.00&0.00&0.00&0.02\\
    NN 4 (71.78\%)&0.00&0.81&19.02&\bf{80.16}&0.00&0.00&0.00&0.00&0.00&0.00&0.00&0.00&0.00&0.00&0.00\\
    NN 5 (72.75\%)&0.14&15.13&16.91&\bf{67.80}&0.00&0.00&0.00&0.00&0.00&0.00&0.00&0.00&0.00&0.00&0.00\\
    NN 6 (71.72\%)&0.00&0.18&9.32&\bf{90.48}&0.00&0.00&0.00&0.00&0.00&0.00&0.00&0.00&0.00&0.00&0.00\\
    NN 7 (72.75\%)&0.00&0.32&18.44&\bf{81.24}&0.00&0.00&0.00&0.00&0.00&0.00&0.00&0.00&0.00&0.00&0.00\\
    NN 8 (74.57\%)&0.00&0.27&6.39&\bf{93.34}&0.00&0.00&0.00&0.00&0.00&0.00&0.00&0.00&0.00&0.00&0.00\\
    NN 9 (73.24\%)&0.00&0.00&1.05&\bf{98.95}&0.00&0.00&0.00&0.00&0.00&0.00&0.00&0.00&0.00&0.00&0.00\\
    NN 10 (74.34\%)&0.00&0.02&0.85&\bf{99.12}&0.00&0.00&0.00&0.00&0.00&0.00&0.00&0.00&0.00&0.00&0.00\\
    \hline
        \multicolumn{16}{c}{Prediction of NN trained with samples at a background level 94\%.} \\
    \hline
    NN 1 (65.28\%)&0.00&1.20&2.26&\bf{96.53}&0.00&0.00&0.00&0.00&0.00&0.00&0.00&0.00&0.00&0.00&0.00\\
    NN 2 (64.11\%)&0.02&2.23&11.51&\bf{86.24}&0.00&0.00&0.00&0.00&0.00&0.00&0.00&0.00&0.00&0.00&0.00\\
    NN 3 (67.83\%)&0.00&0.26&10.26&\bf{89.47}&0.00&0.00&0.00&0.00&0.00&0.00&0.00&0.00&0.00&0.00&0.00\\
    NN 4 (65.53\%)&0.15&24.05&\bf{48.79}&26.89&0.00&0.04&0.02&0.00&0.01&0.03&0.00&0.00&0.00&0.02&0.01\\
    NN 5 (64.17\%)&0.02&4.12&\bf{72.61}&23.23&0.00&0.00&0.01&0.00&0.00&0.00&0.00&0.00&0.00&0.01&0.00\\
    NN 6 (67.44\%)&0.00&1.22&13.92&\bf{84.86}&0.00&0.00&0.00&0.00&0.00&0.00&0.00&0.00&0.00&0.00&0.00\\
    NN 7 (61.85\%)&0.00&5.07&15.90&\bf{79.01}&0.00&0.00&0.00&0.00&0.00&0.00&0.00&0.00&0.00&0.00&0.00\\
    NN 8 (65.73\%)&0.03&1.31&\bf{58.42}&40.21&0.00&0.00&0.02&0.00&0.00&0.00&0.00&0.00&0.00&0.00&0.00\\
    NN 9 (65.34\%)&0.00&2.94&21.80&\bf{75.25}&0.00&0.00&0.00&0.00&0.00&0.00&0.00&0.00&0.00&0.00&0.01\\
    NN 10 (65.74\%)&0.00&0.08&4.56&\bf{95.34}&0.00&0.00&0.00&0.00&0.00&0.00&0.00&0.00&0.00&0.00&0.00\\
    \hline
        \multicolumn{16}{c}{Prediction of NN trained with samples at a background level 96\%.} \\
    \hline
    NN 1 (55.64\%)&0.00&1.29&19.79&\bf{78.92}&0.00&0.00&0.00&0.00&0.00&0.00&0.00&0.00&0.00&0.00&0.00\\
    NN 2 (54.75\%)&0.00&3.38&25.64&\bf{70.97}&0.01&0.00&0.00&0.00&0.00&0.00&0.00&0.00&0.00&0.00&0.00\\
    NN 3 (53.44\%)&0.00&6.57&32.20&\bf{61.23}&0.00&0.00&0.00&0.00&0.00&0.00&0.00&0.00&0.00&0.00&0.00\\
    NN 4 (51.20\%)&0.60&\bf{55.77}&21.65&21.96&0.00&0.00&0.00&0.00&0.00&0.00&0.00&0.00&0.00&0.00&0.00\\
    NN 5 (56.41\%)&0.00&0.27&3.52&\bf{96.20}&0.00&0.00&0.00&0.00&0.00&0.00&0.00&0.00&0.00&0.00&0.00\\
    NN 6 (50.72\%)&0.39&\bf{34.89}&31.27&33.42&0.00&0.00&0.00&0.00&0.00&0.00&0.00&0.00&0.00&0.00&0.00\\
    NN 7 (54.34\%)&0.00&5.16&21.02&\bf{73.82}&0.00&0.00&0.00&0.00&0.00&0.00&0.00&0.00&0.00&0.00&0.00\\
    NN 8 (53.04\%)&0.02&5.04&33.86&\bf{61.08}&0.00&0.00&0.00&0.00&0.00&0.00&0.00&0.00&0.00&0.00&0.00\\
    NN 9 (58.18\%)&0.00&0.11&31.92&\bf{67.97}&0.00&0.00&0.00&0.00&0.00&0.00&0.00&0.00&0.00&0.00&0.00\\
    NN 10 (55.79\%)&0.05&18.82&28.80&\bf{52.32}&0.00&0.00&0.00&0.00&0.00&0.00&0.00&0.00&0.00&0.00&0.00\\
    
    \hline
    \end{tabular}  \label{tab:Output_state}
\end{table}

\clearpage
\section{Neural Network}

We have compared a MLP-based~\cite{MURTAGH1991183}  NN to a ResNet-based~\cite{He2016DeepRL} NN,
which are implemented using  PyTorch~\cite{NEURIPS2019_bdbca288}. Both NN models are trained with the $\{\mathcal{S}^{90}\}$ samples. The Adam~\cite{2014Adam} optimizer is used to solve these models with a combination of the momentum algorithm with the RMSProp algorithm~\cite{2020Adversarial}. For the activation function, the Relu function compared to the Sigmod function, can effectively avoid the problem of gradient disappearance and alleviate potential overfitting in particular when training a deep neutral network, while the Sigmod function is exponential and computationally expensive. A reasonable solution can be obtained after 500 training epochs using an
initial learning rate value of 0.001. A large learning rate  makes convergence of the loss function difficult, while a small learning rate makes too slow convergence. Our strategy is to dynamically reduce it. The initial learning rate of 0.001 is successfully tested, and the next interval is reduced to the $\frac 12$ of the previous one,
controlled by the stepLR method. The interval is set to 100 epochs in our case. To avoid the network falling into the local minimum solution during the training process, the weights of the neurons are randomly initialized with a normal distribution, and the biases of neurons are set to zero. 
\begin{figure}[h]
    \centering
    \includegraphics[width=0.32\textwidth]{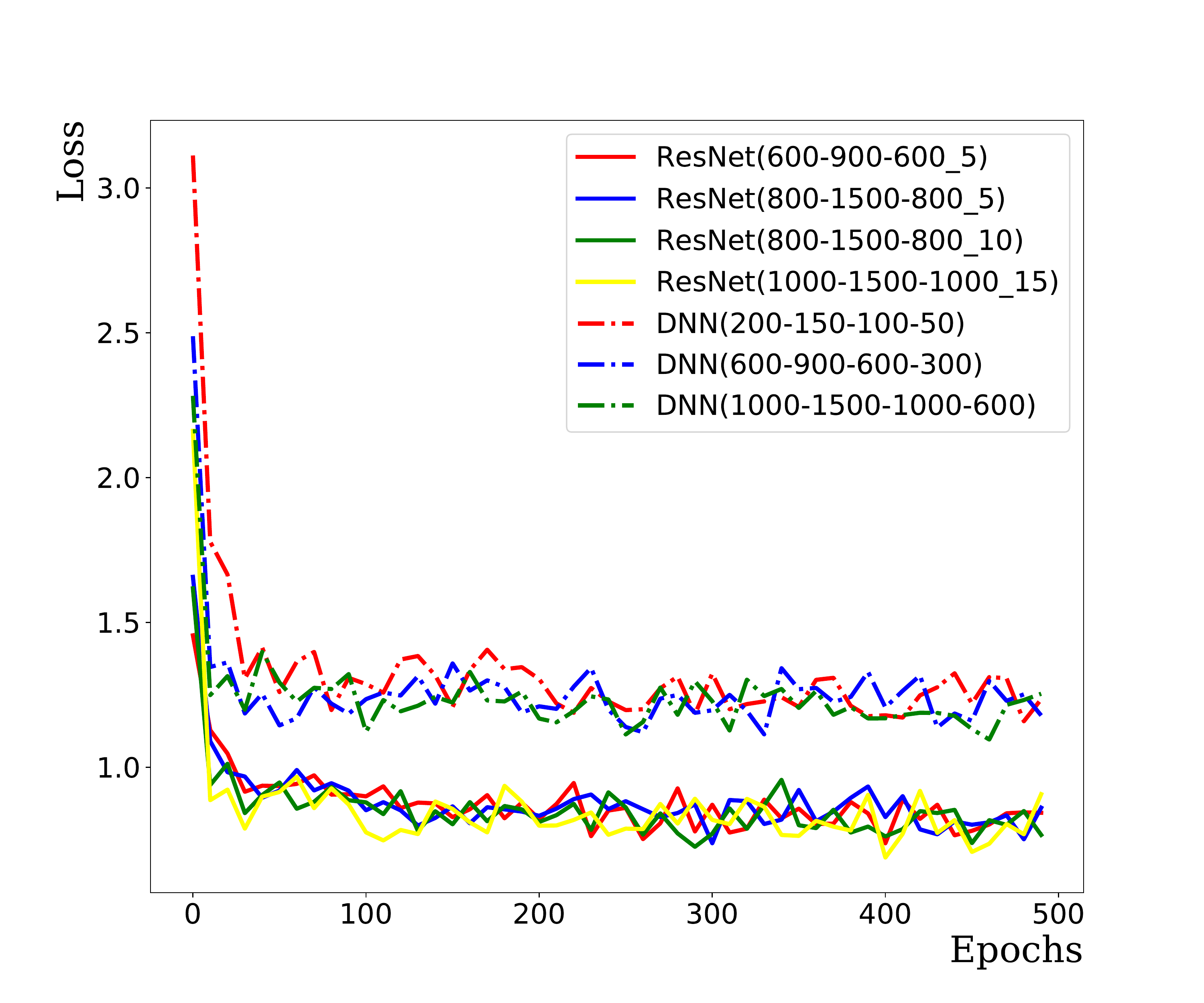}
    \includegraphics[width=0.32\textwidth]{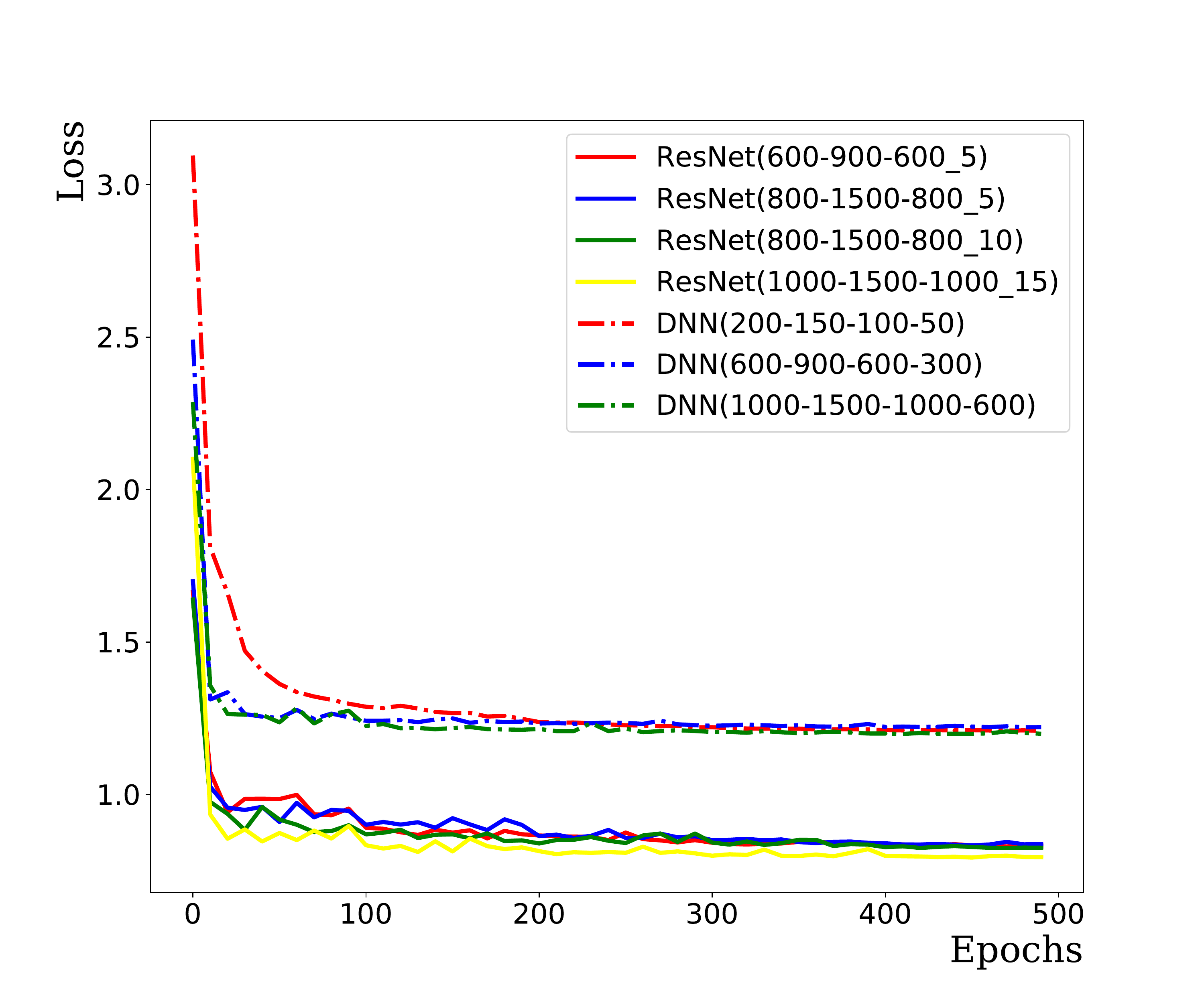}
    \includegraphics[width=0.32\textwidth]{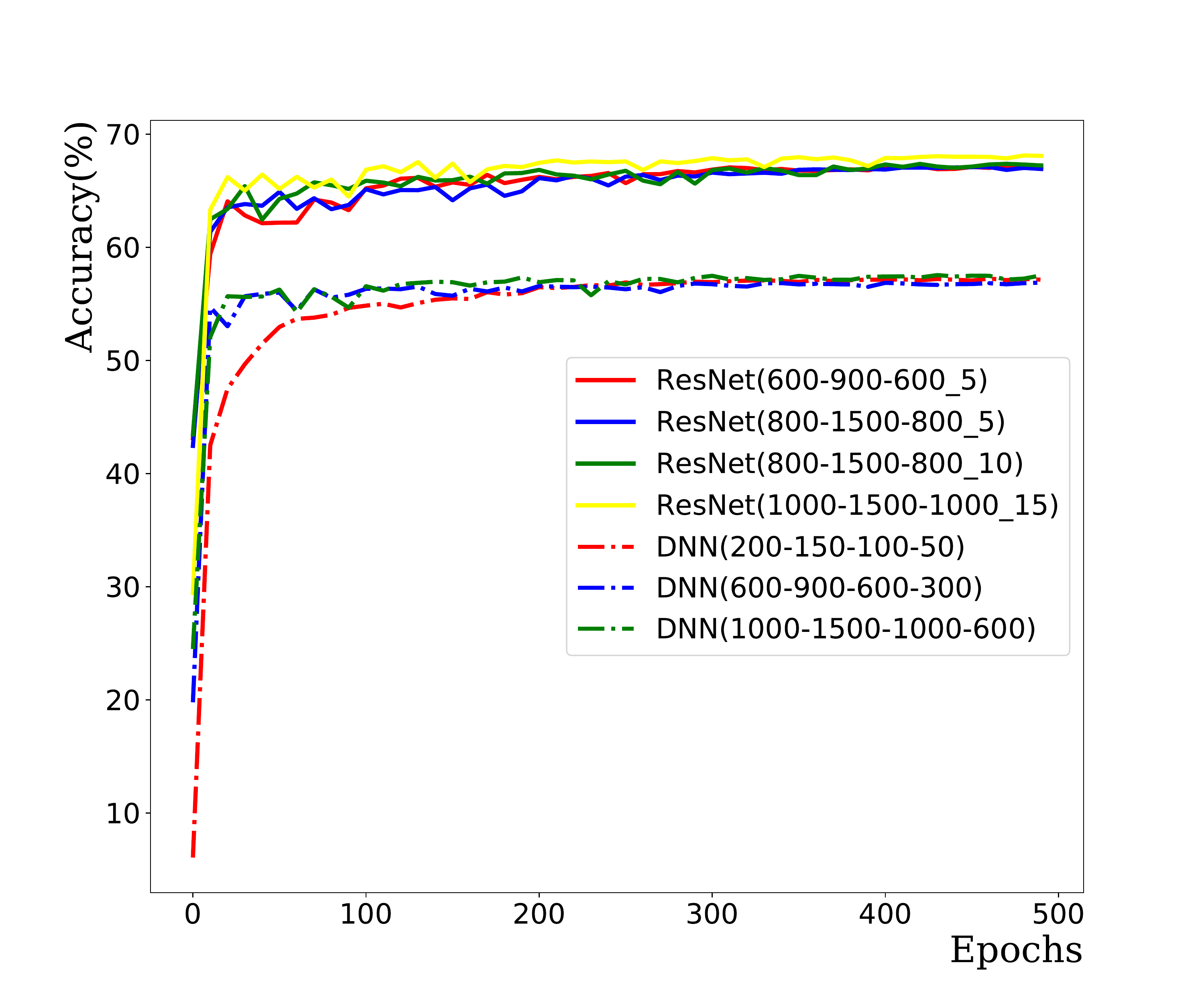}
    \caption{
    (left) For training datasets, the loss functions used in different NNs converge as training
epochs increase. Here, $``$ResNet(600-900-600\_5)$"$ represents a ResNet-based NN consists of five ResBlock, and each block consists of a layer of 600 neurons, a layer of 900 neurons, and a layer of 600 neurons.  $``$DNN(200-150-100-50)$"$ represents a MLP-based NN which consists of  four fully-connected layers, and each layer consists of 200, 150, 100, 50 neurons respectively (the input and output layers are excluded).
(middle) For testing datasets, the loss functions used in different NNs converge as training
epochs increase.
(right) The predicted accuracy increases as training epochs increase.}
    \label{fig:TrainLoss_Comparison_Structure}
\end{figure}
\begin{figure}[ht]
    \centering
    \includegraphics[width=0.42\textwidth]{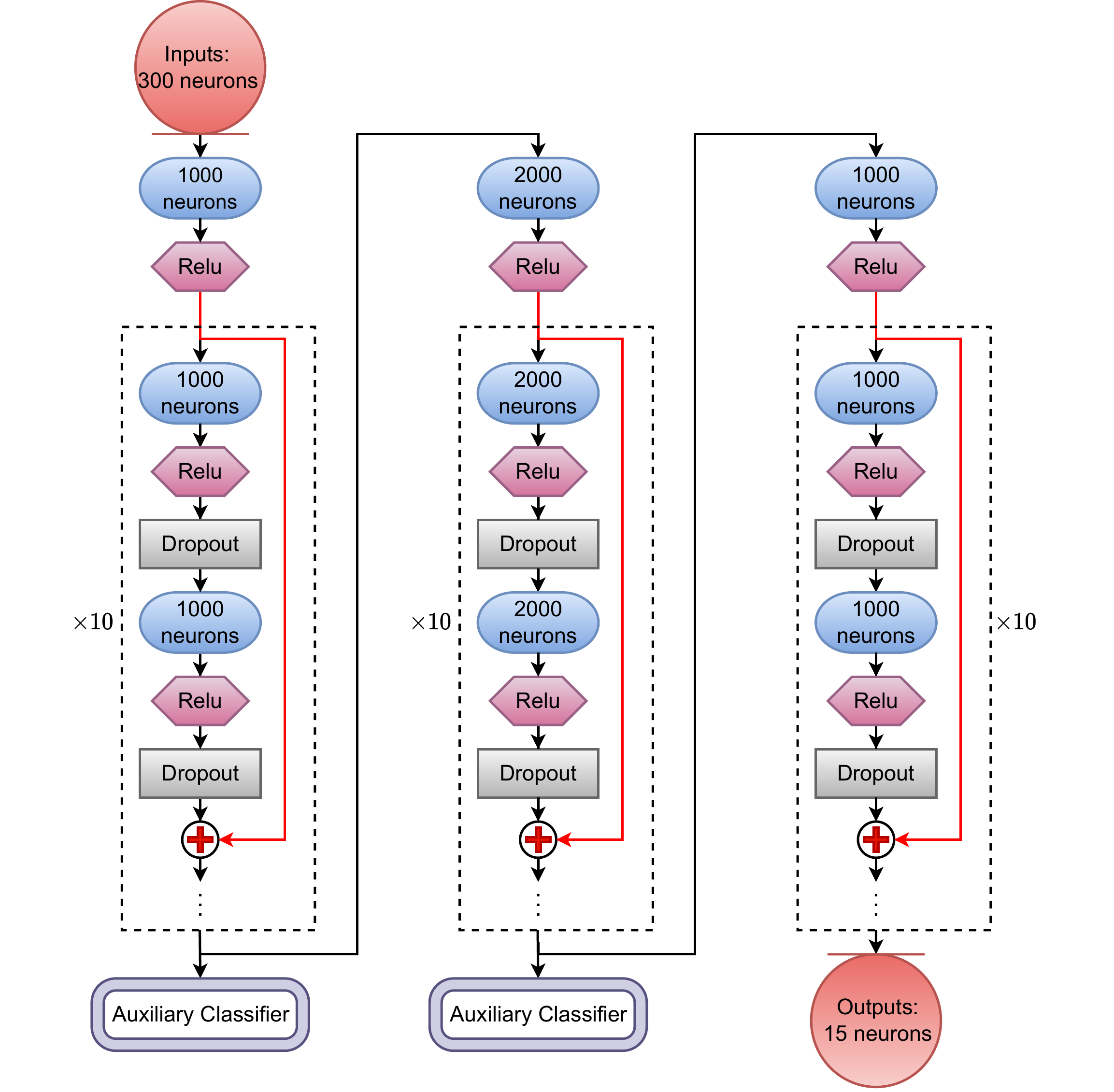}
    \caption{The structure of the ResNet-based NN used in this work.}
    \label{fig:Network1}
\end{figure}
\begin{figure}[ht]
    \centering
    \includegraphics[width=0.32\textwidth]{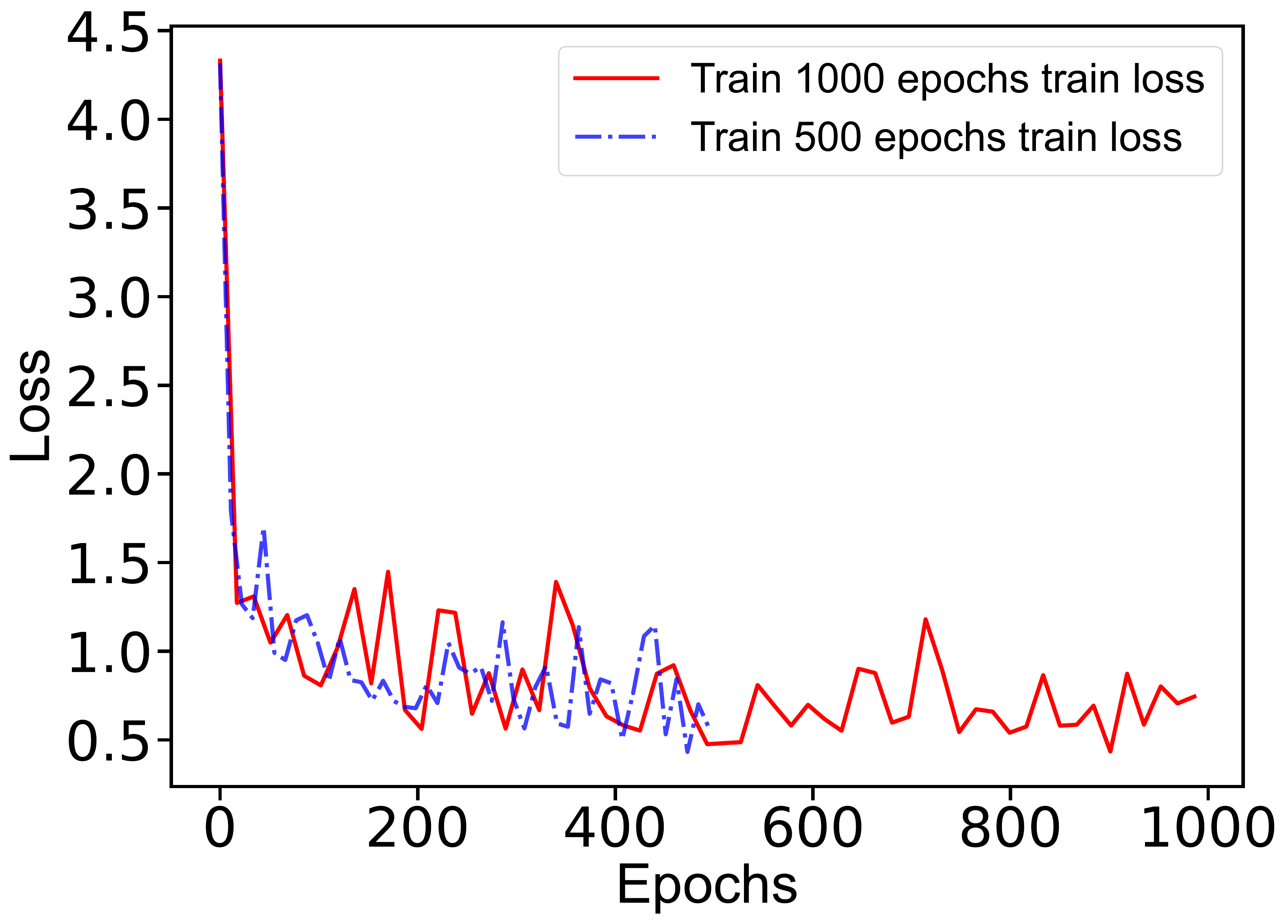}
    \includegraphics[width=0.32\textwidth]{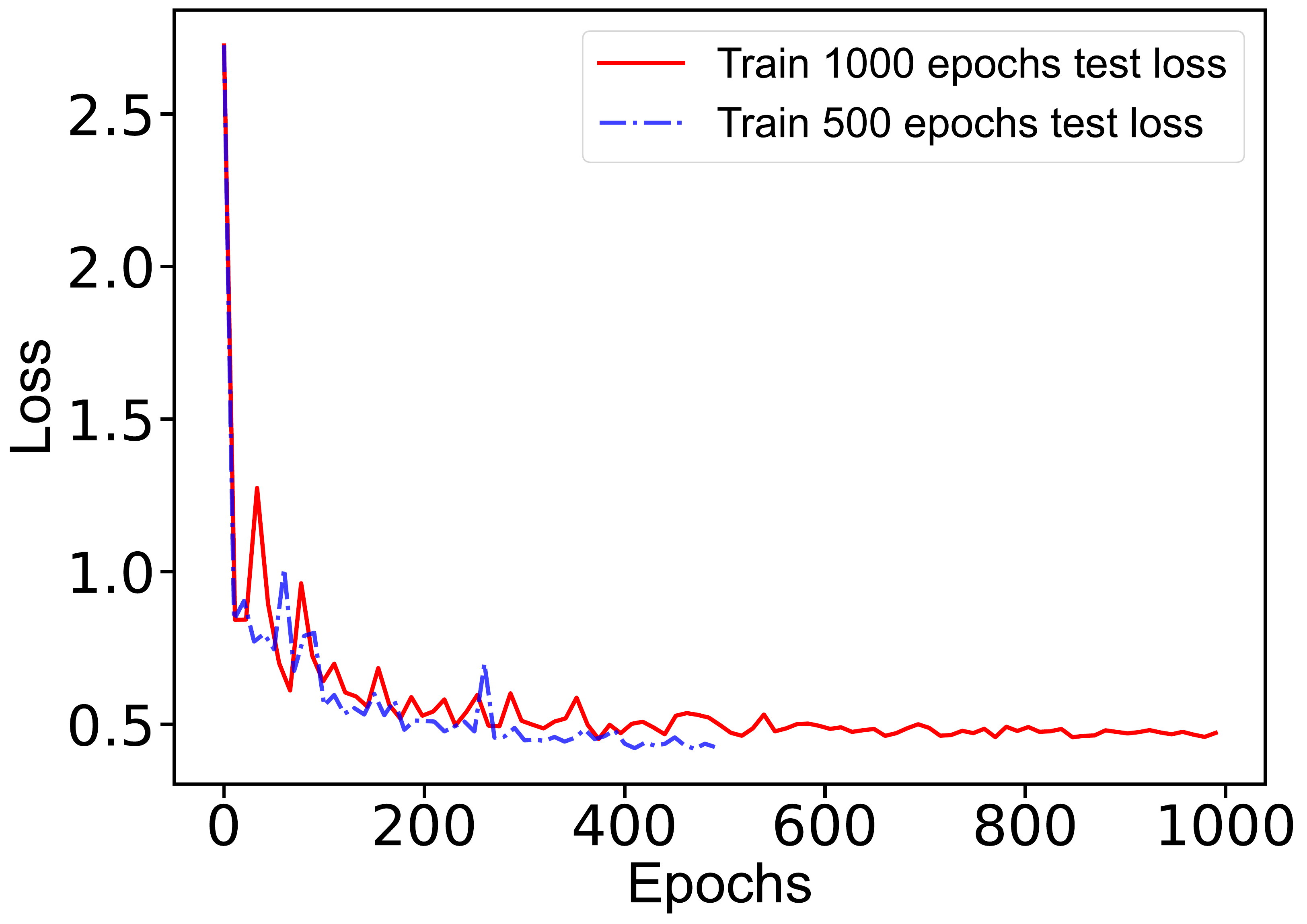}
    \includegraphics[width=0.32\textwidth]{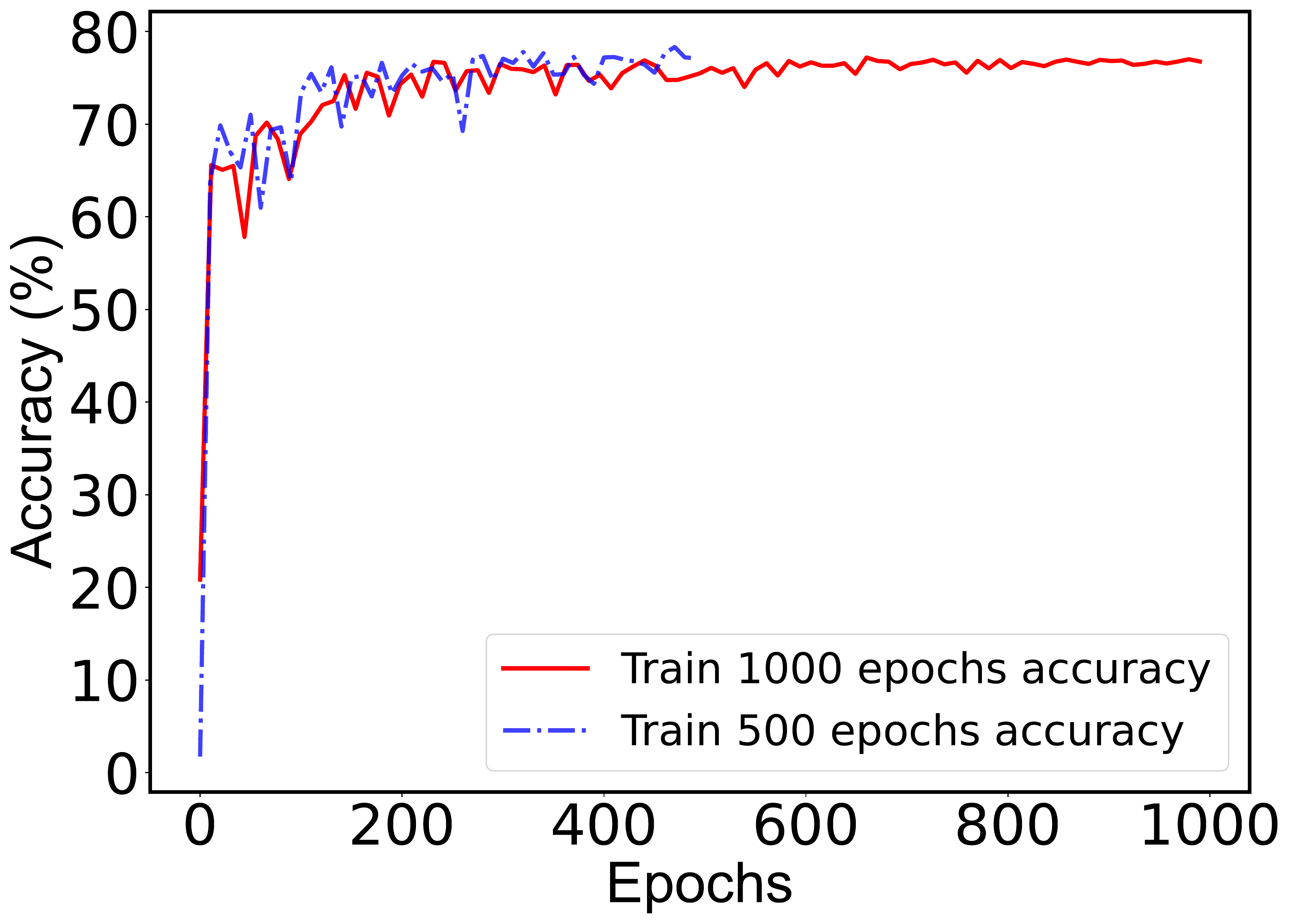}
    \caption{ The dependence of the loss functions on train epochs for training datasets (left) and testing datasets (middle), respectively.
The predicted accuracy increases as training epochs increase (right).}
    \label{fig:TrainLoss}
\end{figure}

To avoid overfitting in the training, we introduce dropout layers in the network. A large dropout probability tends to activate too few neurons and the loss function will not converge; a small dropout probability tends to reduce generalizability and even occur over-fitting. After training, the dropout probability is chosen to 0.3 for ResNet-based NN and 0.2 for MLP-based NN to optimize performance. In case of overfiting, the obtained NN works well on training datasets while probably does not on testing datasets. Thus the loss function calculated with training datasets should obviously differs from the one calculated with testing datasets. However, one cannot see difference in Fig.~\ref{fig:TrainLoss}, which indicates the NN works well both for training and testing datasets. In summary, the CrossEntropyLoss function converges as the training epochs increase,
and the predicted accuracy increases as training epochs increase.
The ResNet-based NN achieves a better solution, as shown in Fig.~\ref{fig:TrainLoss_Comparison_Structure}.
A larger number of neurons and hidden layers require more computing resources and result in difficult training and over-fitting; while a smaller number of neurons and hidden layers result in non-convergence of the loss function and under-fitting. We found an architecture ResNet(1000-1000-2000-2000-1000-1000\_10) which works well after several attempts without optimizing its architecture. The ResNet-based NN is structured as shown in Fig.~\ref{fig:Network1}.

\section{What is the reason for the saturation of NN's accuracy?}
\begin{figure}[ht]
\centering
\includegraphics[width=0.45\textwidth]{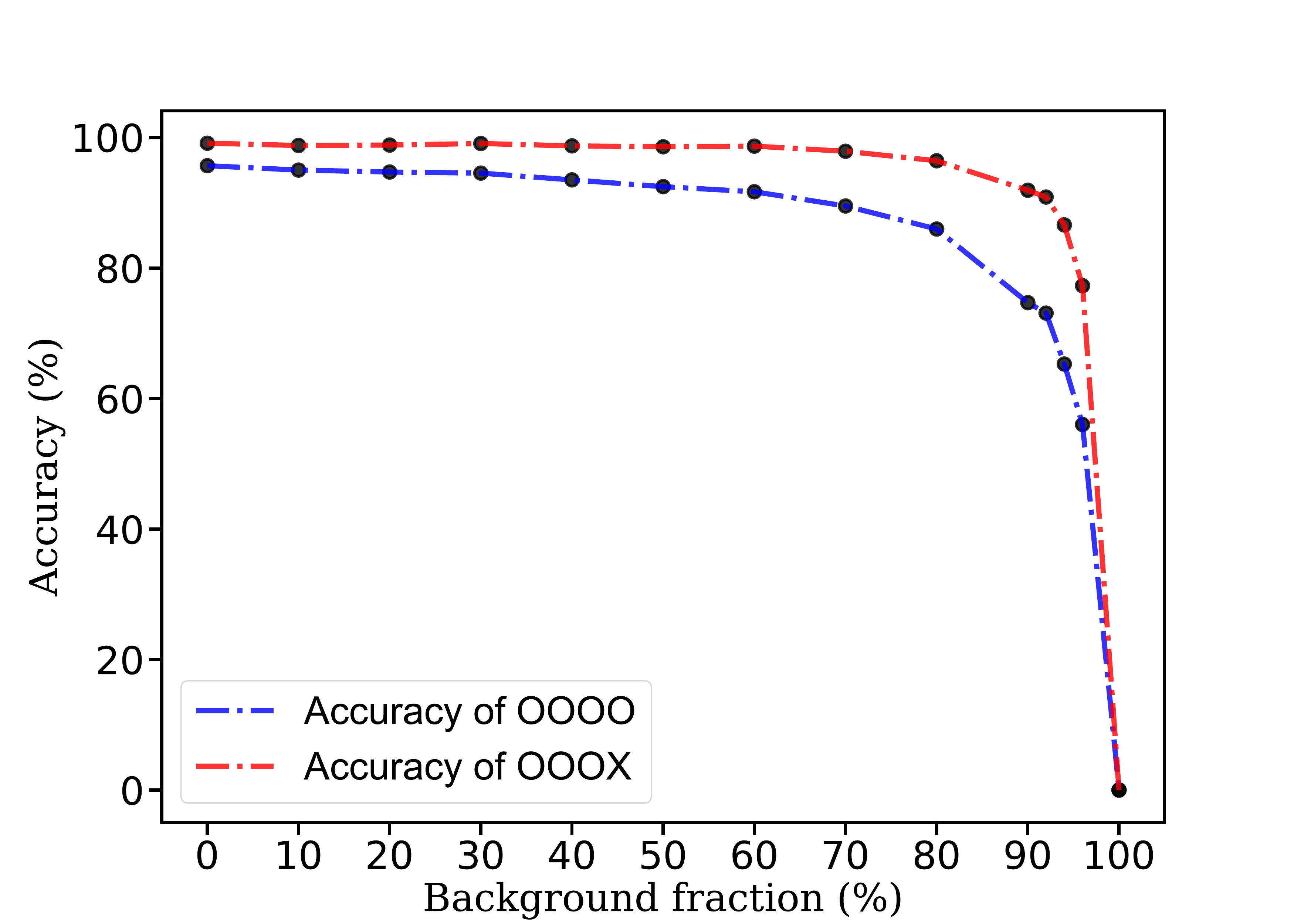}
\includegraphics[width=0.45\textwidth]{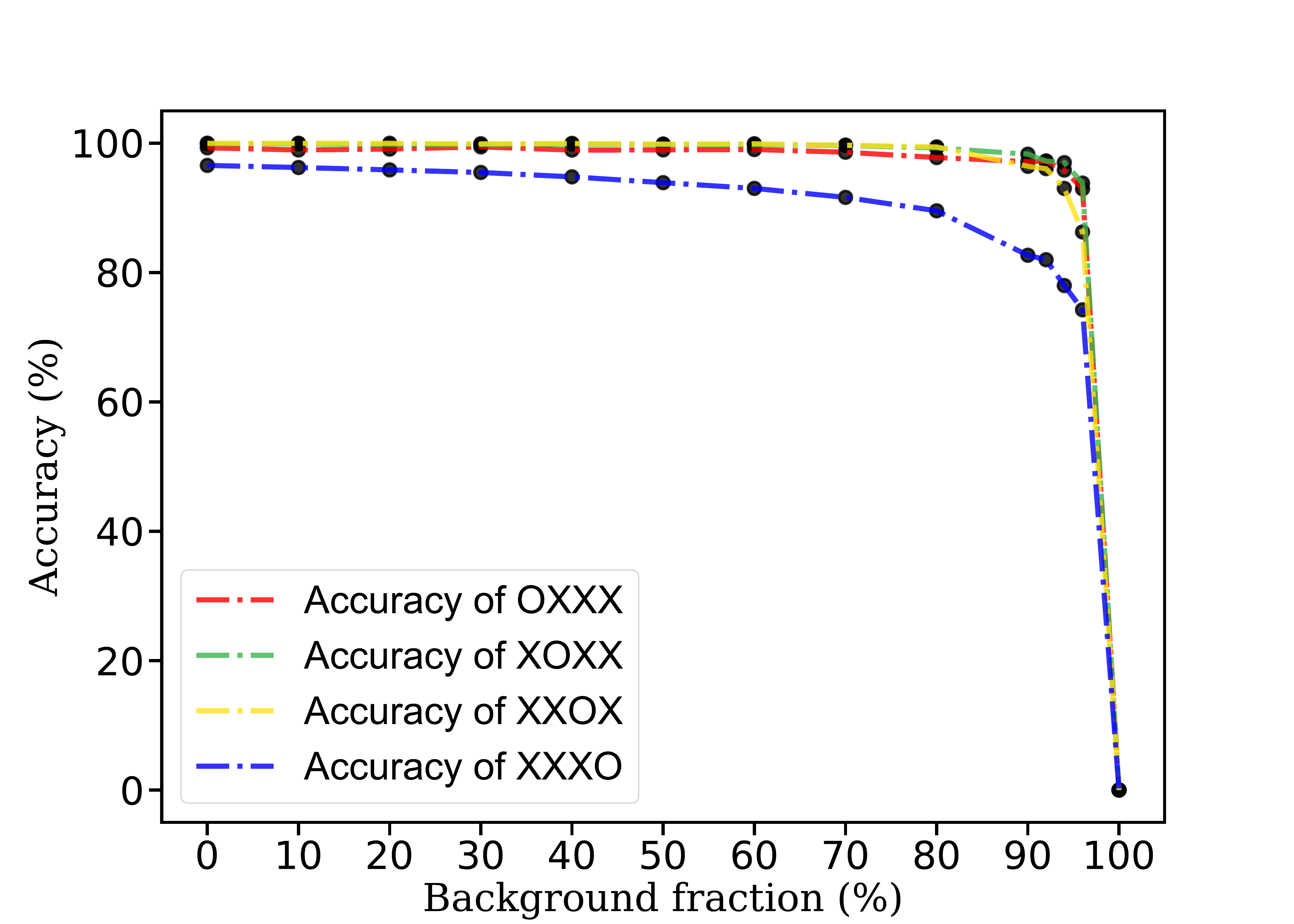}
\caption{The distribution of accuracy with background fraction. Where ``O" means open this label when testing accuracy, ``X" means close this label when testing accuracy.}
	\label{fig:Background_acu}
\end{figure}

The low accuracy is because of two reasons,

(1) Background fraction.
Higher background fractions imply lower prediction accuracy within our expectation.
Clearly at an extreme case 100\% background fraction will result in a zero accuracy.
To make this point clear, we study how the prediction accuracy varies with background fractions, with lots of samples ($\mathcal{S}^{0}$, $\mathcal{S}^{10}$, $\mathcal{S}^{20}$, $\mathcal{S}^{30}$, $\mathcal{S}^{40}$, $\mathcal{S}^{50}$, $\mathcal{S}^{60}$, $\mathcal{S}^{70}$, $\mathcal{S}^{80}$, $\mathcal{S}^{90}$, $\mathcal{S}^{92}$, $\mathcal{S}^{94}$, $\mathcal{S}^{96}$), as summarized in the Tab.~\ref{tab:Background_acu} (1st and 2nd column), and Fig.~\ref{fig:Background_acu} (left).

(2) The forth label, since the peak corresponding to the forth label is submerged in background, the NN fails to predict this label.  Thus when we count correct predictions including this label ``OOOO'' (2nd column), the accuracy drops down. In turn, when we count correct predictions excluding this label ``OOOX'' (3rd column), the accuracy goes up. Please see also Fig.~\ref{fig:Background_acu}. We also show the more results (accuracy of ``OXXX", ``XOXX", ``XXOX", ``XXXO") in Fig.~\ref{fig:Background_acu} (right), and find that the accuracy of ``XXXO" is the lowest.
\begin{table}[ht]
    \centering
    \caption{The value of accuracy with background fraction.  Where ``O" and ``X" same as Fig.~\ref{fig:Background_acu}. ``OOOO" means testing the accuracy of all labels, ``OOOX" means only testing the accuracy of first three labels.}
    \begin{tabular}{ccc}
    \hline
    Background fraction & Accuracy of ``OOOO''(\%)& Accuracy of ``OOOX''(\%)\\
    \hline
     $\mathcal{S}^{0}$   &95.70&99.14\\
     $\mathcal{S}^{10}$   &95.03&98.80\\
     $\mathcal{S}^{20}$   &94.73&98.86\\
     $\mathcal{S}^{30}$   &94.56&99.10\\
     $\mathcal{S}^{40}$   &93.52&98.74\\
     $\mathcal{S}^{50}$   &92.49&98.60\\
     $\mathcal{S}^{60}$   &91.70&98.70\\
     $\mathcal{S}^{70}$   &89.52&97.91\\
     $\mathcal{S}^{80}$   &85.99&96.45\\
     $\mathcal{S}^{90}$   &74.70&91.93\\
     $\mathcal{S}^{92}$   &73.11&90.90\\
     $\mathcal{S}^{94}$   &65.30&86.62\\
     $\mathcal{S}^{96}$   &56.02&77.30\\
    \hline
    \end{tabular}  \label{tab:Background_acu}
\end{table}

\section{SHAP Value}

The SHapley Additive exPlanation (SHAP) value is widely used as a  feature importance metric for well-established models in machine learning. One of prevailing methods for estimating neural network model features is the DeepLIFT algorithm based DeepExplainer. A positive (negative) SHAP value indicates that
a given data point is pushing the NN classification in favor of (against) a given class. A large absolute SHAP value implies a large impact of a given mass bin on the classification. The method requires a dataset as a reference (or denoted as background) for evaluating network features, and we test the SHAP values of 1000 \{$\mathcal{S}^{90}$\} samples and the SHAP values of experimental data with our network based on 3000 \{$\mathcal{S}^{90}$\} samples as a reference. Fig.\ref{fig:shap_value} shows the SHAP values of experimental data corresponding to 15 labels, indicating that data points around the peaks in the mass spectrum have a greater impact. 
Fig.~\ref{fig:SHAP_value} shows the SHAP values of MC samples corresponding to 15 labels.
\begin{figure}[htbp]
    \centering
    \includegraphics[width=0.32\textwidth]{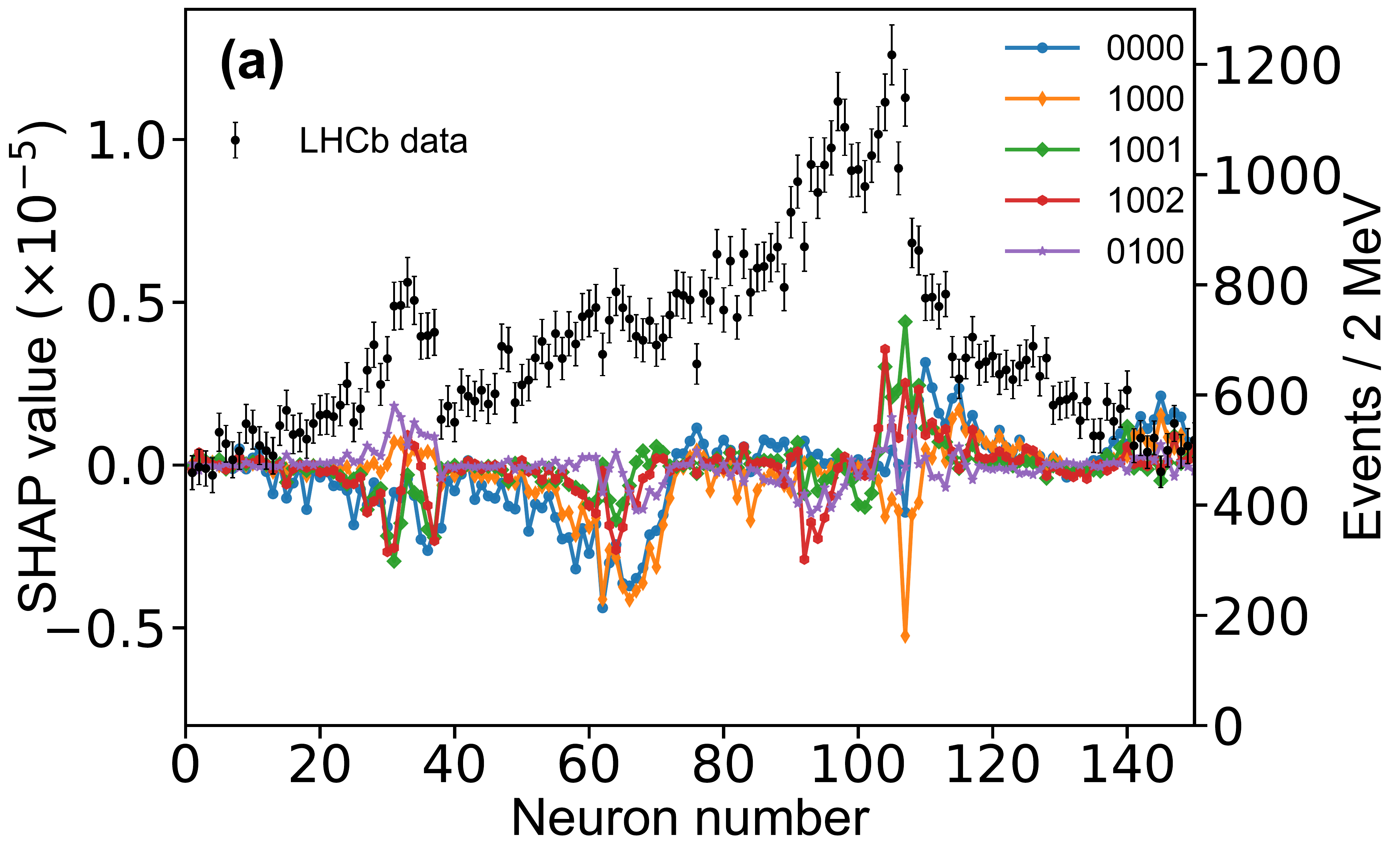}
    \includegraphics[width=0.32\textwidth]{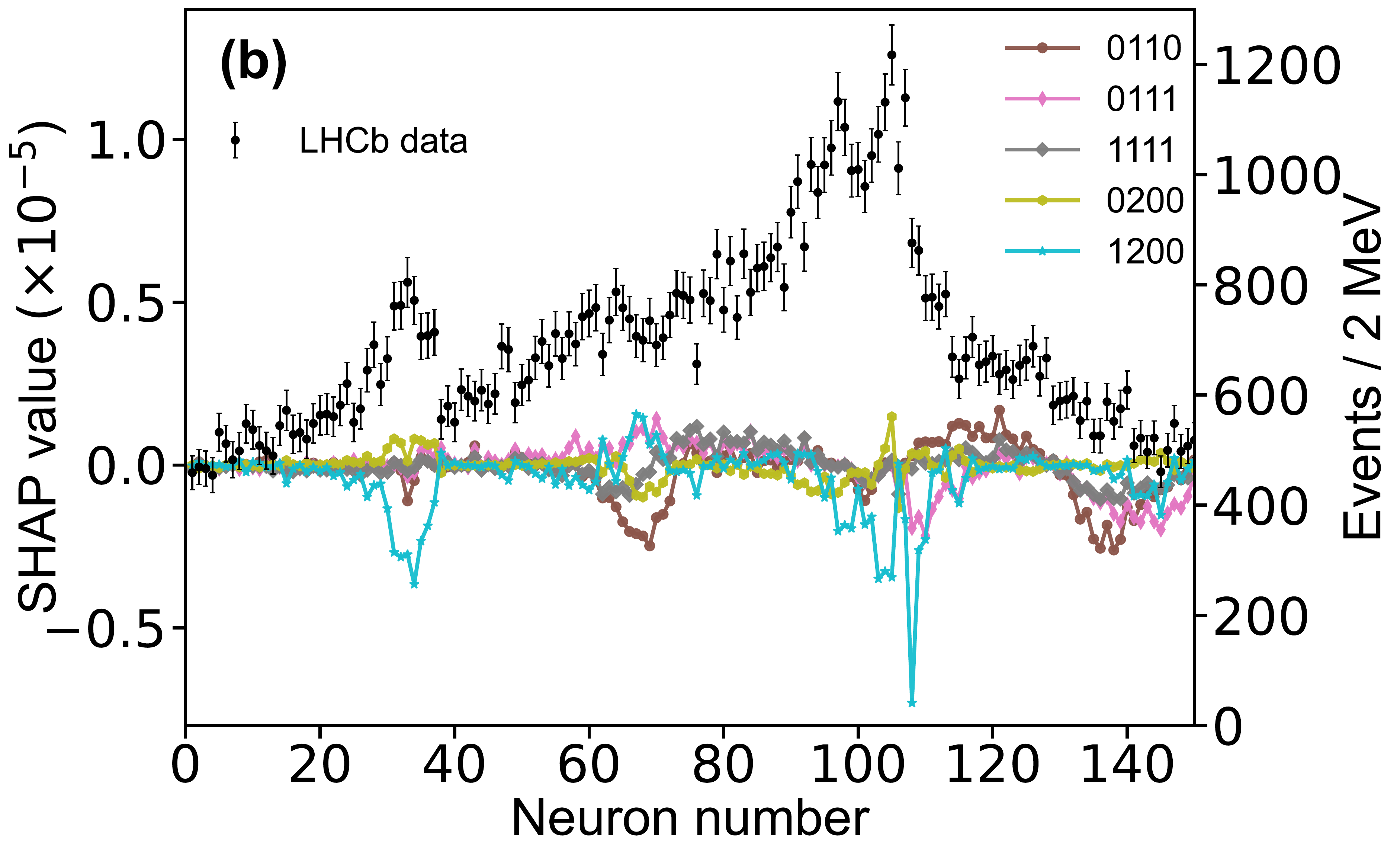}
    \includegraphics[width=0.32\textwidth]{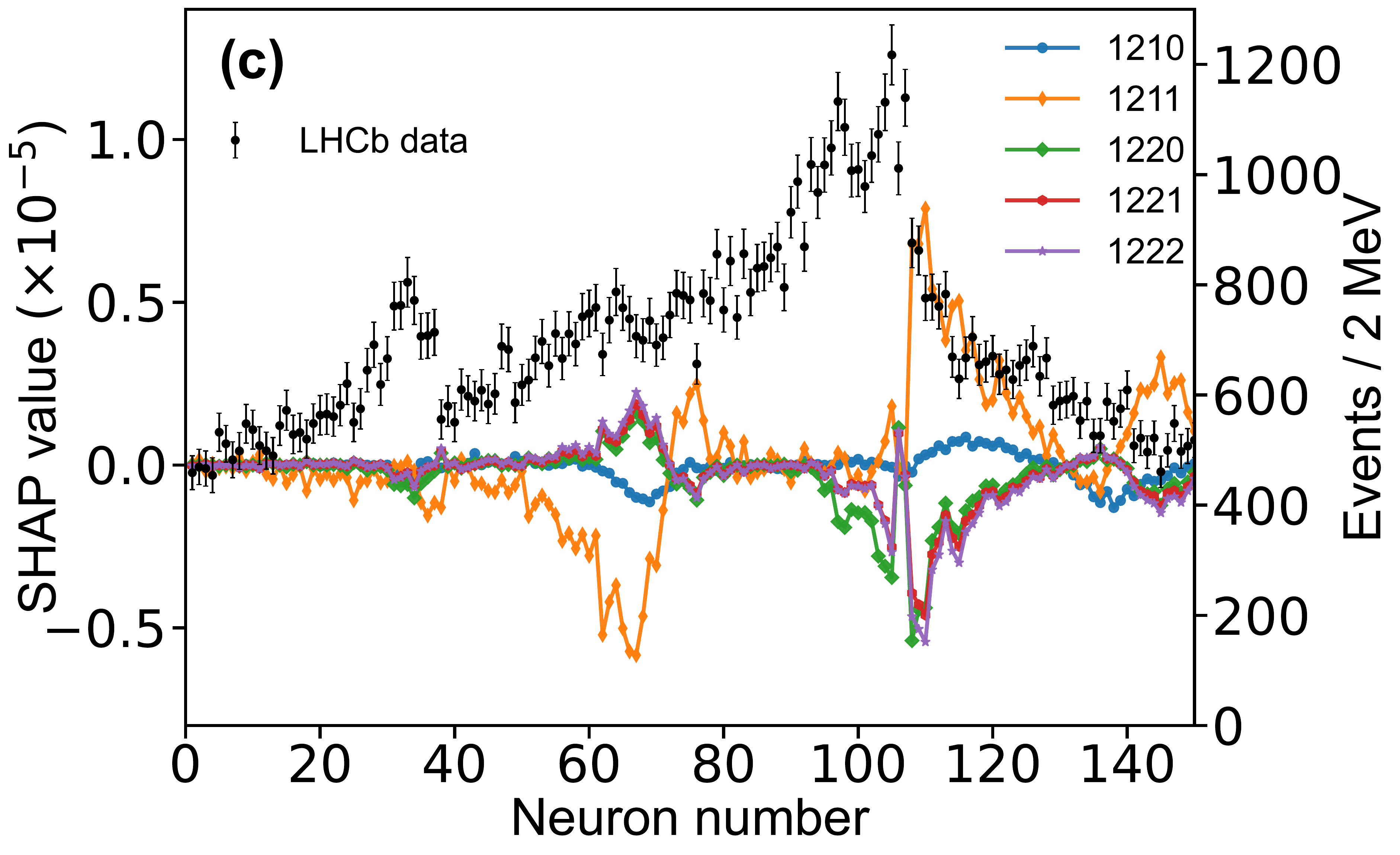}
    \caption{
    (a) Distribution of SHAP values (left axis) in the neurons. The points are the 150 experimental data. Different color lines correspond to five different labels. (b) For the other five different labels. (c) For the other five different labels.}
    \label{fig:shap_value}
\end{figure}
\begin{figure}[htbp]
    \centering
    \includegraphics[width=0.3\textwidth]{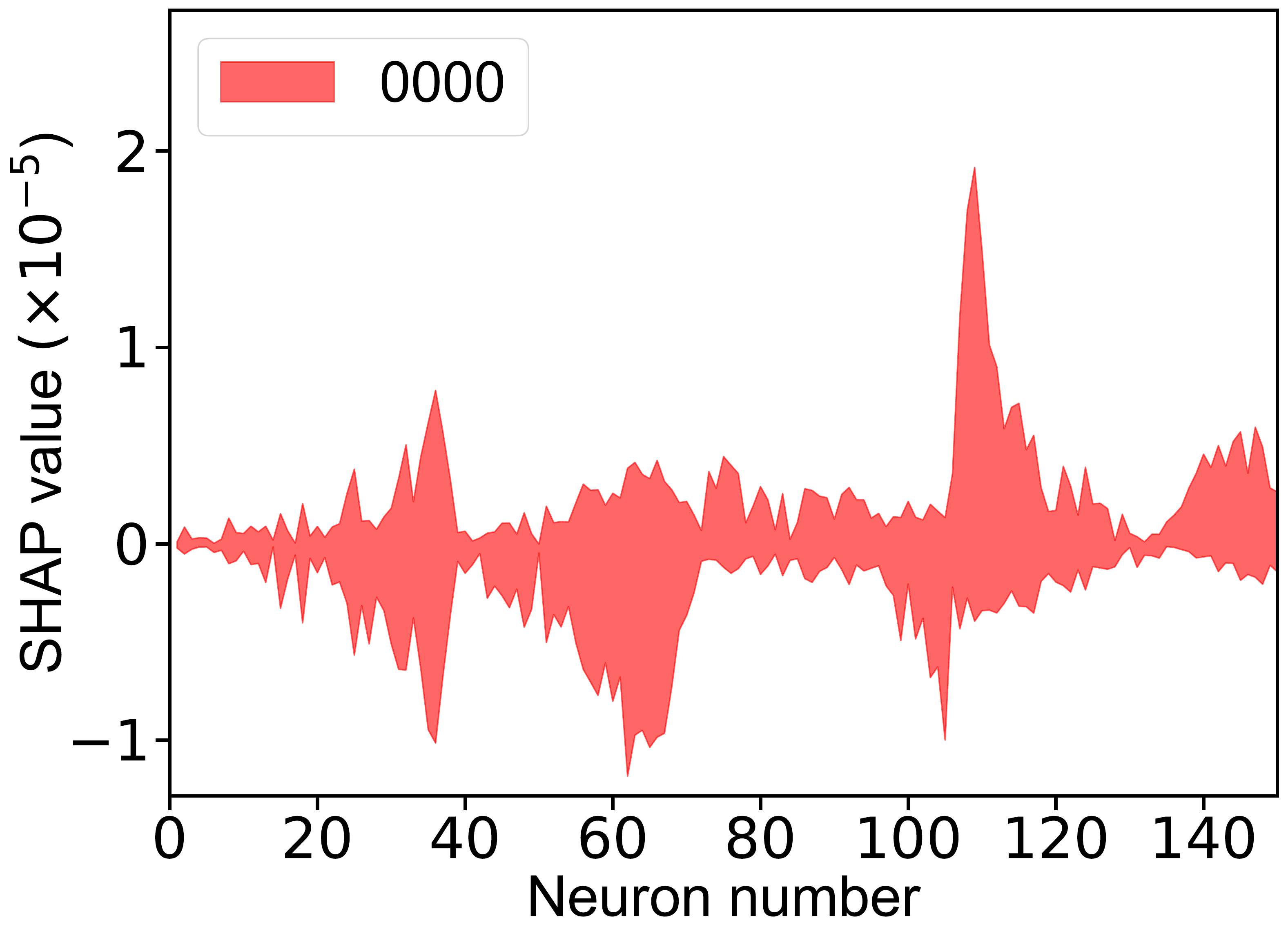}
    \includegraphics[width=0.3\textwidth]{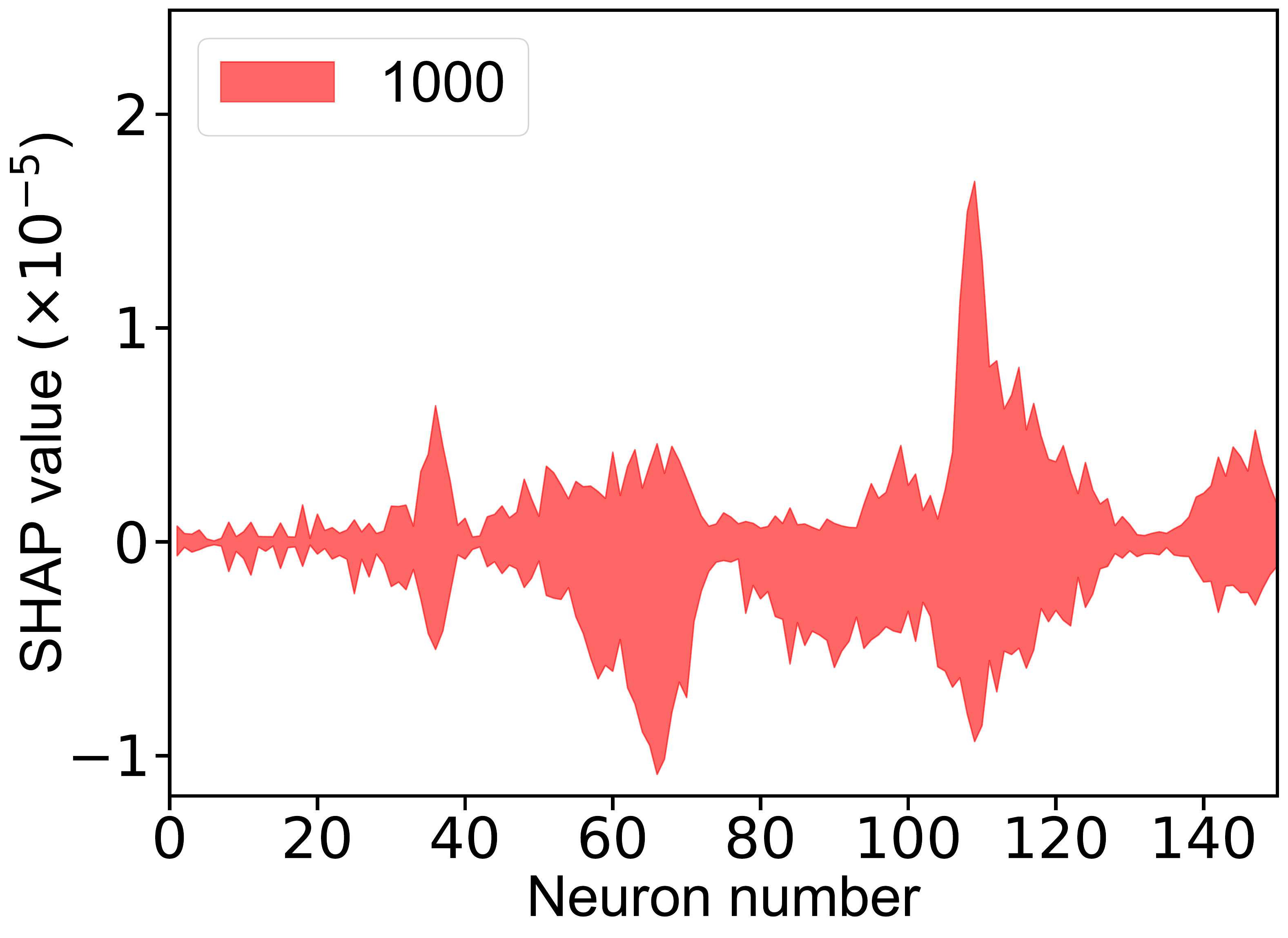}
    \includegraphics[width=0.3\textwidth]{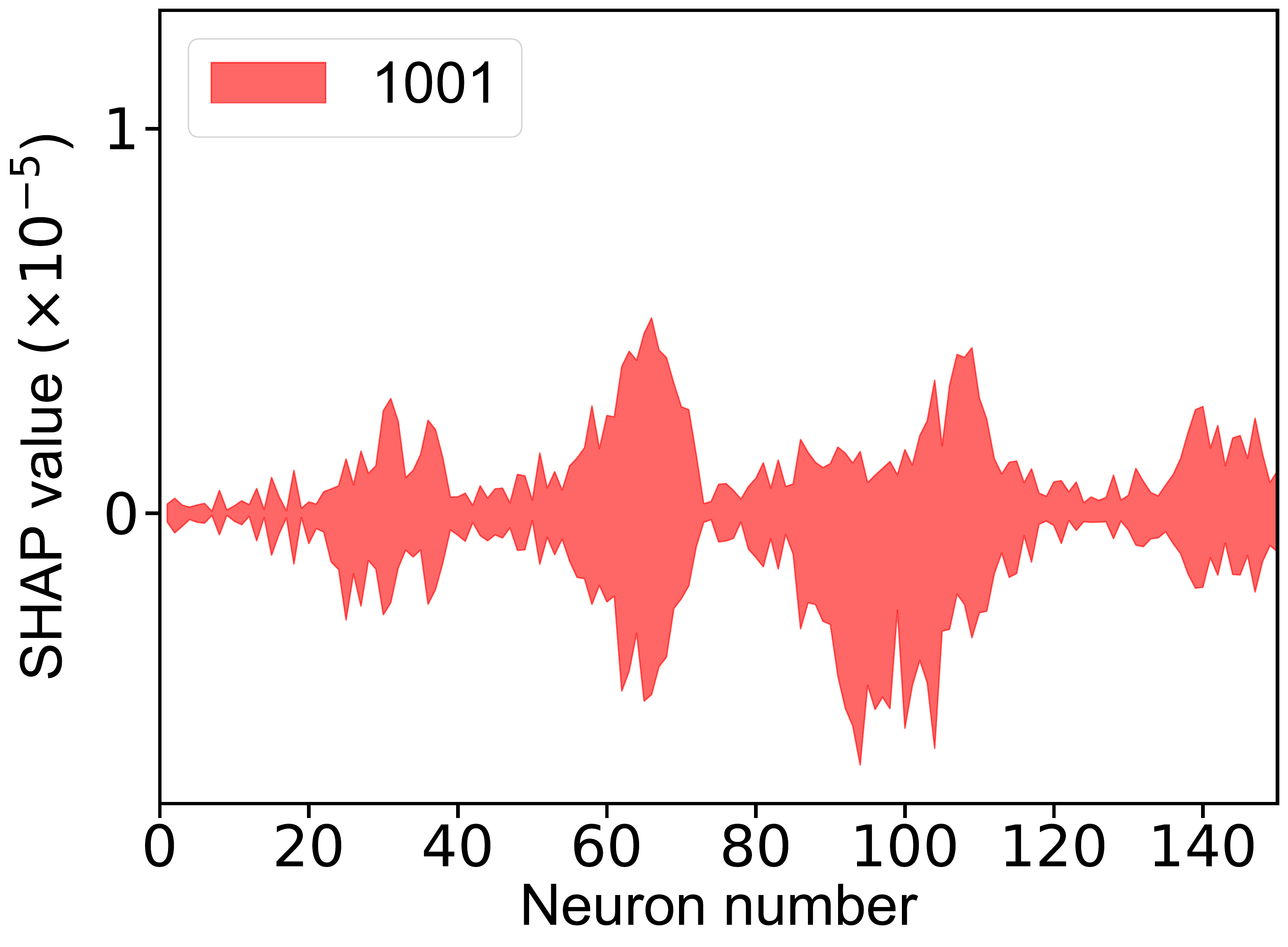}
    \includegraphics[width=0.3\textwidth]{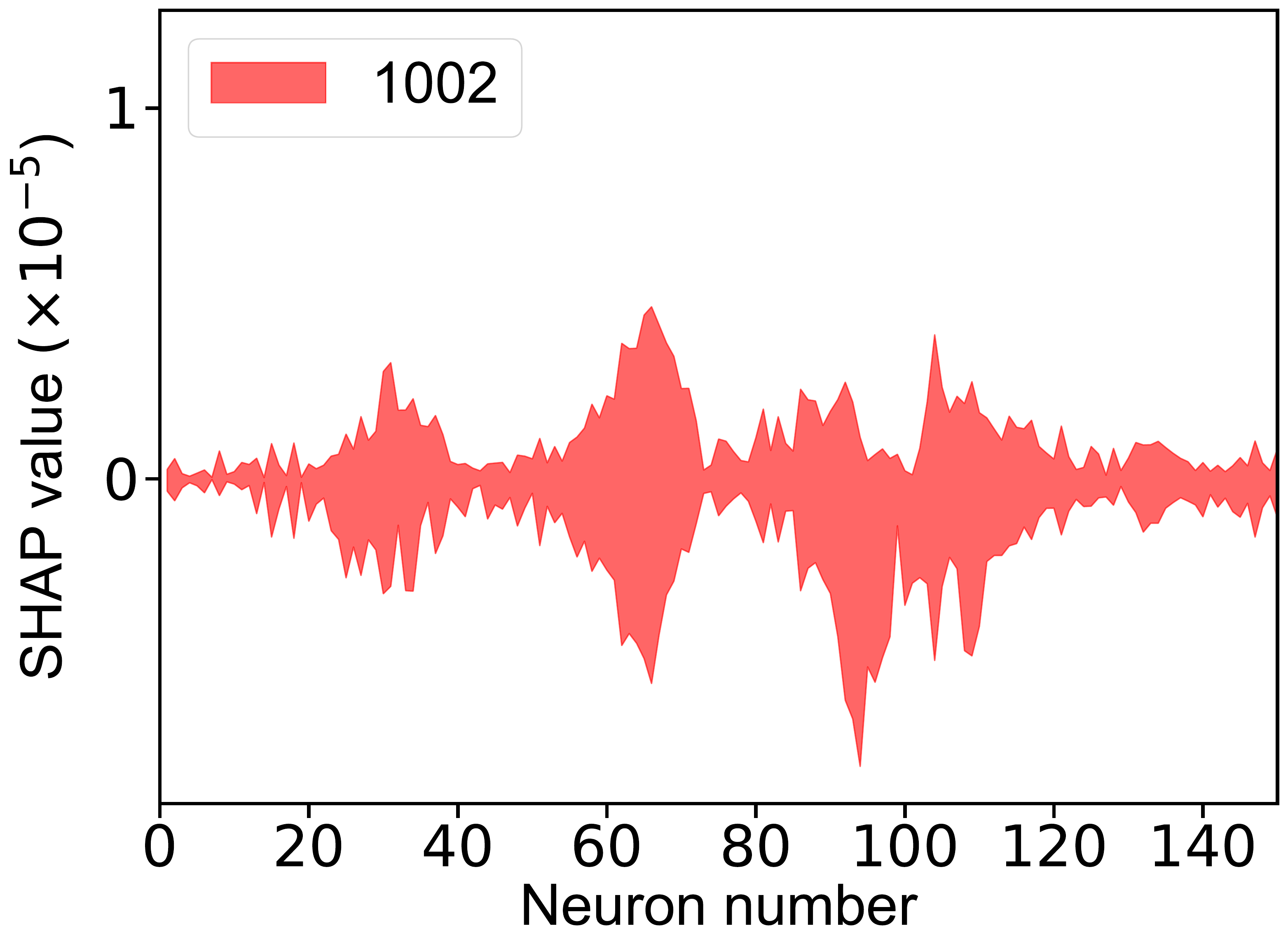}
    \includegraphics[width=0.3\textwidth]{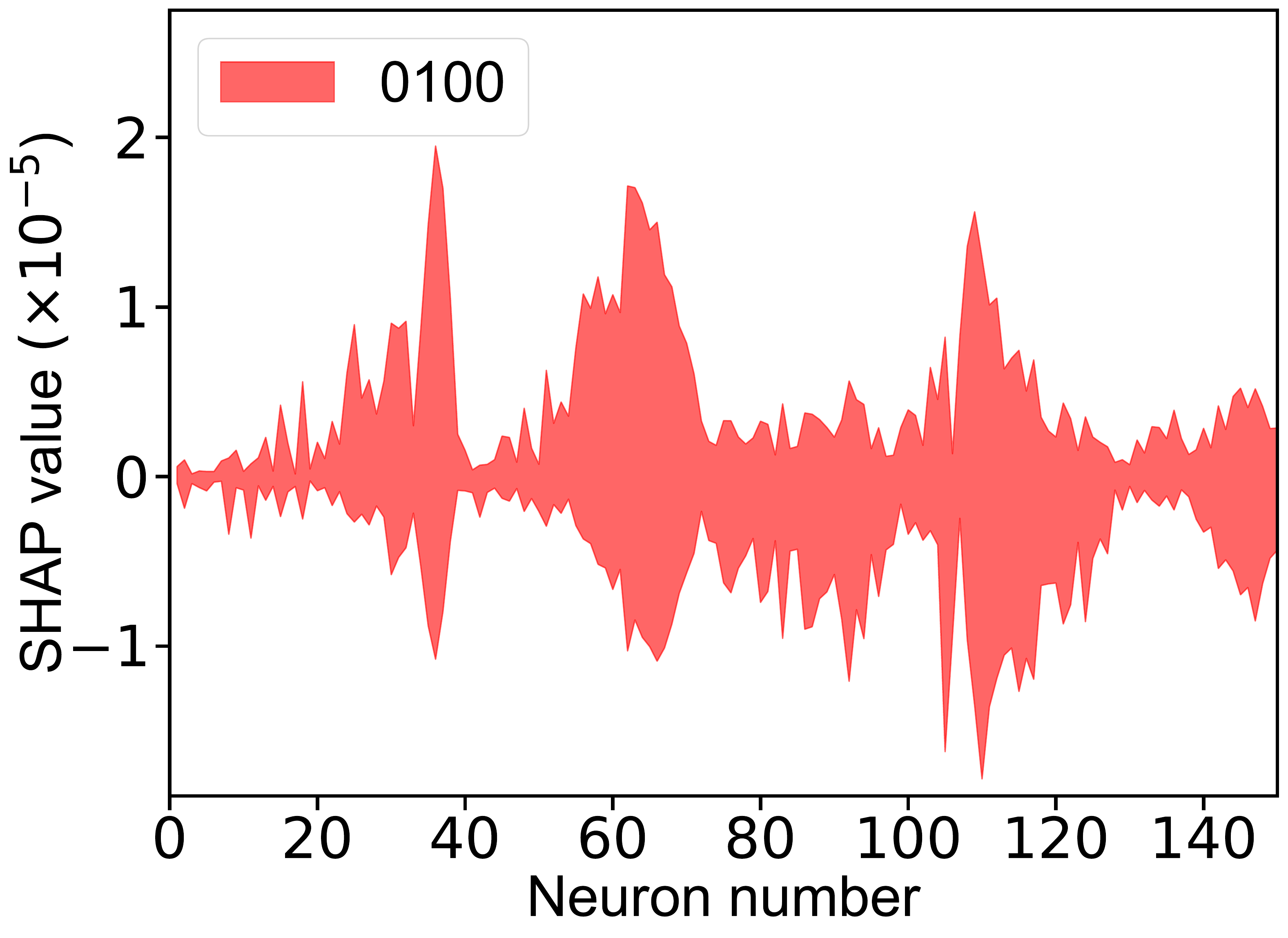}
    \includegraphics[width=0.3\textwidth]{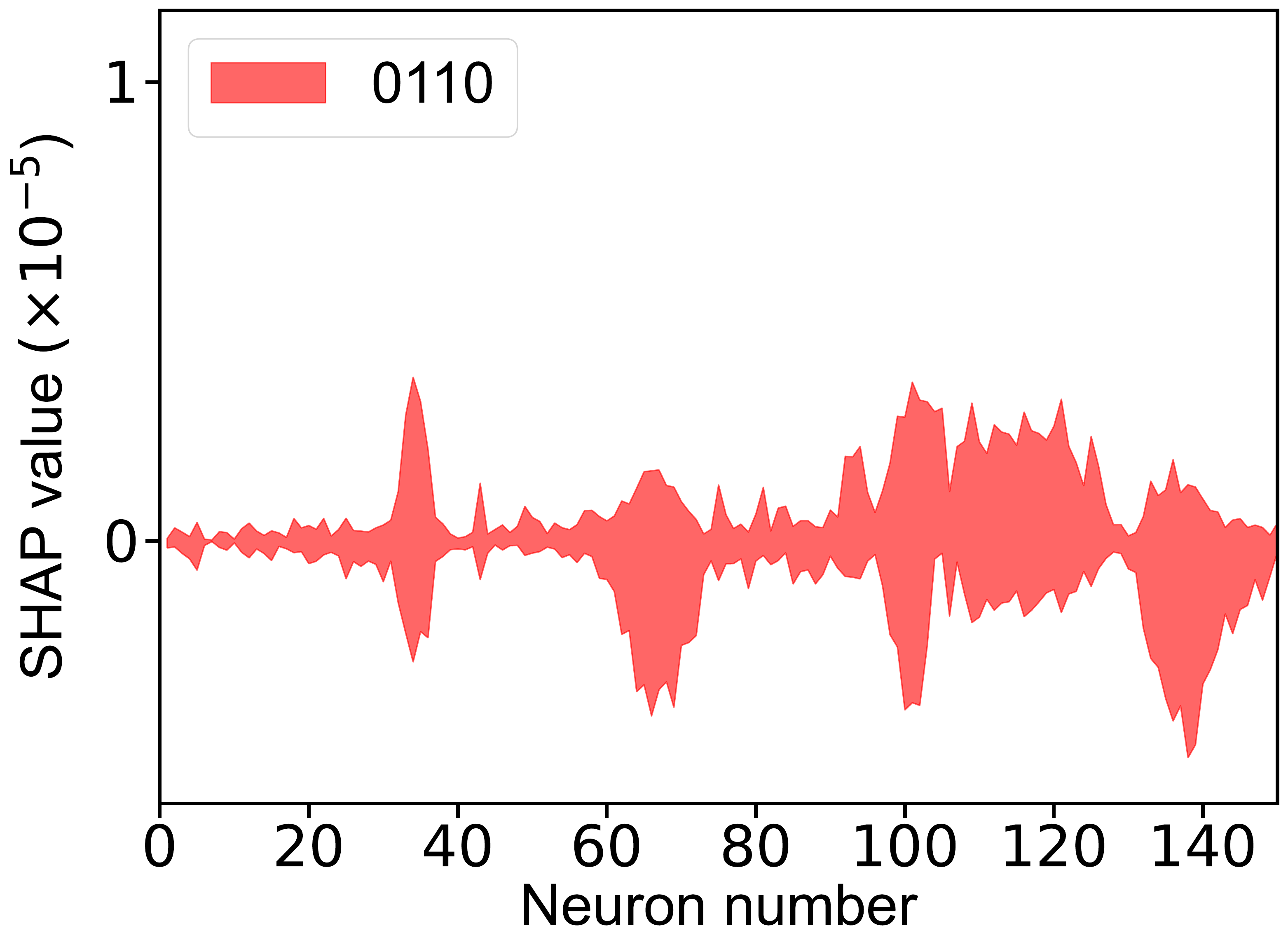}
    \includegraphics[width=0.3\textwidth]{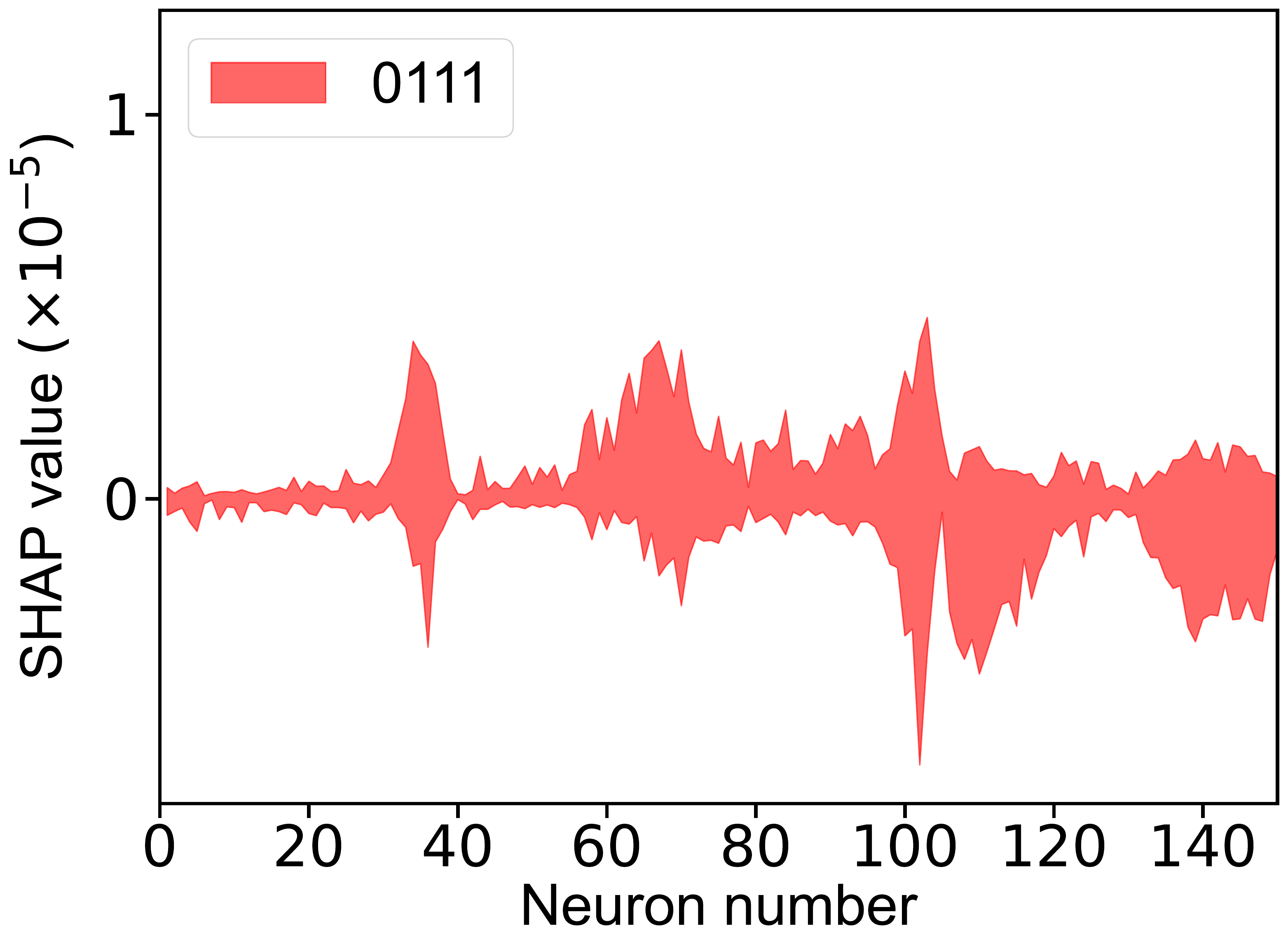}
    \includegraphics[width=0.3\textwidth]{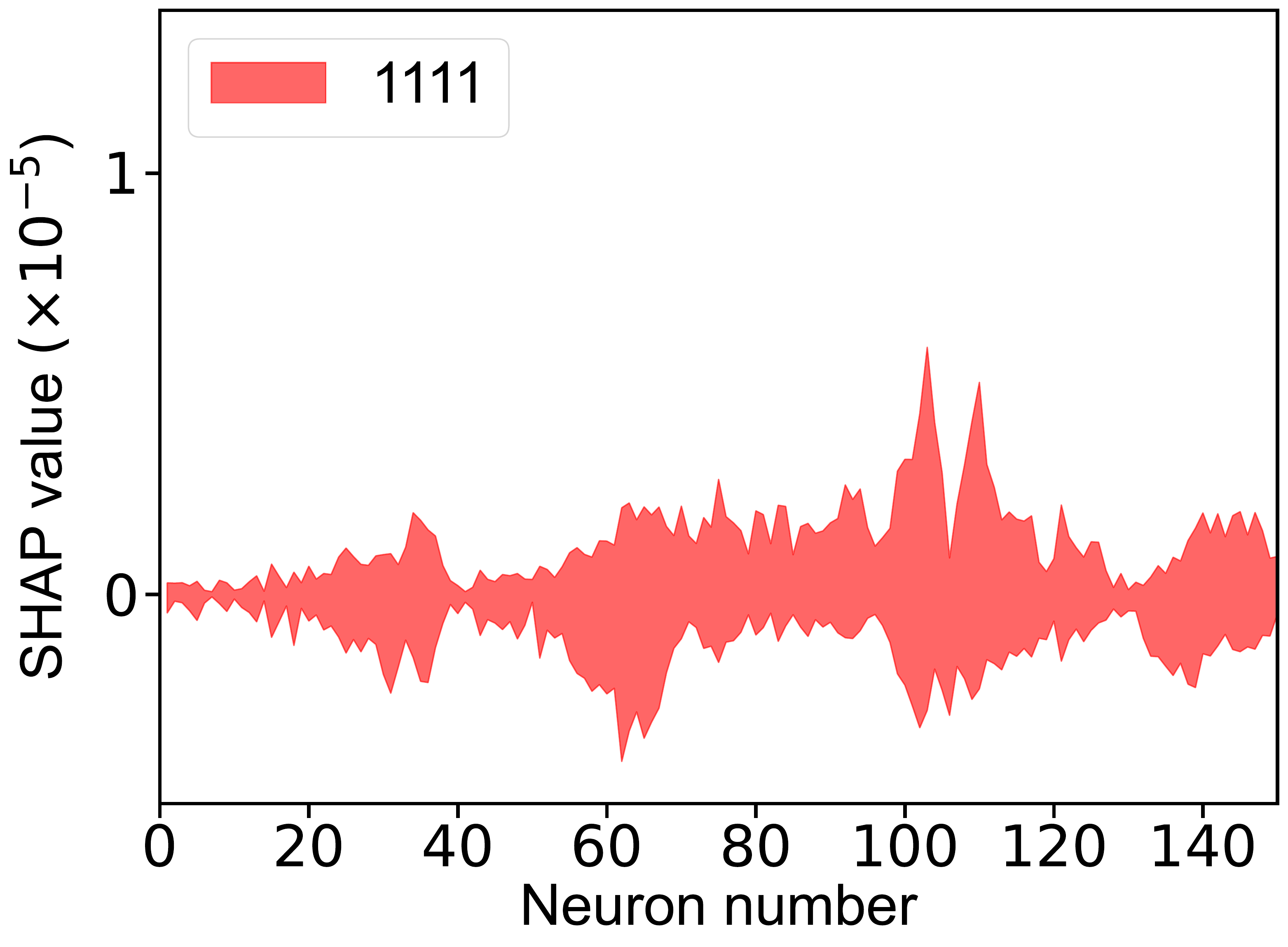}
    \includegraphics[width=0.3\textwidth]{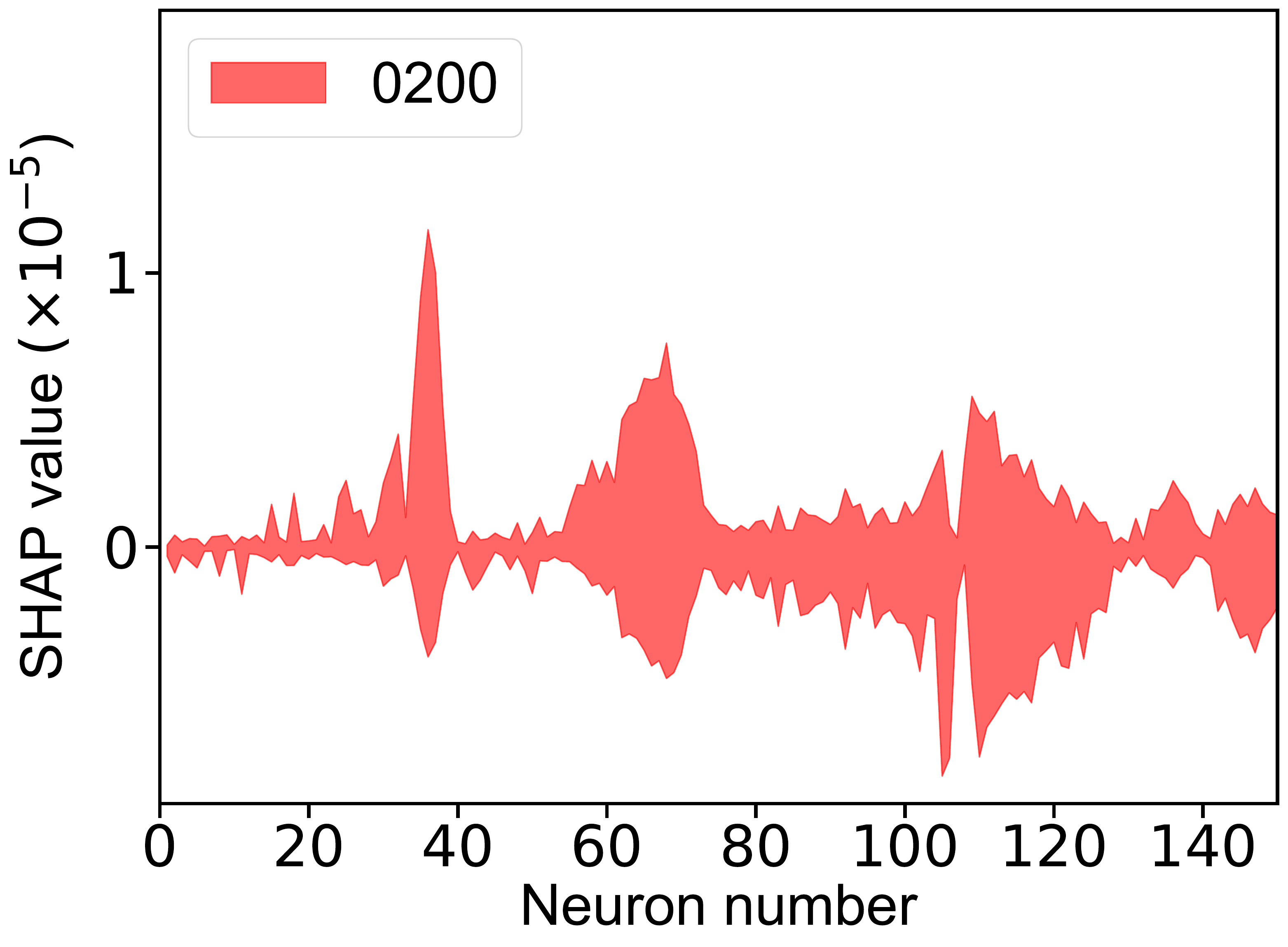}
    \includegraphics[width=0.3\textwidth]{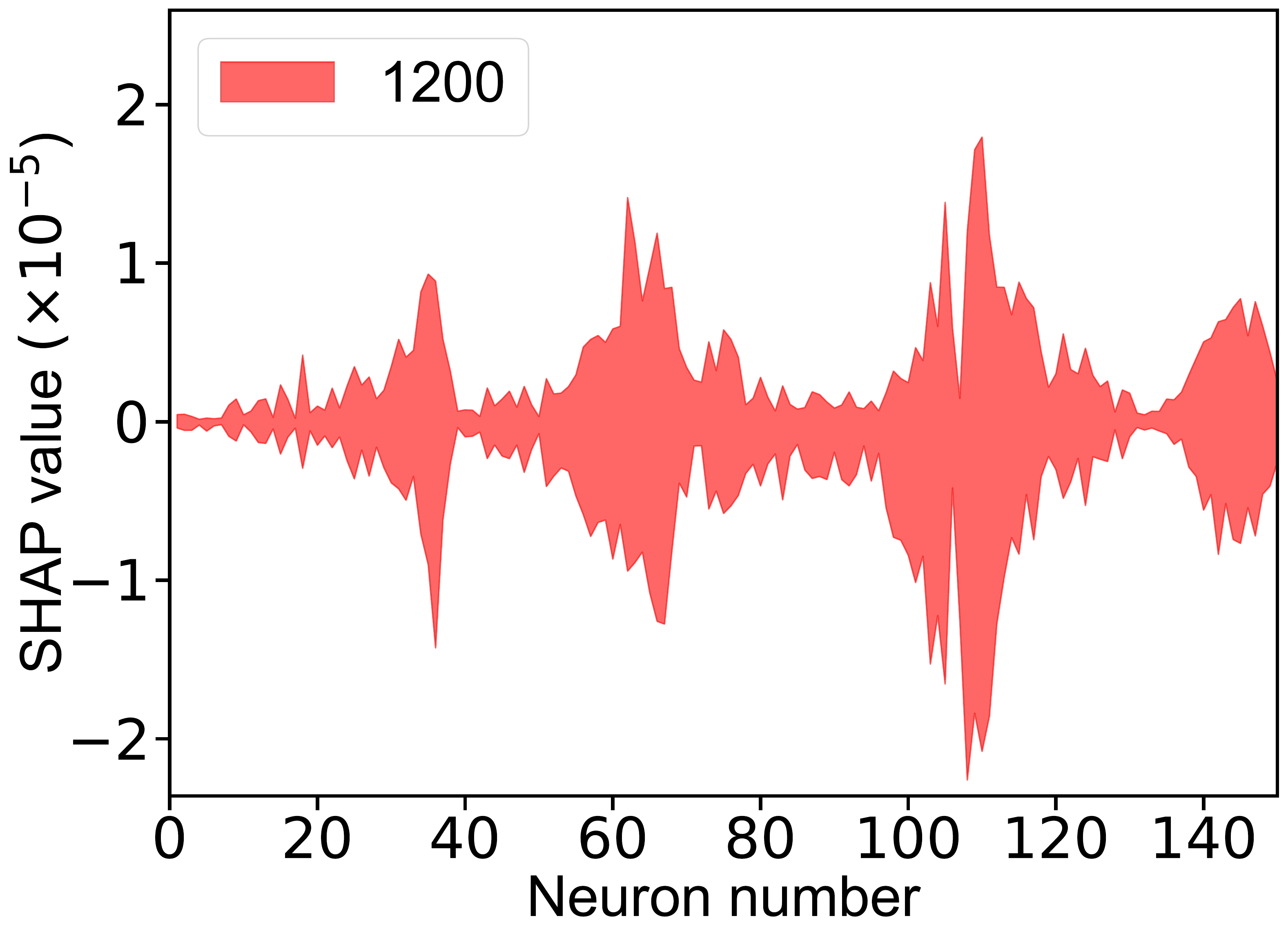}
    \includegraphics[width=0.3\textwidth]{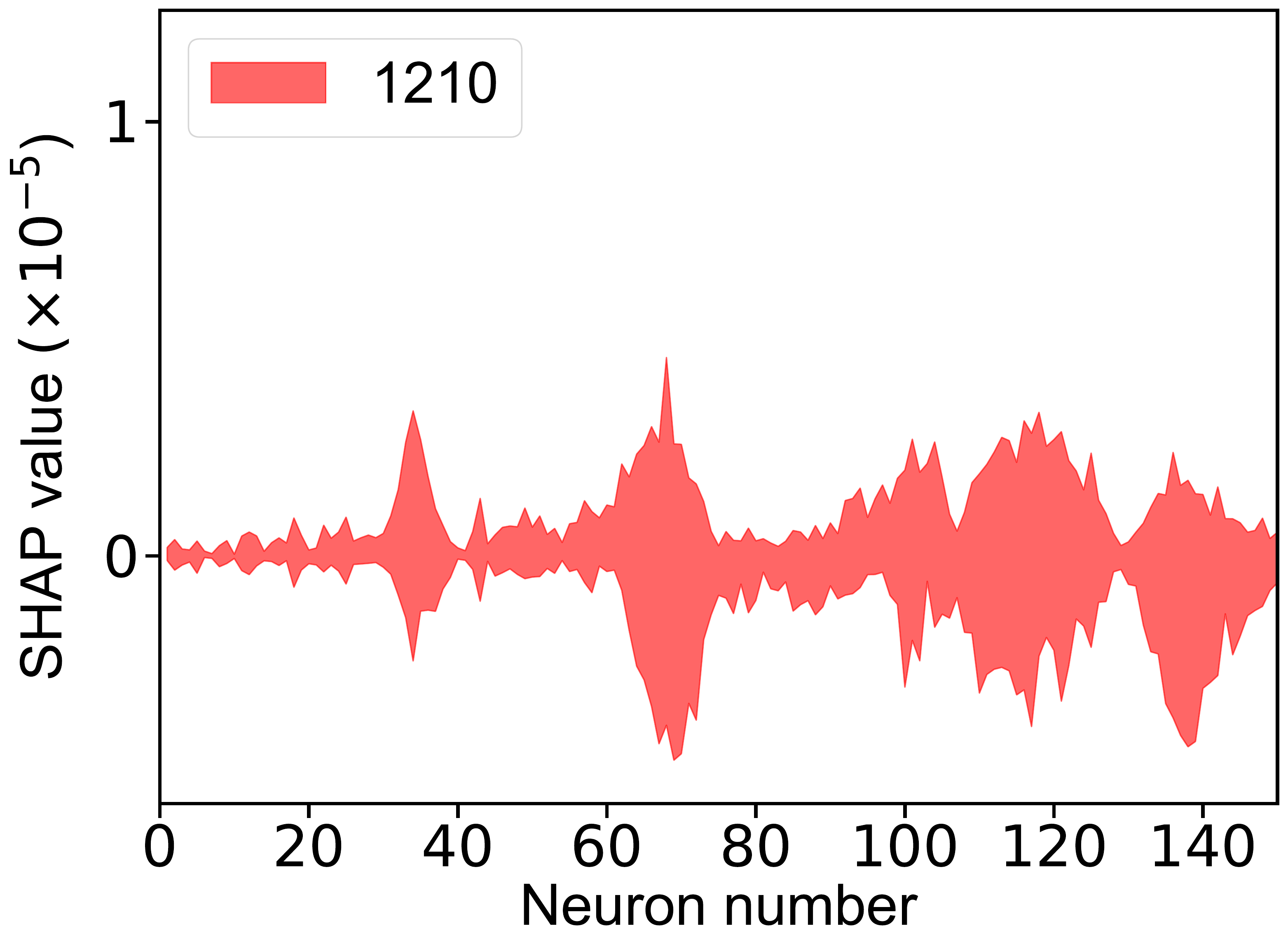}
    \includegraphics[width=0.3\textwidth]{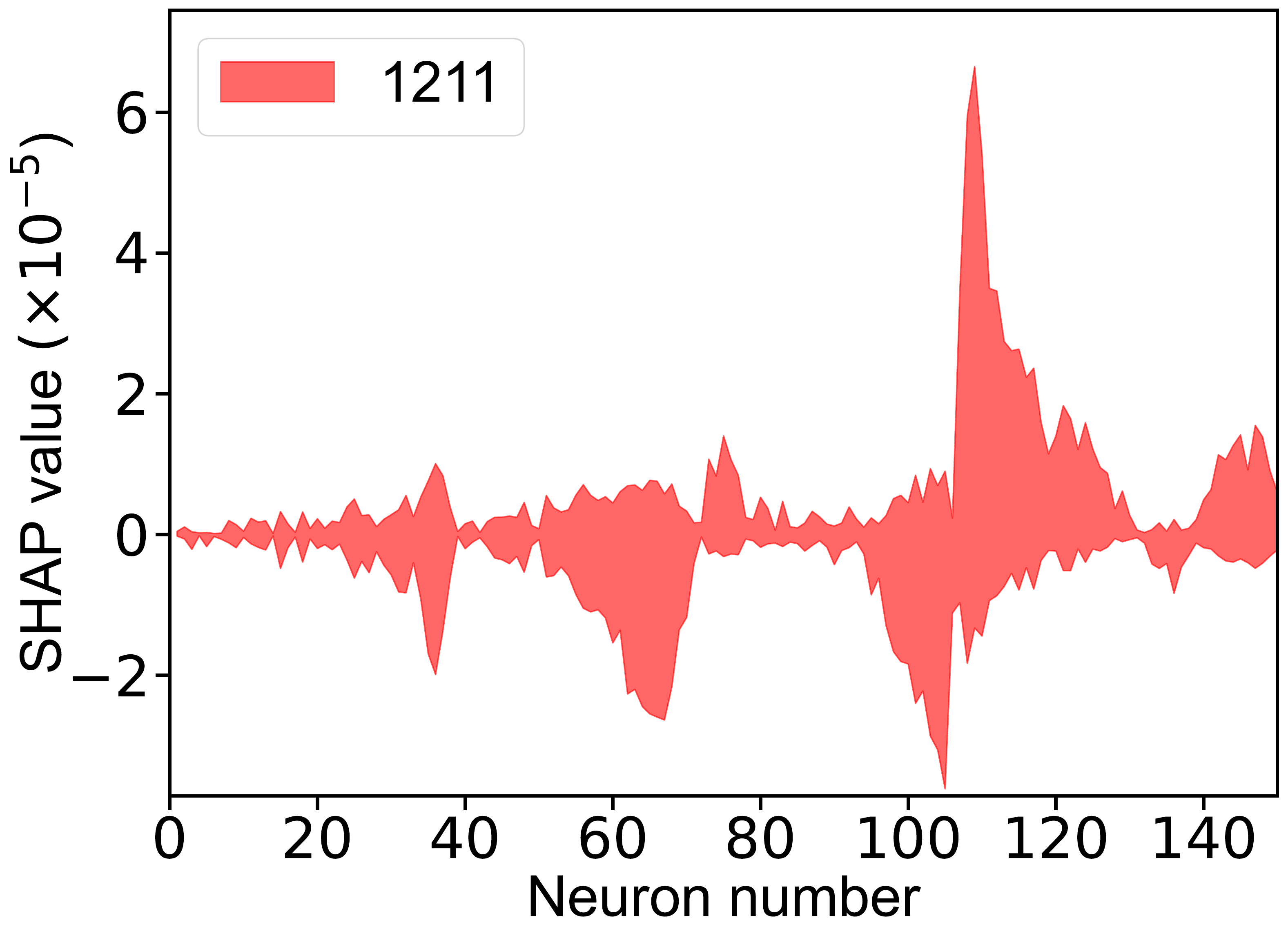}
    \includegraphics[width=0.3\textwidth]{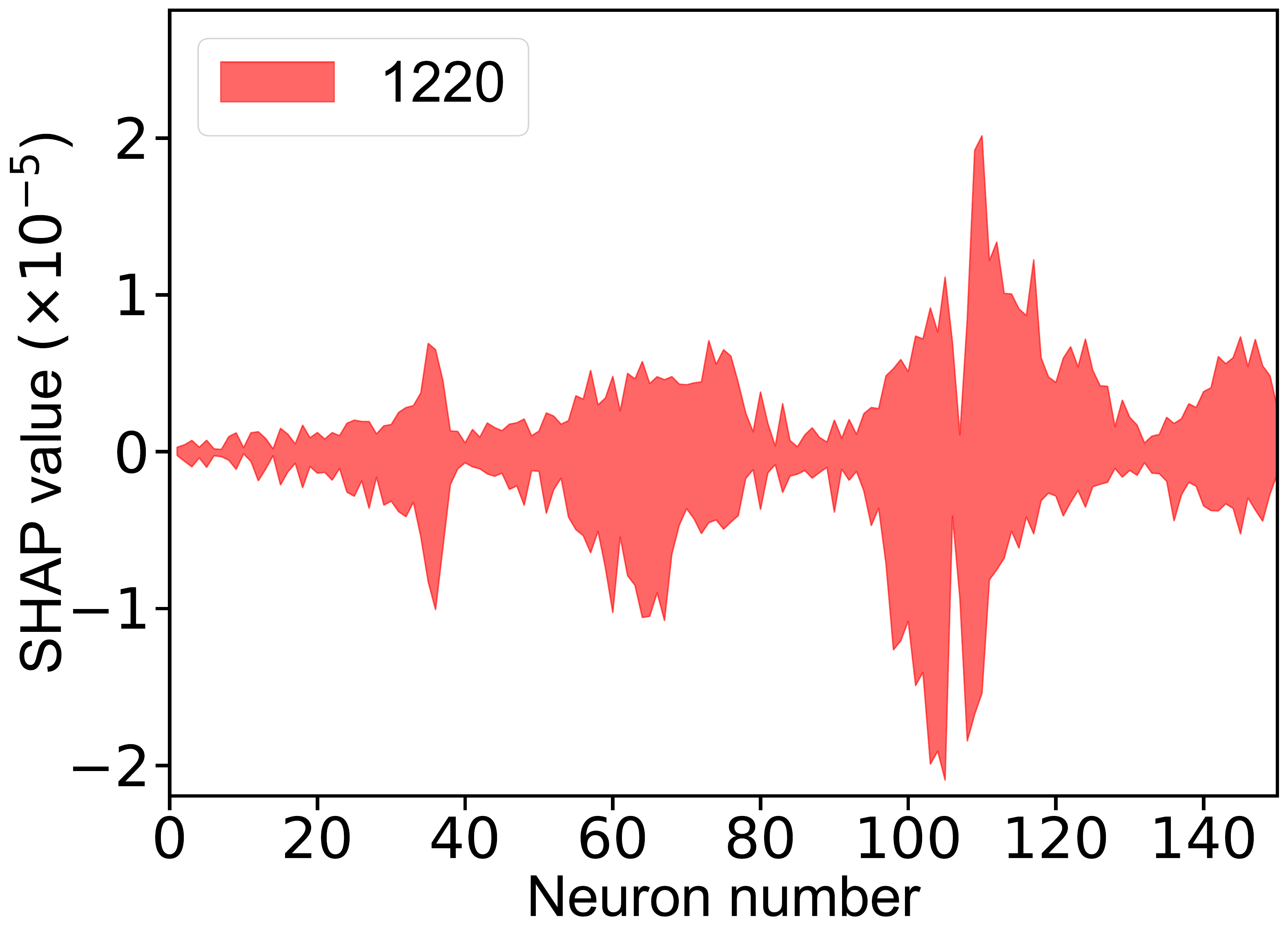}
    \includegraphics[width=0.3\textwidth]{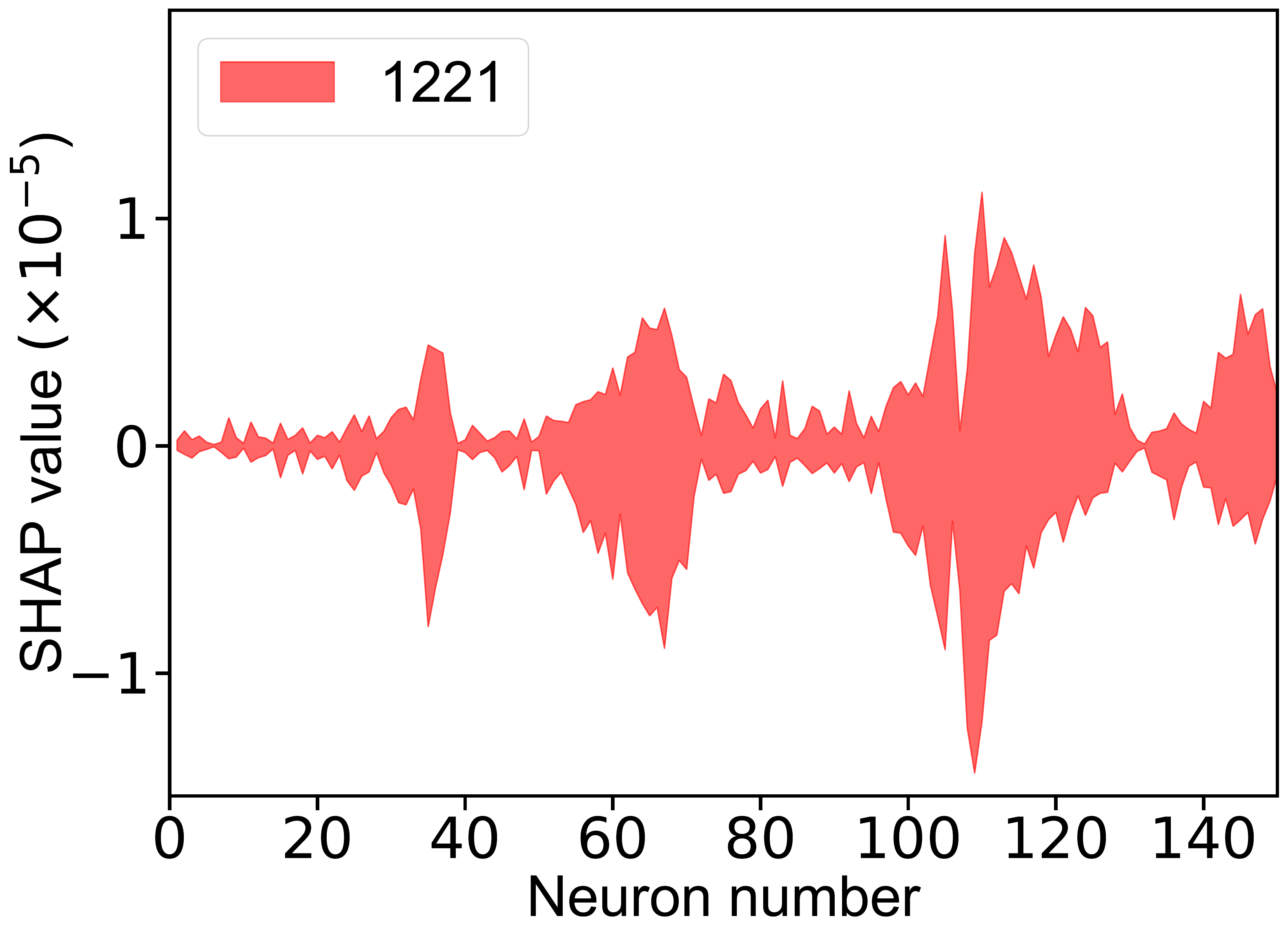}
    \includegraphics[width=0.3\textwidth]{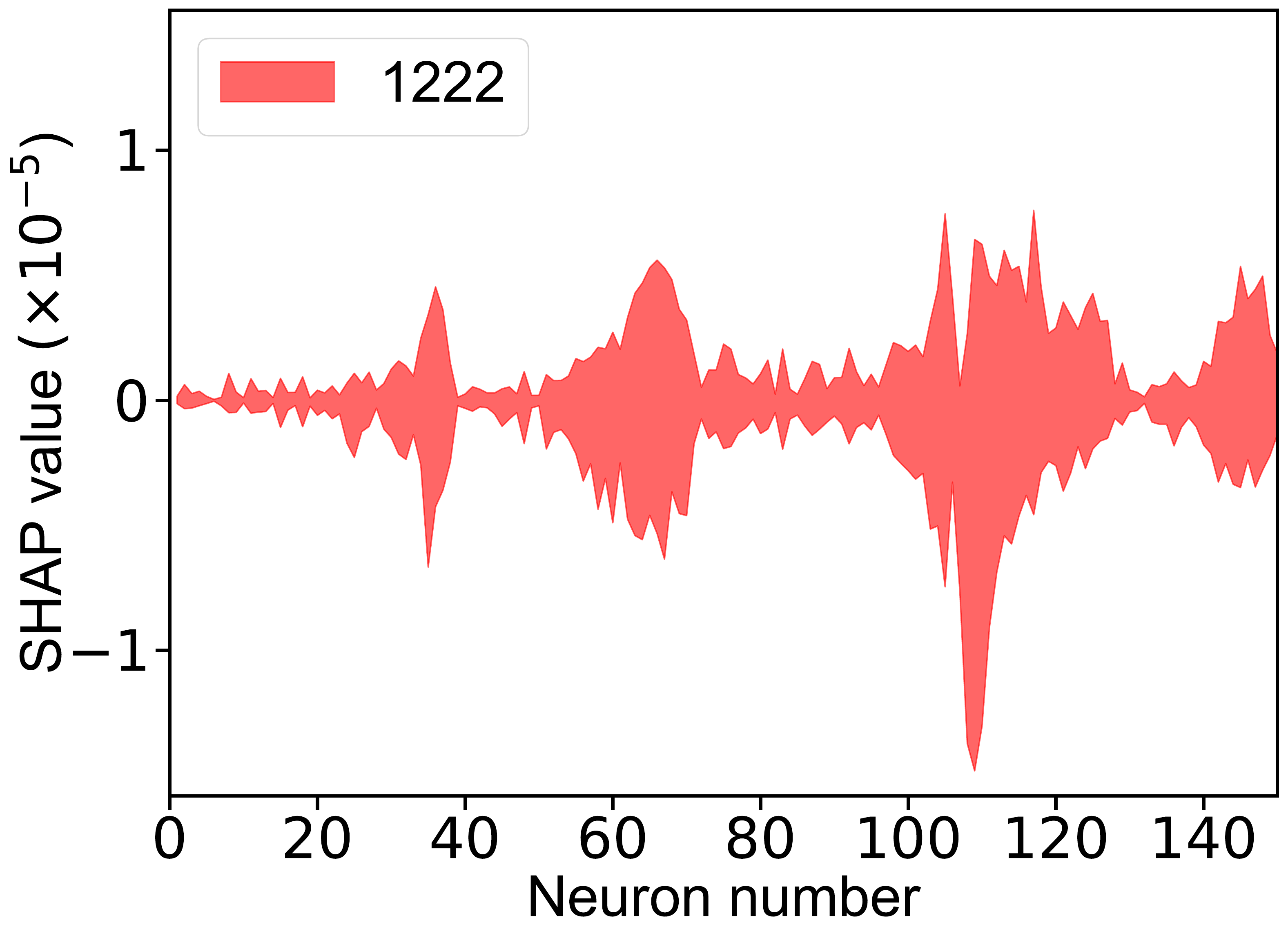}
    \caption{Distribution of SHAP values(left axis) in the neurons for the 15 different labels.}
    \label{fig:SHAP_value}
\end{figure}

\end{document}